\documentclass{article}
\usepackage{comment,lipsum}
\usepackage{multirow,subcaption}
\usepackage[font=small]{caption}

\usepackage{setspace}
\usepackage{amsmath,bm,amssymb}
\usepackage{wrapfig}    
\usepackage{enumitem}
\usepackage{xcolor,graphicx} 
\usepackage{braket}
\usepackage{float}
\usepackage{mathtools}
\usepackage[sort&compress,super,numbers,comma]{natbib}
\usepackage{threeparttable}
\usepackage{hyperref}
\hypersetup{colorlinks,breaklinks,linkcolor=Blue,urlcolor=Blue,anchorcolor=Blue,citecolor=Blue}
\usepackage{color}
\definecolor{DarkBlue}{rgb}{0.0,0.08,0.45}
\definecolor{Blue}{rgb}{0.0,0.0,1.0}
\definecolor{Red}{rgb}{1.0,0.0,0.0}
\definecolor{RedOrange}{rgb}{0.9,0.0,0.2}
\definecolor{dgrn}{RGB}{0,150,0}
\definecolor{dgray}{gray}{0.3}
\usepackage{fullpage}
\usepackage{titlesec}
\titleformat{\section}{\normalfont\bfseries}{\thesection}{1em}{}
\titleformat{\subsection}{\normalfont\bfseries}{\thesubsection}{1em}{}
\titleformat{\subsubsection}{\normalfont\itshape}{\thesubsubsection}{1em}{}

\newcommand{\fig}[2]{\scalebox{#1}{\includegraphics{#2}}}
\newcommand{\fullwidthfig}[1]{\includegraphics[width=6in]{#1}}

\usepackage{dcolumn}
\newcolumntype{.}{D{.}{.}{-1}}

\newcommand{\eg}{\textit{e.g.}}
\newcommand{\etal}{\textit{et~al.}}
\newcommand{\ie}{\textit{i.e.}}
\newcommand{\fns}{\footnotesize}
\newcommand{\mc}[3]{\multicolumn{#1}{#2}{#3}}
\newcommand{\nbd}{\nobreakdash}
\newcommand{\ra}{\rightarrow}
\newcommand{\tworow}[1]{\multirow{2}{*}{#1}}

\newcommand{\br}{\mathbf{r}}
\newcommand{\bR}{\mathbf{R}}
\newcommand{\cmplxi}{\text{i}}
\newcommand{\bhhlyp}{BH{\&}HLYP}

\newcommand{\aufbau}{\textit{aufbau}}
\newcommand{\cHFX}{\alpha_\text{hfx}}
\newcommand{\eval}[1]{\varepsilon_{#1}}
\newcommand{\wRSH}{\mu}

\newcommand{\eHOMO}{{\eval{}}^{}_\text{HOMO}}
\newcommand{\eLUMO}{{\eval{}}^{}_\text{LUMO}}

\newcommand{\xc}{\text{xc}}
\newcommand{\fxc}[1]{\hat{f}_\xc^{#1}}
\newcommand{\fxcker}[1]{f_\xc^{#1}}
\newcommand{\kxc}[1]{\hat{\kappa}_\xc^{#1}}
\newcommand{\kxcker}[1]{\kappa_\xc^{#1}}
\newcommand{\vxc}[1]{\hat{v}_\xc^{#1}}
\newcommand{\vxcpot}[1]{v_\xc^{#1}}

\newcommand{\nocc}{n_\text{occ}}
\newcommand{\nvir}{n_\text{vir}}
\newcommand{\nroots}{n_\text{roots}}
\newcommand{\nbasis}{n_\text{basis}}

\newcommand{\one}{{(1)}}
\newcommand{\zero}{{(0)}}

\title{Density Functional Theory for Electronic Excited States}
\author{
	John M. Herbert\footnote{\href{mailto:herbert@chemistry.ohio-state.edu}{herbert@chemistry.ohio-state.edu}} 
	\\
	\em Department of Chemistry \& Biochemistry \\
	\em The Ohio State University, Columbus, Ohio, 43210~USA \\
}
\date{}

\begin{document}

\maketitle

\begin{abstract}\noindent
This chapter provides a basic introduction to excited-state extensions of density functional theory (DFT), including 
time-dependent (TD-)DFT in both its linear-response and its explicitly time-dependent formulations.   As applied to the Kohn-Sham DFT ground state,
linear-response theory affords an eigenvalue-type problem for the excitation energies in a basis 
of singly-excited Slater determinants, and is widely known
simply as ``TDDFT'' despite its frequency-domain formulation.   This form of TDDFT is the mostly widely-used quantum-chemical 
method for excited states, due to a favorable combination of low cost and reasonable accuracy.  The chapter surveys the 
accuracy of linear-response TDDFT, which is generally more sensitive to the details of the exchange-correlation functional as compared to ground-state DFT, and
also describes some known systemic problems exhibited by this approach.
Some of those problems can be corrected on a case-by-case basis using orbital-optimized, excited-state 
self-consistent field (SCF) calculations, in what is known as excited-state Kohn-Sham theory or 
a ``$\Delta$SCF'' procedure, a class of methods that includes restricted open-shell Kohn-Sham theory. 
Recent successes of these approaches are highlighted, including double excitations and core-level excitations.
Finally, explicitly time-dependent (or ``real-time'') TDDFT involves propagation of the molecular orbitals in time
following an external perturbation, according to the 
Kohn-Sham analogue of the time-dependent Schr\"odinger equation.  The time-dependent approach has been used to model 
strong-field electron dynamics, and in the weak-field limit it provides a route to broadband spectra based on the time evolution of
the dipole moment function.  This is useful for describing high-energy excitations (as in x-ray spectroscopy) and in systems 
where the density of states is high, as demonstrated by a few examples.
\end{abstract}

\tableofcontents

\section{Overview}
\label{sec:Intro}
Following its implementation in molecular quantum chemistry codes in the early 
1990s,\cite{PopGilJoh92,GilJohPop92a,JohGilPop92,JohGilPop93c,MurLamHan92,LamHanAmo93}
density functional theory (DFT) quickly emerged as the most popular tool for ground-state electronic structure calculations
due to its favorable balance of relatively low cost with reasonable accuracy for thermochemistry.
The first excited-state implementations quickly followed,\cite{WeiAhlHas93,BauAhl96a,PetGosGro96,StrScuFri98,HirHea99b,GraPetGro00} 
based on a linear-response (LR) formalism\cite{Cas95a,Cas96,Fur01a} 
that mirrors much earlier work on time-dependent Hartree-Fock theory.\cite{McLBal64}
(The historical development of TDDFT has been summarized elsewhere.\cite{Mai16})
The LR formulation is now known almost universally as 
``time-dependent'' (TD-)DFT, despite its frequency-domain formulation
and implementation.   In its most pedestrian applications, LR-TDDFT produces vertical excitation energies for
closed-shell molecules at ground-state geometries to within a statistical accuracy of $\sim 0.3$~eV,\cite{LauJac13} 
at a cost that is often only a few times more than the cost of the ground-state self-consistent field (SCF) calculation 
and possessing the same formal scaling.\cite{FurRap05}
This is a useful accuracy for electronic absorption spectra.   In view of its low cost, LR-TDDFT has therefore become 
the \textit{de facto} method for computing electronic excitation spectra of finite molecular systems, although some fundamental problems remain in its
application to periodic materials.\cite{HirHeaBar99,Cas09,UllYan16}   
LR-TDDFT is also becoming increasingly popular for photochemical applications,\cite{TapBelVin13,BarCre16,ParRoyFur19} 
despite some problems with the description of conical intersections.\cite{ZhaHer21,HerMan22,HuiNikThi16}  
In part, this popularity is due to a growing recognition that complete active-space (CAS\nbd-)SCF methods cannot be considered
quantitative approaches for excited-state dynamics,\cite{GozHunSch12,GozMelLin13,GozMelVal14,ParLeeHui21} 
due to a lack of dynamical electron correlation effects.  


This chapter provides an overview of TDDFT and other DFT-based methods for computing excitation spectra,
excited-state properties, and for simulating photochemical reactions, emphasizing theory rather than applications but with some molecular
examples to motivate the discussion. For those unfamiliar with the formal underpinnings of TDDFT,
a natural question to ask is ``what does {\em time\/} have to do with excitation energies?''  In fact, one knows from basic quantum 
mechanics that the time evolution of a non-stationary wave function encodes the system's excitation energies 
via the Bohr frequencies, $\omega_{jk} = (E_j - E_k)/\hbar$, therefore in principle 
the time evolution of a quantum system can be used to extract excitation energies.  
The existence of a time-dependent extension of DFT is formally justified by 
the Runge-Gross theorem,\cite{RunGro84,GroUllGos95,van99,GroMai12,Rugvan12,RugPenvan15,BaeKro18} 
which provides something akin to a time-dependent extension of 
the first Hohenberg-Kohn theorem for the ground state,\cite{KocHol01} \ie, a density-to-potential mapping.
In the time-dependent case there are important caveats about initial-state dependence and memory effects.\cite{GroBur06,Mai06,Mai12} 
Those issues have yet to be fully resolved in a computationally feasible way, but this 
has not stymied the practical development and application of TDDFT.

Following a perturbation to the ground state, which creates a superposition of energy eigenstates, the Fourier
components of the time-dependent dipole moment are precisely the Bohr frequencies.  A Fourier transform of
the dipole moment function is itself an absorption spectrum,\cite{McH99} 
\begin{equation}\label{eq:I(w)}
	I(\omega)  = \frac{1}{2\pi}\int_{-\infty}^{+\infty} \big\langle \bm{\mu}(0)\bm{\cdot}\bm{\mu}(t)\big\rangle e^{-\cmplxi\omega t} dt \; .
\end{equation}
Excitation energies are also encoded in the isotropic 
frequency-dependent polarizability, $\alpha(\omega)$, which has a sum-over-states expression 
\begin{equation}\label{eq:alpha(omega)} 
	\alpha(\omega) = \frac{e^2}{m_e} \sum_{n>0} \frac{f_{0n}}{\omega_{n0}^2 - \omega^2}
\end{equation}
where $m_e$ is the electron mass, $\omega_{n0} = (E_n - E_0)/\hbar$, and 
\begin{equation}\label{eq:f0n}
	f_{0n}= \left(\frac{2m_e\omega_{n0}}{3e^2\hbar}\right) |\langle 0 | \hat{\bm{\mu}} | n \rangle|^2 
\end{equation}	
is the dimensionless dipole oscillator strength for the $|0\rangle \ra |n\rangle$ transition.\cite{McH99}
The poles of response function $\alpha(\omega)$ 
therefore encode excitation energies, with residues that encode oscillator strengths.\cite{FurRap05}
In the early days of quantum chemistry, Eq.~\eqref{eq:alpha(omega)} was actually used to compute excitation energies
for atoms and atomic ions,\cite{MoiMuk72,RoyGupMuk75} by computing $\alpha(\omega)$ as the response to an applied field,
and a version of this approach would eventually reappear in the form of ``real-time" TDDFT.\cite{YabBer99b,YabNakIwa06}
The poles of the Kohn-Sham response function also serve this purpose,\cite{ZanSov80,PetGosGro96,Ull09} and the LR formalism 
applied to the Kohn-Sham ground state turns this idea into a robust computational paradigm,
in the form of an eigenvalue-type problem for the excitation energies.\cite{Cas95a,Cas96,Fur01a}   
Although the LR formulation exists strictly in the frequency or energy domain, the time-dependent origins
of the phenomenology have persisted in the name ``TDDFT".


Despite its overwhelming popularity, LR-TDDFT excitation energies do tend to be
more sensitive to the details of the exchange-correlation (XC) functional as compared to ground-state properties computed with DFT.
In some sense, the statistical accuracy of $\sim 0.3$~eV that is quoted above should therefore be interpreted as representative of the 
best-case scenario with state-of-the-art 
functionals, and assuming that certain systemic pathologies can be avoided.   LR-TDDFT may not be the theory that we want, but it remains
the best theory that we have for excited states of large and even medium-size molecules.   This theory is introduced formally in 
Section~\ref{sec:TDDFT} and that discussion constitutes the most substantial part of this chapter, just as LR-TDDFT 
occupies the most significant place amongst excited-state DFT methods.  It holds that position because it is easy to use, not significantly more expensive
than ground-state DFT, and provides a slew of excited states in an automated way, starting from a ground-state SCF solution.

While the accuracy of LR-TDDFT is often quite reasonable, certain systemic problems have been identified and 
excited-state Kohn-Sham procedures have been developed to circumvent these.   Rather than applying LR to the ground state, 
these methods look for an excited-state (non-\aufbau) solution to the SCF equations, and for this reason the excited-state Kohn-Sham approach
is often called a ``$\Delta$SCF'' method.  Although not formally justified by the Runge-Gross theorem, the $\Delta$SCF approach
has an admirable record of rectifying the deficiencies of LR-TDDFT, again at a cost comparable to that of a ground-state DFT calculation.
What is lost in the $\Delta$SCF approach is the ability to compute a whole spectrum of states at once, making the 
state-specific $\Delta$SCF procedure much more labor-intensive for the user.    This approach is described in Section~\ref{sec:DeltaSCF}.

Finally, it is possible to take the time dependence in TDDFT at face value and to propagate Kohn-Sham molecular orbitals (MOs) in time, following a 
perturbation applied to the ground-state density.    This is accomplished by solving the time-dependent Kohn-Sham (TDKS) equation, 
\begin{equation}\label{eq:TDKS}
	\cmplxi\hbar\frac{d}{dt}\psi_{k\sigma}(\br,t) = \hat{F}_\sigma\psi_{k\sigma}(\br,t)
\end{equation}
which is a one-electron analogue of the time-dependent Schr\"odinger equation.   (Here, $\sigma \in\{\alpha,\beta\}$ is a spin index.) 
The one-electron effective Hamiltonian in Eq.~\eqref{eq:TDKS} is the Fock operator $\hat{F}_\sigma$ that comes from the 
ground-state Kohn-Sham eigenvalue problem that determines the MOs:
\begin{equation}\label{eq:KS-SCF}
	\hat{F}_\sigma\psi_{k\sigma}(\br) = \eval{k}\psi_{k\sigma}(\br) \; .
\end{equation}
The ``real-time'' approach to TDDFT,\cite{ProIsb16,LiGovIsb20}
which is described in Section~\ref{sec:TDKS},
consists in solving Eq.~\eqref{eq:TDKS} by propagating the MOs in time following a perturbation to the ground-state 
that creates a time-evolving density,
\begin{equation}\label{eq:rho(r,t)}
	\rho_\sigma(\br,t) = \sum_{k}^\text{occ} \big| \psi_{k\sigma}(\br,t) \big|^2 \; ,
\end{equation}
expressed here for $\sigma$-spin electrons.   (The total charge density is $\rho = \rho_\alpha + \rho_\beta$.)
This approach can be used to simulate strong-field electron dynamics,\cite{UllBan12}
which is a topic of contemporary interest in attosecond molecular science.\cite{Lep12,NisDecCal17,PalMar20,RamLeoNeu16} 
It also provides a route to broadband spectra via Fourier transform of the time-dependent dipole moment function, in a direct
realization of Eq.~\eqref{eq:I(w)}.

This chapter assumes a basis familiarity with ground-state DFT, as represented by the SCF eigenvalue problem in
Eq.~\eqref{eq:KS-SCF}, which will serve as our starting point.   It should therefore be familiar that the Fock operator takes the form
\begin{equation}\label{eq:Fock}
	\hat{F}_\sigma = -\tfrac{1}{2}\hat{\nabla}^2 + v_\text{ext} + v^{}_\text{H} + \vxc{\sigma}
\end{equation}
in atomic units.  The quantities $v_\text{ext}$, $v^{}_\text{H}$, and $ \vxc{\sigma}$ are known as the external, Hartree, and XC potentials, 
respectively.  In the field-free case, the external potential is simply the interaction of the electrons with the nuclei,\cite{Cap06}
\begin{equation}	
	v_\text{ext}(\br) = -\sum_A \frac{Z_A}{ \| \br - \bR_A \| } \; .
\end{equation}
More generally, $v_\text{ext}(\br)$ might also contain 
a field-dependent contribution such as $-\bm{\mathcal{E}}(\br,t)\bm{\cdot}\br$ in the presence of an electric field
$\bm{\mathcal{E}}(\br,t)$.  The Hartree (or Coulomb) potential $v^{}_\text{H}(\br)$ 
describes self-repulsion of the electrons,\cite{Cap06} equivalent to what is often called 
``$J$'' in Hartree-Fock theory.\cite{SzaOst82,KocHol01} It is a functional of the density, given by 
\begin{equation}
	v^{}_\text{H}[\rho](\br) = \int \frac{\rho(\br,\br')}{ \| \br - \br' \| } \; d\br' \; .
\end{equation}
The final component of $\hat{F}_\sigma$ is $\vxc{\sigma} = \delta E_\text{xc}/\delta\rho_\sigma$, 
the XC operator for $\sigma$-spin electrons.   In ``pure'' Kohn-Sham theory, this quantity should be a local potential 
$\vxcpot{\sigma}(\br)$ rather than an operator, but herein we allow the possibility for mixing some nonlocal 
Hartree-Fock exchange (HFX), as is done in the hybrid density functionals that are most useful in molecular DFT.  For hybrid functionals 
$\vxc{\sigma}$ is a nonlocal operator and this scenario is sometimes called 
\textit{generalized Kohn-Sham theory},\cite{BaeKro18} although the use of hybrid functionals can no longer be considered exotic in molecular DFT.

The textbook by Koch and Holthausen\cite{KocHol01} is a good resource for ground-state DFT (though not for TDDFT), as
are several literature overviews.\cite{von04,Cap06} Updated ground-state benchmarks, relative to the rather dated ones
in Ref.~\citenum{KocHol01}, can be found elsewhere.\cite{MarHea17,JacCav17}
For TDDFT, the textbook by Ullrich,\cite{Ull12} or else overviews by Gross and co-workers,\cite{MarGro03,MarGro04,GroMai12}
provide the rigorous foundations of the theory, which are mostly omitted here.
Several other reviews cover LR-TDDFT in a pedagogical way.\cite{DreHea05,EllFurBur09,CasHui12}
For overviews of molecular applications of LR-TDDFT, see reviews by Jacquemin and 
co-workers,\cite{JacPerCio09,AdaJac13,LauAdaJac14,JacAda16,SanJac16}
who have also reviewed the accuracy benchmarks\cite{LauJac13} and functional selection.\cite{ChaPlaMen13}


\section{Linear-Response (``Time-Dependent'') DFT}
\label{sec:TDDFT}
This section describes the formalism and application of LR-TDDFT, commonly known simply as ``TDDFT".  The starting point is the 
TDKS equation [Eq.~\eqref{eq:TDKS}] that describes how the ground-state MOs $\psi_{k\sigma}$ evolve in time following a perturbation that is 
applied at $t=0$.   If that perturbation is taken to be a time-oscillating field at frequency $\omega$, 
\begin{equation}\label{eq:V(t)}
	V(t) = \tfrac{1}{2}\big(\mathcal{E} e^{-\cmplxi \omega t} + \mathcal{E}^\ast e^{+\cmplxi \omega t}\bigr) \; ,
\end{equation}
then in the weak-field limit ($\mathcal{E}\rightarrow 0$), the response of the ground state can be computed exactly using first-order 
perturbation theory.\cite{Fur01a}   Formally, one ought to show that the poles of the frequency-dependent response function can be 
obtained from those of the independent-particle (Kohn-Sham) response function,\cite{PetGosGro96}
but for that exercise the reader is referred to reviews by Marques and Gross.\cite{MarGro03,MarGro04} 
For a derivation of LR-TDDFT based on a variational principle, see Ref.~\citenum{ZieSetKry09}.

\subsection{Theoretical Formalism}
\label{sec:TDDFT:Formalism}

The derivation from perturbation theory 
starts from the equivalent Liouville-von Neumann (LvN) form of the TDKS equation, which is 
\begin{equation}\label{eq:LvN}
	\cmplxi\hbar \frac{d\hat{P}_\sigma}{dt} = \hat{F}_\sigma\hat{P}_\sigma - \hat{P}_\sigma\hat{F}_\sigma
\end{equation}
where 
\begin{equation}
	\hat{P}_{\sigma}(t) = \sum_{k}^\text{occ} \bigl|\psi_{k\sigma}(t)\big\rangle\big\langle\psi_{k\sigma}(t)\big|
\end{equation}
is the time-evolving one-electron density operator for $\sigma$-spin electrons.
Expanding Eq.~\eqref{eq:LvN} to first order in the perturbed Fock and density matrices, in the presence of the perturbation $V(t)$,  
one obtains the unperturbed LvN equation at zeroth order.  This is equivalent to the ground-state Kohn-Sham
eigenvalue problem in Eq.~\eqref{eq:KS-SCF}.    Working equations for LR-TDDFT are obtained by equating the first-order 
terms,\cite{HirHea99b,Fur01a,DreHea05} as elaborated below.

\subsubsection{Linear Response Theory}
\label{sec:TDDFT:Formalism:LR}
To consider this in more detail, recognize that the perturbation $V(t)$ in Eq.~\eqref{eq:V(t)} is a one-electron operator 
whose spatial part can be expanded in the MO basis, leaving the time dependence to be carried by $e^{\pm\cmplxi\omega t}$.
Introducing a set of unknown coefficients $z_{pq\sigma}$ and $z_{qp\sigma}^\ast$, representing real and 
imaginary parts of the first-order response, the first-order perturbed density matrix can be expressed as 
\begin{equation}
	P_{pq\sigma}(t) 
	= P_{pq\sigma}^{(0)}  + P_{pq\sigma}^\one(t)
	= P_{pq\sigma}^{(0)} +
		\tfrac{1}{2}\big(z_{pq\sigma} e^{-\cmplxi \omega t} + z_{qp\sigma}^\ast e^{\cmplxi \omega t}\big) \; ,
\end{equation}
where $P_{pq\sigma}^{(0)}$ is the unperturbed density matrix at $t=0$.    This change in the density matrix is accompanied by a corresponding change
in the Fock matrix.   Through first order, the Fock matrix is\cite{HirHea99b}
\begin{equation}\label{eq:F(1)} 
	F_{pq\sigma}(t) =  F^\zero_{pq\sigma} + V_{pq} +
	\sum_{rs\tau} 
	\left(
		\frac{ \partial F_{pq\sigma} }{ \partial P_{rs\tau} } 
	\right)
	P^\one_{rs\tau}(t) \; ,
\end{equation}
where the unperturbed Fock operator $\hat{F}_\sigma^{(0)}$ has the form given in Eq.~\eqref{eq:Fock}.   
The first-order response of the density matrix is thus coupled to a term of the form\cite{HirHea99b}
\begin{align}\label{eq:dF/dP}
\begin{aligned}
	\frac{ \partial F_{pq\sigma} }{ \partial P_{rs\tau} } &= 
	\big(p_\sigma q_\sigma \big| s_{\tau} r_{\tau}\big) +
		\left(p_\sigma q_\sigma \left| 
			\frac{ \delta^2 E_\text{xc} }{ \delta \rho_\sigma \delta\rho_{\tau} }
		\right| s_{\tau} r_{\tau}\right) 
\\
	&= (pq|sr) 
	+ \big( p_\sigma q_\sigma \big| \fxc{\sigma\tau} \big| s_{\tau} r_{\tau} \big) \; .
\end{aligned}
\end{align}
The first term, $(p_\sigma q_\sigma | s_{\tau} r_{\tau}) = (pq|sr)$, is a Coulomb integral expressed in Mulliken notation,\cite{SzaOst82}
while $\fxc{\sigma\tau} = \delta^2 E_\text{xc}/\delta\rho_\sigma\delta\rho_{\tau}$.   The latter quantity is discussed in more detail below.

So far, the MO indices $p, q, r, s$ are arbitrary and could refer either to occupied or virtual orbitals.   In fact, the idempotency 
condition $\hat{P}^2_{\sigma} = \hat{P}_{\sigma}$ imposes restrictions.   Through first order, the idempotency condition is 
\begin{equation}\label{eq:idemp}
	\sum_r \big(P^\zero_{pr\sigma} P^\one_{rq\sigma} + P^\one_{pr\sigma}P^\zero_{rq\sigma} \big)
	= P^\one_{pq\sigma}
\end{equation}
since $\hat{P}_{\sigma}^\zero\hat{P}_{\sigma}^\zero = \hat{P}_{\sigma}^\zero$.   As a matrix, $\mathbf{P}_{\sigma}^\zero$ 
contains only occupied--occupied and virtual--virtual blocks because the occupied--virtual block vanishes as a 
condition of SCF convergence.\cite{HerHea04}   Using $i, j, \ldots$ to index occupied MOs and $a, b, \ldots$ for virtual MOs,
this means $P^\zero_{ia\sigma} = 0 = P^\zero_{ai\sigma}$, so the constraint in Eq.~\eqref{eq:idemp} implies that the only non-vanishing coefficients
in $P^\one_{pq\sigma}$ are $z_{ia\sigma}$ and $z_{ai\sigma}$.\cite{HirHea99b,DreHea05}  
Conventional LR-TDDFT notation is obtained by relabeling these coefficients as
\begin{subequations}
\begin{align}
	x_{ia\sigma} & = z_{ai\sigma}
\\
	y_{ia\sigma} &= z_{ia\sigma} \; .
\end{align}
\end{subequations}
Collecting these unknowns into vectors $\mathbf{x}$ and $\mathbf{y}$, one may 
rewrite the first-order terms in the LvN equation in matrix form as\cite{DreHea05,EllFurBur09,CasHui12}    
\begin{equation}\label{eq:LR-TDDFT}
	\left(\begin{array}{ll}
		\mathbf{A} & \mathbf{B} \\
		\mathbf{B}^\ast & \mathbf{A}^\ast 
	\end{array}\!\!\right)
	\left(\begin{array}{c}
		 \mathbf{x}^{(n)} \\ 
		 \mathbf{y}^{(n)}
	\end{array}\!\right)
	= \omega_n
	\left(\begin{array}{cr}
		\bm{1} & \bm{0} \\
		\bm{0} & -\bm{1}
	\end{array}\right)
	\left(\begin{array}{c}
		 \mathbf{x}^{(n)} \\
		 \mathbf{y}^{(n)}
	\end{array}\!\right) \; .
\end{equation}
This represents a system of equations for the excitation energies $\omega_n$ and the amplitudes $x_{ia\sigma}^{(n)}$ and $y_{ia\sigma}^{(n)}$, 
and constitutes
the basic working equations of LR-TDDFT.    (The index $n$, which counts excited states, will usually be omitted for compactness.)
The system in Eq.~\eqref{eq:LR-TDDFT} is often called the \textit{Casida equations},\cite{Cas95a,Cas96} 
although these are formally identical to the equations of time-dependent Hartree-Fock theory.\cite{McLBal64}
The matrices $\mathbf{A}$ and $\mathbf{B}$ are known as \textit{orbital Hessians},\cite{FurRap05}
for reasons that are discussed below, and they derive from the derivative of $\hat{F}_\sigma$ with respect to $\hat{P}_{\tau}$ in 
Eq.~\eqref{eq:dF/dP}.   In the canonical MO basis, the matrix elements of $\mathbf{A}$ and $\mathbf{B}$ are\cite{DreHea05,FurRap05}
\begin{subequations}\label{eq:A-B-matrix}
\begin{align}
\label{eq:A-matrix}
	A_{ia\sigma,jb\tau} &= (\eval{a\sigma} - \eval{i\sigma})\delta_{ij}\delta_{ab}\delta_{\sigma\tau} + (ia|jb)
	- \cHFX (ij|ab) \delta_{\sigma\tau} + (1-\cHFX) (ia|\kxc{\sigma\tau}|jb) 
\\
\label{eq:B-matrix}
	B_{ia\sigma,jb\tau} &= (ia|bj) - \cHFX (ib|aj)\delta_{\sigma\tau}
	 + (1-\cHFX)(ia|\kxc{\sigma\tau}|bj) 
\end{align}
\end{subequations}
where $\kxc{\sigma\tau} = \fxc{\sigma\tau} -\cHFX(\delta^2 E_\text{HFX}/\delta\rho_\sigma\delta\rho_{\tau})$.
The quantity $\cHFX$ will be used throughout this chapter to mean the 
coefficient of HFX (often called ``exact exchange'') contained in the functional $E_\xc[\rho]$, with $0 \leq \cHFX \le 1$.   
For example, $\cHFX = 0.2$ for B3LYP, meaning meaning 20\% HFX and 80\% semilocal exchange. 
We have chosen to separate the HFX terms in Eq.~\eqref{eq:A-B-matrix}, which can be expressed in terms of electron repulsion integrals
$(ij|ab)$ and $(ib|aj)$, leaving $\kxc{\sigma\tau}$ as the second functional derivative of the semilocal contribution, $E_\text{xc} - E_\text{HFX}$.

The solution $(\mathbf{x},\mathbf{y})$ of Eq.~\eqref{eq:LR-TDDFT} parameterizes the \textit{transition density matrix} 
for the excitation in question.   In real space, this quantity is\cite{StrScuFri98,Fur01a,FurRap05} 
\begin{equation}\label{eq:TDM}
	T(\mathbf{r},\mathbf{r}') 
		= \sum_{ia\sigma} \Bigl[ x_{ia\sigma} \, \psi^{}_{a\sigma}(\mathbf{r}) \, \psi^{\ast}_{i\sigma}(\mathbf{r}') 
		+ y_{ia\sigma}\, \psi^{}_{i\sigma}(\mathbf{r}) \, \psi^{\ast}_{a\sigma}(\mathbf{r}') \Bigr] \; .
\end{equation}
The unknowns $\mathbf{x}$ and $\mathbf{y}$ satisfy a bi-orthogonal normalization condition,\cite{Fur01a}
\begin{equation}\label{eq:normalization}
	\sum_{ia\sigma} (x_{ia\sigma}^2 - y_{ia\sigma}^2) = 1 \; ,
\end{equation}
which is also a feature of the much older \textit{time-dependent Hartree-Fock} (TDHF) theory.\cite{McLBal64} 
For historical reasons, TDHF is also known as the \textit{random phase approximation} (RPA),\cite{Row68,OddJorYea84} 
because it can be derived within an equation-of-motion formalism for the single-particle excitation operators,\cite{CasHui12}
similar to the historical RPA.\cite{Row68}  However, TDHF\slash RPA can also be considered to be a special case of LR-TDDFT
corresponding to the Hartree-Fock functional, \ie, $\cHFX=1$ and $\kxc{\sigma\tau}\equiv0$.

The number of unknown amplitudes in Eq.~\eqref{eq:LR-TDDFT} is $2\nocc\nvir$, hence to solve this equation for all of the
excitation energies $\omega$ would incur sixth-order cost, $\mathcal{O}(\nocc^3\nvir^3)$.
Because matrix--vector products such as $\mathbf{Ax}$ or $\mathbf{By}$ can be computed with only fourth-order cost, 
in practical calculations Eq.~\eqref{eq:LR-TDDFT} is solved iteratively for just the lowest few eigenvalues 
($\nroots$).\cite{OlsJenJor88,WeiAhlHas93,StrScuFri98,FurKruNgu16,VecBraSha17,ZhoPar21} 
The cost of that calculation scales as $\nroots \times \mathcal{O}(\nocc^2\nvir^2)$,\cite{FurRap05} which 
is typically not significantly greater than the cost of the ground-state SCF calculation if $\nroots \sim 10$.  Therefore if  
ground-state DFT is feasible then LR-TDDFT is probably tractable also, at least for the lowest few excited states.   It is worth noting, however, 
that the memory footprint to solve Eq.~\eqref{eq:LR-TDDFT} 
is $\nroots\times\mathcal{O}(\nocc\nvir)$, which is significantly more than the ground-state memory requirement.
This can become a problem for large systems if a large number of excited states is desired, \eg, in models of semiconductors
where a band structure is emerging.\cite{GaoPauSch15}    For such applications, the real-time approach that is described in Section~\ref{sec:TDKS} 
offers a low-memory alternative to LR-TDDFT.

Some alternative forms of the basic LR-TDDFT equation are also worth considering.   We first note that the matrices
$\mathbf{A}$ and $\mathbf{B}$ in Eq.~\eqref{eq:A-B-matrix} can be rewritten as 
\begin{subequations}\label{eq:A-B-matrix-v2}
\begin{align}
	A_{ia\sigma,jb\tau} &= (\eval{a\sigma} - \eval{i\sigma})\delta_{ij}\delta_{ab}\delta_{\sigma\tau}  + K_{ia\sigma,jb\tau}
\\
	B_{ia\sigma,jb\tau} &= K_{ia\sigma,bj\tau} 
\end{align}
\end{subequations}
where $\mathbf{K}$ is a coupling matrix,\cite{Cas95a,Cas02}
\begin{equation}
	K_{ia\sigma,jb\tau} = \int \int \psi_{i\sigma}(\br) \; \psi_{a\sigma}(\br)
	\left(
		\frac{1}{ \| \br - \br' \|} + \fxcker{\sigma\tau}(\br,\br')
	\right)
	\psi_{j\tau}(\br') \; \psi_{b\tau}(\br') \; d\br \; d\br'
\end{equation}
with a Hartree--XC kernel.\cite{EllFurBur09} In practice, this looks like the energy-transfer coupling\cite{YouHsu14} between transition densities 
$\rho_{ia\sigma}(\br) = \psi_{i\sigma}(\br) \; \psi_{a\sigma}(\br)$ and $\rho_{bj\tau}(\br') = \psi_{b\tau}(\br') \; \psi_{j\tau}(\br')$.
One can therefore consider that solution of the LR equations 
reveals how the zeroth-order, independent-particle excitations $\psi_{i\sigma}\rightarrow\psi_{a\sigma}$
are coupled to obtain excited states of the interacting system.

Assuming that the orbitals are real, so that $\mathbf{A}^\ast = \mathbf{A}$ and $\mathbf{B}^\ast = \mathbf{B}$, then Eq.~\eqref{eq:LR-TDDFT}
is equivalent to a pair of equations
\begin{equation}\label{eq:pseudo-eigen}
	(\mathbf{A}\pm\mathbf{B})(\mathbf{x}\pm \mathbf{y}) = \omega(\mathbf{x}\mp \mathbf{y}) \; ,
\end{equation} 
which makes it clear that Eq.~\eqref{eq:LR-TDDFT} is not a conventional eigenvalue problem.  
However, upon introducing new variables 
\begin{equation}\label{eq:z=x_pm_y}
	\mathbf{z}_\pm = \sqrt{\omega} (\mathbf{A}\mp\mathbf{B})^{-1/2}(\mathbf{x}\pm\mathbf{y}) \; ,
\end{equation}
which satisfy the more conventional normalization condition $\mathbf{z}_\pm^\dagger\mathbf{z}_\pm = 1$,\cite{FurAhlWac00,LuzZhi10} 
the LR-TDDFT equations can be transformed into either of two equivalent, Hermitian eigenvalue 
problems.\cite{UllRow71,McCCus71,FurAhlWac00,LuzZhi10,VecBraSha17}
These are 
\begin{equation}\label{eq:Hermitian}
	\bm{\Omega}_\pm \mathbf{z}_\pm = \omega^2 \mathbf{z}_\pm 
\end{equation}
where
\begin{equation}
	\bm{\Omega}_\pm = (\mathbf{A}\mp\mathbf{B})^{1/2} (\mathbf{A}\pm\mathbf{B})(\mathbf{A}\mp\mathbf{B})^{1/2} \; .
\end{equation}
The $\bm{\Omega}_+$ version of Eq.~\eqref{eq:Hermitian} is especially convenient for semilocal functionals ($\cHFX=0$), 
because in that case $\mathbf{A}-\mathbf{B}$ is diagonal and one obtains a Hermitian 
eigenvalue problem with half the dimension of the original pseudo-eigenvalue problem in Eq.~\eqref{eq:LR-TDDFT}.

For a closed-shell (spin-restricted) ground state, another important transformation is 
\begin{subequations}\label{eq:x-y-spin}
\begin{align}
	x_{ia}^\pm &=  (x_{ia\alpha} \pm x_{ia\beta})/\sqrt{2}
\\
	y_{ia}^\pm &=  (y_{ia\alpha} \pm y_{ia\beta})/\sqrt{2} \; ,
\end{align}
\end{subequations}
which affords amplitudes for singlet ($+$) and triplet ($-$) spin functions.  Making use of the unitary transformation\cite{OddJorYea84}   
\begin{equation}
	\left(\begin{array}{cc}
		\mathbf{A} + \mathbf{B} & \bm{0} \\
		\bm{0} & \mathbf{A} - \mathbf{B} \\
	\end{array}\right)
	= \frac{1}{2}
	\left(\begin{array}{rr}
		\bm{1} & \bm{1} \\
		-\bm{1} & \bm{1} \\
	\end{array}\right)	
	\left(\begin{array}{cc}	
		\mathbf{A} & \mathbf{B} \\
		\mathbf{B} & \mathbf{A} 
	\end{array}\right)		
	\left(\begin{array}{rr}	
		\bm{1} & -\bm{1} \\
		\bm{1} & \bm{1} \\	
	\end{array}\right)	\; ,
\end{equation}
in addition to Eq.~\eqref{eq:x-y-spin}, one obtains singlet and triplet versions of 
$\mathbf{A}\pm\mathbf{B}$ that function as \textit{stability matrices}.\cite{WeiAhlHas93,BauAhl96a,FurAhlWac00}
In other words, these are Hessian matrices whose eigenvalues characterize whether the ground state is stable with respect to orbital rotations.
For example, the \textit{singlet stability matrix} is\cite{FurAhlWac00}
\begin{equation}\label{eq:sing-stab}
	(\mathbf{A}^+ + \mathbf{B}^+)_{ia,jb} = (\eval{a} - \eval{i})\delta_{ij}\delta_{ab} + 4 (ia|jb) 
	+ 2 (ia | \big(\fxc{\alpha\alpha} + \fxc{\beta\beta}\big) | jb ) \; .
\end{equation}
A negative eigenvalue in $\mathbf{A}^+ + \mathbf{B}^+$ indicates an 
instability, which looks like a negative excitation energy from the standpoint of LR-TDDFT.   This correspondence is a consequence of the 
\textit{Thouless theorem},\cite{Tho60} which states that 
orbital rotations (and therefore orbital relaxation) can always be parameterized as single excitations.
Along similar lines, eigenvalues of the \textit{triplet instability matrix}\cite{FurAhlWac00}
\begin{equation}\label{eq:trip-stab}
	(\mathbf{A}^- + \mathbf{B}^-)_{ia,jb} = (\eval{a} - \eval{i})\delta_{ij}\delta_{ab} 
	+ 2 (ia | \big(\fxc{\alpha\alpha} - \fxc{\alpha\beta}\big) | jb )
\end{equation}
 indicate whether the ground-state solution is stable with respect to spin symmetry breaking 
 (restricted $\rightarrow$ unrestricted).\cite{CizPal67}
In the presence of an unstable reference state, the transformation in Eq.~\eqref{eq:z=x_pm_y} may become problematic,
which can lead to failure of certain iterative LR-TDDFT algorithms.

\subsubsection{Adiabatic Approximation}
\label{sec:TDDFT:Formalism:AA}

We have not yet discussed the key ingredient in the orbital Hessian matrices that makes LR-TDDFT different from 
TDHF\slash RPA, namely, the \textit{exchange-correlation kernel}, $\fxc{\sigma\tau}$.    
A more careful application of LR theory would note that the quantities $\mathbf{A}(\omega)$
and $\mathbf{B}(\omega)$ are themselves functions of the excitation energy $\omega$.\cite{Cas95a,Cas96,Fur01a}
In wave function terms, this can be understood based on the fact that any exact theory of many-electron excitation energies that is 
formulated as an effective single-particle theory must ultimately invoke an effective Hamiltonian 
that is energy-dependent, in order to encapsulate the effects of higher-order excitations.\cite{Cas96,CasHui16,AutLoo20}   
(In many-body theory this energy-dependent contribution is sometimes called the ``self-energy''.\cite{Ort17}) 
Proof-of-concept models for an energy-dependent kernel $\fxcker{\sigma\tau}(\br,\br',\omega)$ have been put 
forward,\cite{MaiZhaCav04,Mai06,Mai12,AutLoo20} which have close connections to  
many-body perturbation theory and the Bethe-Salpeter equation.\cite{CasHui16,AutLoo20}
However, there are no production models for molecular Hamiltonians at present.

To appreciate the nature of the approximation in neglecting the energy dependence of $\fxc{\sigma\tau}$, 
consider that this quantity arises in Eq.~\eqref{eq:dF/dP} as the second 
functional derivative of the XC energy with respect to the density, or the first derivative of the XC potential.
For a time-evolving density $\rho_\sigma(\br,t)$, this means\cite{GroBur06}
\begin{equation}\label{eq:AA1}
	\fxcker{\sigma\tau}(\br,\br',t-t') = 
	\frac{ \delta \vxc{\sigma}[\rho](\br,t) }{ \rho_{\tau}(\br',t') } \; .
\end{equation}
(This expression leaves $\vxc{\sigma}$ in the form of an operator, which technically makes this an example of 
generalized Kohn-Sham theory.\cite{BaeKro18})
The time dependence in $\fxcker{\sigma\tau}(\br,\br',t-t')$ means that this quantity depends on the whole history of the time evolution of 
the density,\cite{GroBur06,Mai06,Mai12} imparting a frequency dependence upon Fourier transformation:
$\fxcker{\sigma\tau}(\br,\br',\omega)$.    For practical purposes,
it is basically a requirement to invoke the \textit{adiabatic approximation},\cite{Cas95a,BurWerGro05,CasHui12}
which assumes locality in time and therefore differentiates with respect to the instantaneous density:\cite{Cas95a}
\begin{equation}\label{eq:AA2}
	\frac{ \delta \vxc{\sigma}[\rho](\br,t) }{ \rho_{\tau}(\br',t') } \approx
	\delta(t-t') \frac{ \delta \vxc{}[\rho](\br) }{ \rho_\tau(\br') } \; .
\end{equation}
The ``memory'' of the kernel is thereby neglected, tantamount to assuming that $\vxc{}[\rho](\br,t)$ can be approximated using a 
conventional ground-state energy functional $E_\xc[\rho]$, whose functional derivative is evaluated using the time-evolving 
density:\cite{BurWerGro05}
\begin{equation}
	\vxc{\sigma}(\br,t) = \left.\frac{ \delta E_\xc[\rho] }{ \delta \rho_\sigma(\br) }\right|_{\rho_\sigma(\br) = \rho_\sigma(\br,t)}  \; .
\end{equation}
Time dependence is thus carried entirely by the density and not by the functional.
The frequency dependence of $\fxc{\sigma\tau}$ disappears and conventional (ground-state) density functionals are all that is required
for LR-TDDFT within the adiabatic approximation.

One immediate ramification of this approximation is that the LR-TDDFT equations have precisely 
$2\nocc\nvir$ solutions for the excitation energy $\omega$, coinciding with the number of unknown amplitudes $x_{ia\sigma}$ and $y_{ia\sigma}$.
In wave function language, these are the ``one-particle, one-hole'' (1p1h) states, as in conventional configuration interaction with
single excitations (CIS).  States with significant double-excitation character (2p2h states) 
are either absent altogether,\cite{TozHan00,MaiZhaCav04,BurWerGro05,ThiKum09,EllGolCan11} or at best severely shifted.\cite{ThiKum09} 
The latter are therefore 
generally considered to be out of reach within the adiabatic approximation that is ubiquitous in practical LR-TDDFT calculations.\cite{EllGolCan11}

\subsubsection{Tamm-Dancoff Approximation}
\label{sec:TDDFT:Formalism:TDA}

Given a ground-state functional $E_\xc[\rho]$, all that is required for LR-TDDFT within the adiabatic approximation are second functional derivatives
\begin{equation}
	\fxc{\sigma\tau}(\br,\br') = \frac{ \delta^2 E_\xc }{ \delta \rho_\sigma(\br)\; \delta\rho_{\tau}(\br') } \; ,
\end{equation}
from which the matrix elements of $\mathbf{A}$ and $\mathbf{B}$ can be evaluated.   Upon solution of 
Eq.~\eqref{eq:LR-TDDFT} or one of its equivalent forms, it is often found that the amplitudes $y_{ia\sigma}$ are $10^2$--$10^3$ times 
smaller than the largest $x_{ia\sigma}$.  Invoking
the approximation $y_{ia\sigma} \approx 0$, one obtains a conventional Hermitian eigenvalue problem 
\begin{equation}\label{eq:TDA}
	\mathbf{Ax} = \omega\mathbf{x} 
\end{equation}
whose dimension is half that of the original LR-TDDFT pseudo-eigenvalue problem, and where the matrix $\mathbf{B}$ does not appear.   
The basis for this approximation can be understood from the fact that the matrix elements of $\mathbf{A}$ are typically much larger (at least
along the diagonal) as compared to the matrix elements of $\mathbf{B}$, because the leading contribution to $\mathbf{A}$ is a difference
in one-particle energy levels ($A_{ia\sigma,ia\sigma} = \eval{a\sigma} - \eval{i\sigma} + \cdots$).     For historical reasons that are related to a similar
approximation that is made in nuclear physics,\cite{Ull12} neglect of $\mathbf{y}$ is known as the  \textit{Tamm-Dancoff approximation} (TDA).
For hybrid functionals, the reduction in dimension leads to a concomitant reduction in cost although for semilocal functionals the same 
reduction in dimensionality can be achieved using the $\bm{\Omega}_+$ version of Eq.~\eqref{eq:Hermitian}.  

For the Hartree-Fock functional ($\cHFX=1$ and no correlation), Eq.~\eqref{eq:TDA} is equivalent to the CIS eigenvalue
equation.\cite{ForHeaPop92}   Excited-state wave functions in CIS 
are linear combinations of singly-excited Slater determinants $|\Psi_{i\sigma}^{a\sigma}\rangle$,
\begin{equation}
	|\Psi\rangle = \sum_i^\text{occ}\sum_a^\text{vir}\sum_\sigma x_{ia\sigma} |\Psi_{i\sigma}^{a\sigma}\rangle \; ,
\end{equation}
and for this reason we identify the variables $x_{ia\sigma}$ as excitation amplitudes.   The neglected amplitudes $y_{ia\sigma}$ 
represent de-excitation, insofar as TDHF\slash RPA was originally introduced in the nuclear and many-body physics literature as a means
to add correlation to the ground state.\cite{McLBal64,Row68}
In fact, LR-TDDFT in the form of Eq.~\eqref{eq:LR-TDDFT} was introduced in molecular quantum chemistry as the 
``DFT random phase approximation''.\cite{CasJamBoh96,JamCasSal96}
For that reason, solution of Eq.~\eqref{eq:LR-TDDFT} or its equivalents, \textit{without} invoking the TDA, is sometimes called ``RPA''.
However, in view of a resurgence of interest in using the RPA formalism as a means for correlating the ground
state,\cite{ScuHenSor08,JanLiuAng10,HesGor11,RenRinJoa12,EshBatFur12,ScuHenBul13,CheVooAge17,Ber18}
it is better to refer to Eq.~\eqref{eq:LR-TDDFT} as ``full'' LR-TDDFT, if there is a need to distinguish it from the TDA version in Eq.~\eqref{eq:TDA}.

Quantitatively, the impact of the TDA on excitation energies is often $<0.1$~eV,\cite{HirHea99b} though a 
potentially detrimental impact is that oscillator strengths within the TDA no longer satisfy the Thomas-Reiche-Kuhn sum rule,\cite{McH99}
\begin{equation}\label{eq:TRK}
	\sum_{n>0} f_{0n} = N \; .
\end{equation}
This constraint is satisfied by the full LR-TDDFT approach, at least in the complete-basis limit.\cite{Har69b,Cas95a,Fur01a} 
Some small-molecule tests suggest that errors incurred by the TDA are relatively mild ($<15$\%),\cite{HerJac11b}
and perhaps not noticeable in band shapes once vibrational broadening is taken into account.\cite{ChaLauAda13}

A more important consequence of the TDA is that it decouples the excitation energy problem from the ground-state stability problem, 
whereas for full LR-TDDFT a triplet instability in the ground state manifests as a negative excitation energy, or as an imaginary
root of the Hermitian eigenvalue problem in Eq.~\eqref{eq:Hermitian}.   This may cause problems for eigensolvers that implicitly
assume $\omega > 0$.   Note that triplet instabilities are associated with spin-symmetry breaking, \ie, with the emergence of an
unrestricted solution that is lower in energy than the restricted solution.    Where they appear, these instabilities cause significant artifacts in potential energy 
surfaces computed using LR-TDDFT, including the 
appearance of spurious cusps.\cite{CaiRei00,CasGutGua00,Cas02,MynCas17,HaiRetHea19}  In contrast, the 
variational nature of the CIS-type equation, as opposed to the pseudo-eigenvalue problem that characterizes 
full LR-TDDFT, prevents this from happening within the TDA.\cite{CasHui12}

Along similar lines, it has been appreciated for a long time that TDHF specifically is prone to triplet 
instabilities.\cite{Tho60,BalMcL64,Kou67,CizPal67,DunMcK68,Jor73,MatIto75,NiySan78,ChaLevMil78} 
In fact, the appearance of imaginary excitation energies at equilibrium geometries of small molecules led Furche and
Ahlrichs to conclude that this method is ``rather useless$\ldots$ for the investigation of excited potential energy surfaces''.\cite{FurAhl02}
In contrast, spin-symmetry breaking near the equilibrium geometry is often significantly 
mitigated when DFT is substituted for Hartree-Fock theory.\cite{BauAhl96b} 
Because most molecular LR-TDDFT calculations use hybrid functionals that include some fraction of HFX, it can be expected
that problems with triplet instabilities may increase as that fraction increases, which is precisely what is found in 
practice.\cite{GodOrl99,OrlGod00,OrlGod02a,OrlGod02b,LutHelJas10}
Similarly, in calculations using range-separated functionals, 
which incorporate HFX at long range in the electron--electron Coulomb potential, these instabilities are sensitive to the 
length scale on which that mixing is introduced.\cite{CuiYan10,SeaKoeZha11,PeaWilToz11,PeaToz12,PeaWarToz13,BokGreBok15}  
Invoking the TDA thus improves the accuracy of triplet excitation energies.\cite{PeaWilToz11,PeaToz12,PeaWarToz13}
For photochemical problems, where exploration of excited-state potential energy surfaces is paramount, Casida \etal\ suggest 
that the TDA is effectively mandatory,\cite{CasIpaCor06} in order to avoid excitation energies that drop to zero (and then become imaginary) 
as the system moves through a Coulson-Fischer point where spin-symmetry breaking occurs.   Instabilities appear to proliferate as one
moves away from the ground-state geometry on an excited potential surface.\cite{CasIpaCor06,CorDorIpa07,TapTavRot08}  
For example, in the photochemical ring-opening reaction of oxirane (C$_2$H$_4$O), 51\% of configuration space is estimated to exhibit an instability
with semilocal DFT, as compared to 93\% of space with B3LYP.\cite{CasIpaCor06}    

\subsubsection{Analytic Gradients}
\label{sec:TDDFT:Formalism:Grad}
Photochemical simulations require analytic excited-state gradients.  That formalism, which is closely connected to response theory for optical 
properties,\cite{ParFur18} is not discussed here but can be found elsewhere.\cite{VanAmo99,VanAmo00,FurAhl02,FurRap05,LiuGanSha10,RapHut12} 
Nonadiabatic or ``derivative coupling'' vectors between excited states,\cite{HerZhaMor16,HerMan22} 
which are needed for nonadiabatic molecular dynamics simulations,\cite{TapBelVin13,BarCre16,ParRoyFur19,ZhaHer21,HerMan22}
have also been formulated.\cite{ZhaHer14b,ZhaHer15a,LiLiu14,OuFatAlg14,OuBelFur15,AlgOuSub15,OuAlgSub15,WanWuLiu21}
Evaluation of the nonadiabatic couplings has the same formal complexity as evaluation of the excited-state gradient.\cite{ZhaHer14b,ParRoyFur19}


Here, we comment briefly regarding how the gradient formalism bears on static properties of the excited state, such as the dipole moment or 
atomic population analysis.  The density matrix for the excited state can be written 
\begin{equation}\label{eq:Pexc}
	\mathbf{P}_\text{rlx} = \mathbf{P}_\text{unrlx} + \mathbf{Z} 
	=\mathbf{P}_0 + \Delta\mathbf{P} 
	+ \mathbf{Z} \; ,
\end{equation}
which is sometimes called the ``relaxed'' density matrix, from which an ``unrelaxed'' contribution 
$\mathbf{P}_\text{unrlx} = \mathbf{P}_0 + \Delta\mathbf{P}$ is separated out.  Here, 
$\mathbf{P}_0$ represents the ground-state density matrix and the unrelaxed change upon excitation ($\Delta\mathbf{P}$) 
can be obtained from the amplitudes $\mathbf{x}$ and $\mathbf{y}$.   The remaining contribution 
($\mathbf{Z}$) represents orbital relaxation.\cite{FurRap05,LiuGanSha10,RapHut12}   

The unrelaxed density change $\Delta\mathbf{P}$ can be separated into particle (electron) and hole contributions,
\begin{equation}\label{eq:DeltaP}
	\Delta\mathbf{P} = \Delta\mathbf{P}^\text{elec} + \Delta\mathbf{P}^\text{hole} \; ,
\end{equation}
which are given by\cite{FurAhl02,IpaCorDor09,RapHut12,CheLiuMa14}
\begin{subequations}\label{eq:dP_v1}
\begin{align}
	\Delta\mathbf{P}^\text{elec} &= \frac{1}{2}\Bigl[
		(\mathbf{x}+\mathbf{y})^\dagger(\mathbf{x}+\mathbf{y})
		+(\mathbf{x}-\mathbf{y})^\dagger(\mathbf{x}-\mathbf{y})
	\Bigr]
\\
	\Delta\mathbf{P}^\text{hole} &= -\frac{1}{2}\Bigl[
		(\mathbf{x}+\mathbf{y})(\mathbf{x}+\mathbf{y})^\dagger
		+(\mathbf{x}-\mathbf{y})(\mathbf{x}-\mathbf{y})^\dagger
	\Bigr] \; .
\end{align}
\end{subequations}
These expressions can be rearranged to afford\cite{IpaCorDor09}
\begin{subequations}\label{eq:dP_v2}
\begin{align}\label{eq:dP-elec}
	(\Delta\mathbf{P}^\text{elec})_{ab\sigma} &= \sum_{i} (x^\ast_{ia\sigma} x_{ib\sigma} + y^\ast_{ia\sigma} y_{ib\sigma})
\\ \label{eq:dP-hole}
	(\Delta\mathbf{P}^\text{hole})_{ij\sigma} &= - \sum_{a} (x_{ia\sigma} x^\ast_{ja\sigma} + y_{ia\sigma} y^\ast_{ja\sigma}) \; .
\end{align}
\end{subequations}
Note that $\Delta\mathbf{P}$ only contains 
occupied--occupied ($\Delta\mathbf{P}^\text{hole}$) and virtual--virtual ($\Delta\mathbf{P}^\text{elec}$)
contributions to the excited-state density matrix, not occupied--virtual contributions.   The latter are contained in 
$\mathbf{Z}$,\cite{FurAhl02,FurRap05} the evaluation of which requires solution of so-called $Z$-vector equations.\cite{HanSch84} 
This has the same formal complexity as an excited-state gradient calculation.  

Excited-state properties
should be computed using the relaxed density matrix $\mathbf{P}_\text{rlx}$, because $\Delta\mathbf{P}$ and $\mathbf{Z}$ 
make similar contributions.\cite{ForHeaPop92,RonAngBel14,MasCamFri18} 
Especially when the change in density is large, as for an excitation with significant charge-transfer (CT) character, 
the use of the unrelaxed density matrix may lead to unacceptable errors in excited-state properties.   For example,
excited states of $p$-nitroaniline involving intramolecular CT character exhibit 
relaxed and unrelaxed dipole moments (computed using $\mathbf{P}_\text{rlx}$ versus $\mathbf{P}_\text{unrlx}$, respectively) 
that differ by more than 10~D in some cases!\cite{RonAngBel14}

\subsection{Performance and Practice}
\label{sec:TDDFT:Application}
As discussed above, the formal scaling of LR-TDDFT is $\nroots\times\mathcal{O}(\nbasis^4)$ for hybrid functionals.\cite{FurRap05} 
In practical terms, where only a few low-lying excited states are desired, this means that LR-TDDFT is generally feasible 
if the corresponding ground-state calculation is practical, perhaps up to about 400 atoms for single-point calculations or
150--250 atoms where excited-state geometry optimization is required,\cite{AdaJac13,LauAdaJac14} with more severe limitations
where excited-state frequency calculations are required.\cite{JacPlaAda12,LauAdaJac14}     These estimates are appropriate where
basis sets of double-$\zeta$ quality are used, which is generally adequate.
Triple-$\zeta$ basis sets may be considered to be converged.\cite{EllFurBur09}  Systems that possess 
a dense manifold of excited states (\eg, semiconductors) can create significant storage bottlenecks
for the trial vectors that are required by the iterative eigensolver. In such cases, the ``real-time'' TDDFT approach
of Section~\ref{sec:TDKS} may be advantageous, although Section~\ref{sec:TDDFT:Application:Trunc} describes ways to reduce 
the cost while staying within the confines of LR-TDDFT.    Following a discussion of XC functional approximations in 
Section~\ref{sec:TDDFT:Application:XC}, the accuracy of LR-TDDFT excitation energies is addressed in 
Section~\ref{sec:TDDFT:Application:Benchmarks} by considering its performance for small-molecule benchmarks.
Finally, techniques for visualizing and understanding the states that are computed are discussed in Section~\ref{sec:TDDFT:Application:Visualization}.

\subsubsection{Restriction of the Excitation Manifold}
\label{sec:TDDFT:Application:Trunc}
\begin{figure}[t]
\centering
\fullwidthfig{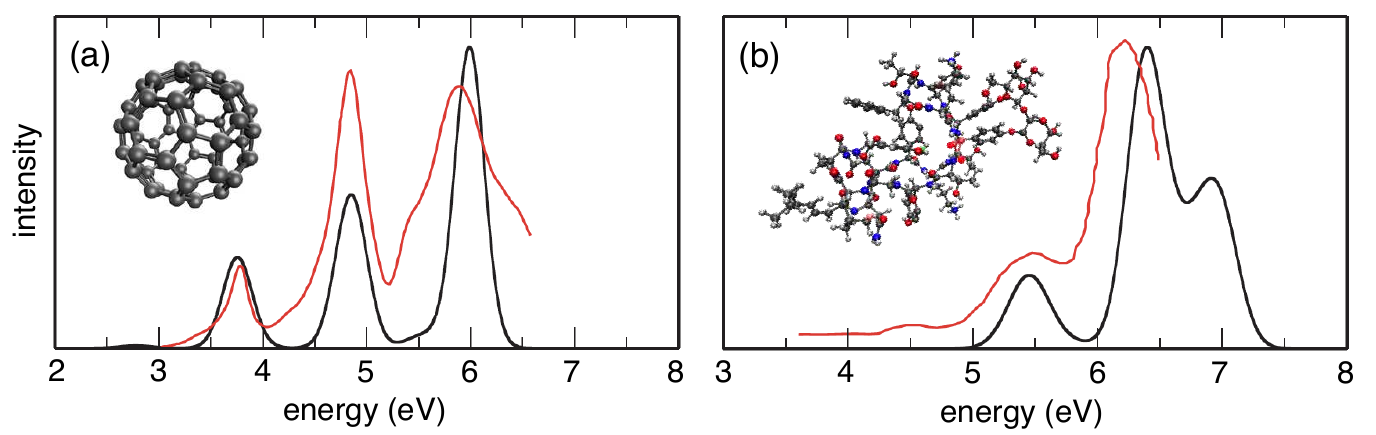}
\caption{
	Electronic absorption spectra of 
	(a) C$_{60}$ (PBE\slash 6-31G* level) and 
	(b) the antibiotic ramoplanin (271 atoms and 2,483 basis functions, CAM-B3LYP\slash 6-31G* level).
	Experimental spectra are shown in red and the spectra in black are computed
	from LR-TDDFT\slash TDA excitation energies with 0.2~eV Gaussian broadening.  Core orbitals and 70\% of virtual 
	orbitals are excluded from each calculation.
	Adapted from Ref.~\protect\citenum{HanGeoBes18a}; copyright 2018 Taylor \& Francis.
}\label{fig:TDDFT-large-Besley}
\end{figure}

Significant cost reduction in LR-TDDFT calculations for large molecules can 
be achieved by neglecting some of the amplitudes $x_{ia\sigma}$, in addition to neglecting \textit{all} of the amplitudes $y_{ia\sigma}$.
Figure~\ref{fig:TDDFT-large-Besley} shows examples
of excitation spectra computed for C$_{60}$ and for $\rm C_{119}H_{154}ClN_{21}O_{40}$, 
performed by excluding over 70\% of the virtual orbitals (based on orbital energies $\eval{a\sigma}$) 
without adverse effects on the spectral envelope.\cite{HanGeoBes18a}

Similar truncations of the excitation manifold can be used to access core-excited states.\cite{Bes16,ZhuAlaHer21,KasSteJen20,Bes21}
There is significant interest in core excitations in contemporary quantum 
chemistry,\cite{NorDre18,Bes20,BokKuh20,KasSteJen20,Bes21,RanPen21}
driven by the recent availability of tabletop laser sources with femtosecond time 
resolution.\cite{DepOliGau15,KleEkiGol19,SchElsHol19,GenMarGug19,SmiBalCha20}
However, core-to-valence excitations lie embedded in an ionization continuum and, at a practical level, lie above all of the valence-excited
states, such that the use of iterative eigensolvers is prohibitively expensive if the spectrum must be computed starting from the lowest 
excitation energies.  By retaining only those amplitudes $x_{ia\sigma}$ for which $i$ is a core orbital on the atom of interest, 
core-excited states emerge as the smallest eigenvalues and can be computed directly.   This ``frozen valence occupied'' approximation has 
historically been called ``core--valence separation'',\cite{CedDomSch80,BarCed81,HerFra20}  
and it introduced negligible error for K-edge excitations where $\psi_{i\sigma}$ is a 1s orbital.\cite{HerFra20}
Another strategy to access core-level excitations is energy windowing,\cite{LiaFisFri11,Bes16} in which the amplitudes 
$x_{ia\sigma}$ are excluded unless $\eval{a\sigma}-\eval{i\sigma}$ lies within the window of interest.

\subsubsection{Exchange-Correlation Functionals}
\label{sec:TDDFT:Application:XC}
Before considering the accuracy of LR-TDDFT it is useful to introduce a paradigm for classifying various density-functional
approximations, for which we use the taxonomy of ``Jacob's ladder''.\cite{PerSch01,PerRuzTao05,Per13}  
At each rung on this metaphorical ladder, the functional dependence of
$E_\xc$ grows more intricate, incorporating more sophisticated functionality depending on the density, its gradients, the Laplacian, etc.:
$E_\xc[\rho_\sigma, \hat{\bm{\nabla}}\rho_\sigma, \hat{\nabla}^2\rho_\sigma, \tau_\sigma,\{\psi_{a\sigma}\}]$.
In a statistical sense (and \textit{only} in a statistical sense), 
it is true that the best functionals on the higher rungs of the ladder outperform the best 
functionals on the lower rungs.\cite{MarHea17,GoeHanBau17}    These rungs map onto various inputs 
$\rho_\sigma,\hat{\bm{\nabla}}\rho_\sigma,\hat{\nabla}^2\rho_\sigma,\ldots$ as follows.
\begin{itemize}
	\item\textbf{Rung~1:} \textit{Local Density Approximation} (LDA).  The baseline LDA functional comes from the uniform electron gas model
	in which $E_\xc$ is a functional of $\rho(\br)$ only, or of
	$\rho_\alpha(\br)$ and $\rho_\beta(\br)$ if the system is spin-polarized.   This approach does not afford useful accuracy
	for molecular quantum chemistry, with errors of 60--100~kcal/mol for atomization energies\cite{ErnScu99,MarHea17}
	and $\sim20$~kcal/mol for barrier heights.\cite{MarHea17}
	
	\item\textbf{Rung~2:}  \textit{Generalized Gradient Approximations} (GGAs).   This class of functionals include a dependence on the density
	gradients $\hat{\bm{\nabla}}\rho_\sigma(\br)$.    These are often called ``semilocal'' approximations, to distinguish them from LDA while
	acknowledging that in their mathematical form, GGAs remain local 
	in the sense that $\vxcpot{\sigma}(\br)$ is a multiplicative potential.    GGA functionals 
	significantly improve thermochemistry relative to LDA; typical errors are 
	10--20~kcal/mol for atomization energies\cite{ErnScu99,MarHea17} and 5--15~kcal/mol for barrier 
	heights.\cite{MarHea17,GoeHanBau17}
	
	\item\textbf{Rung~3:} \textit{Meta-GGAs} (mGGAs).  These are also semilocal but incorporate additional
	derivatives including $\hat{\nabla}^2\rho_\sigma$ and the \textit{kinetic energy density},
	\begin{equation}
		\tau_\sigma(\br) = \sum_i^\text{occ}\big\|\hat{\bm{\nabla}}\psi^{}_{i\sigma}(\br)\big\|^2 \; .
	\end{equation}
	The function $\tau_\sigma(\br)$ is related to the electron localization function,\cite{BecEdg90} and together with $\hat{\nabla}^2\rho_\sigma$
	it can be used to express the noninteracting kinetic energy.\cite{PerCon07}   The best mGGA functionals
	improve upon GGA thermochemistry, with errors of 5--10~kcal/mol for atomization energies\cite{MarHea17} and 
	3--6~kcal/mol for barrier heights.\cite{MarHea17,GoeHanBau17}   It is worth noting that some mGGAs introduce a considerable number of
	parameters,\cite{MarHea17} and exhibit basis-set and grid sensitivities suggesting that they may be 
	overfitted.\cite{MarHea13,Goe15,MarHea16,DasHer17}

	\item\textbf{Rung~4:} \textit{Hyper-GGAs}.  As originally defined by Perdew \etal,\cite{PerSch01,PerRuzTao05}
	this category consists of functionals that incorporate ``exact exchange and compatible correlation''.\cite{PerSch01}   
	A few genuine hyper-GGAs have been put forward,\cite{PerStaTao08,HauOdaScu12} but
	it has proven difficult to construct correlation functionals that work well with 100\% HFX.  As such, the fourth rung on Jacob's ladder 
	has effectively been redefined to mean \textit{hybrid functionals},\cite{Per13} which 
	incorporate some fraction of HFX ($0 < \cHFX < 1$), in conjunction with a fraction $1-\cHFX$ of semilocal exchange.
	These functionals are sometimes further categorized as either 
	hybrid GGAs or hybrid mGGAs, depending on the nature of the semilocal contribution.
	The best hybrid functionals exhibit errors $<5$~kcal/mol for atomization energies\cite{MarHea17} and
	2--4~kcal/mol in barrier heights.\cite{MarHea17,GoeHanBau17}   
	This has made hybrids the \textit{de facto} choice for molecular quantum chemistry.
	
	\item\textbf{Rung~5:} \textit{Double hybrid functionals}.\cite{GoeGri14} 
	These incorporate a fraction of the second-order M{\o}ller-Plesset (MP2) correlation 
	energy in addition to fractional HFX, which introduces a dependence on the virtual MOs $\{\psi_{a\sigma}\}$ whereas functionals
	on the lower rungs depend only on the occupied MOs.   Although double hybrids exhibit some of the best accuracy in contemporary DFT, 
	with errors of 1--3~kcal/mol for barrier heights,\cite{GoeGri14,GoeHanBau17}
	the introduction of MP2 correlation brings with it the
	much slower basis-set convergence of wave function methods, as well as a question of whether the orbital-dependent MP2 term
	should be self-consistently optimized.  (Typically, it is not.\cite{GoeGri14})
\end{itemize}
A comprehensive list of functionals, up-to-date as of 2017, can be found in Ref.~\citenum{MarHea17}.
The aforementioned error statistics pertain to ground-state thermochemistry whereas accuracy for vertical excitation energies computed
using LR-TDDFT is considered in the next section.  For double hybrid functionals, the formulation of LR-TDDFT\cite{GriNee07} begins
to look more like CIS with perturbative double excitations, a method known as CIS(D),\cite{HeaRicOum94} which incurs a formal scaling of
$\nroots\times\mathcal{O}(\nbasis^5)$ and often requires a form of quasi-degenerate perturbation 
theory.\cite{HeaOumMau99,CasRheHea08}   For these reasons, the application of double hybrids to excited-state
problems is still in its infancy and is not discussed in this chapter.

\begin{figure}
\centering
\fig{1.0}{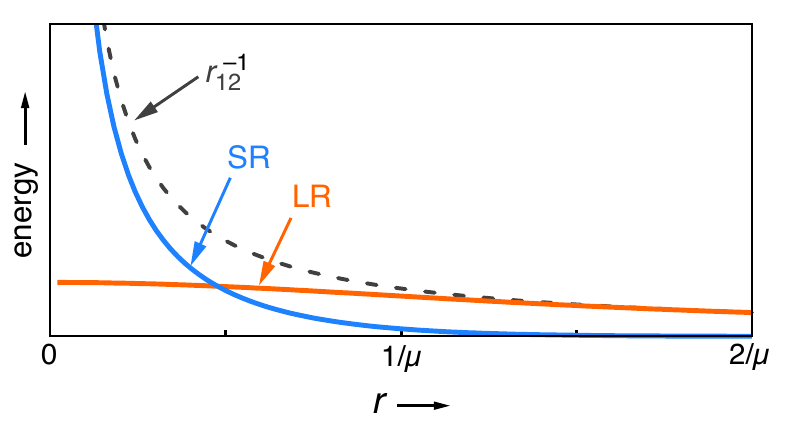}
\caption{
	Partition of the electron--electron Coulomb potential $r_{12}^{-1}$ into short-range (SR) and long-range (LR) components on a
	length scale $\sim \wRSH^{-1}$, according to Eq.~\protect\eqref{eq:r12-partition} with $\alpha + \beta = 1$.
}\label{fig:r12-partition}
\end{figure}

In the context of LR-TDDFT there is one further category of functionals that warrants mention, namely, 
\textit{range-separated hybrid} (RSH) functionals.    These partition the electron--electron Coulomb interaction 
($r_{12}^{-1}$) into a short-range (SR) component and a long-range (LR) background, typically using the error function (erf):
\begin{equation}\label{eq:r12-partition}
	\frac{1}{r^{}_{12}} = 
	\underbrace{
		\frac{1-[\alpha + \beta \,\text{erf}(\wRSH r^{}_{12})]}{r^{}_{12}}
	}_{
		\text{SR}
	}
	+
	\underbrace{
		\frac{\alpha + \beta \,\text{erf}(\wRSH r^{}_{12})}{r^{}_{12}} 
	}_{
		\text{LR}
	}	\; .
\end{equation} 
This partition introduces parameters $\alpha$, $\beta$, and $\wRSH$.
The latter is the \textit{range separation parameter} that determines the length scale ($\sim \wRSH^{-1}$) of the separation between the 
SR and LR components of $r_{12}^{-1}$; see Fig.~\ref{fig:r12-partition}.   Consider a GGA or hybrid functional of the form 
\begin{equation}\label{eq:Exc}
	E_\xc = \alpha E_\text{x,HF} + (1-\alpha)E_\text{x,GGA} + E_\text{c}^\text{GGA} \; ,
\end{equation}
where $E_\text{x,GGA}$ is the semilocal GGA exchange functional and $\alpha\equiv \cHFX$ is the coefficient of HFX.
The RSH functional corresponding to Eq.~\eqref{eq:Exc} is
\begin{equation}\label{eq:RSH}
	E^\text{RSH}_\text{xc} = 
	\alpha E_\text{x,HF}^\text{SR}
	+ (\alpha + \beta)E_\text{x,HF}^\text{LR}
	+ (1-\alpha)E_\text{x,GGA}^\text{SR}
	+  (1-\alpha-\beta) E_\text{x,GGA}^\text{LR}
	+ E^\text{GGA}_\text{c} 
\end{equation}
where 
\begin{equation}\label{eq:GGAx-LR}
	E_\text{x,GGA}^\text{LR} = E^\text{GGA}_\text{x} -E_\text{x,GGA}^\text{SR} \; .
\end{equation}
Quantities labeled ``SR'' or ``LR'' in these equations are evaluated using the corresponding component of $r_{12}^{-1}$.
The idea is to correct the asymptotic behavior of a semilocal potential $\vxcpot{\sigma}(\br)$ at long range (using HFX), while
attempting to have minimal impact on the SR behavior of the GGA or hybrid GGA in question, since that functional is responsible for
the favorable thermochemical predictions in the ground state.  RSH functionals have become popular enough that
traditional hybrids such as PBE0 or B3LYP are often called ``global hybrid'' (GH) functionals in contemporary parlance, to emphasize that 
HFX is added at all length scales in these cases.
The definition of $E_\text{x,GGA}^\text{LR}$ in Eq.~\eqref{eq:GGAx-LR}
is consistent with other literature,\cite{KroKum18,ManRefRei18,BhaCheGev18}
although it is worth noting that this quantity is not truly long-ranged.  In fact, the reason that RSH functionals were introduced
in the first place is to address the fact that semilocal exchange falls off too rapidly with distance,\cite{IikTsuYan01,TawTsuYan04,SonHirTsu07}  
leading to an insufficiently attractive interaction potential between a well-separated electron and hole.\cite{DreWeiHea03,TawTsuYan04,SonHirTsu07} 
This is further discussed in Section~\ref{sec:TDDFT:Problems:CT}.

Nomenclature and usage for RSH functionals has become somewhat muddled and the remainder of this section attempts to clarify it.
An RSH functional is any 
that uses a partition of $r_{12}^{-1}$ into SR and LR components, with Eq.~\eqref{eq:r12-partition} as the most common partition  
although other forms have been 
explored,\cite{BaeNeu05,SonTokSat07,AkiTen08,ChaHea08c,SonYamHir11,SonYamHir12}
including variants with a three-way partition of $r_{12}^{-1}$ into short\nbd-, middle\nbd-, and long-range 
contributions.\cite{HISSa,SonWatNak08,SonWatHir09,BesPeaToz09,BesAsm10,ChaKawHir18}   
For many of these functionals, the range-separation parameter(s)
are optimized or fitted alongside other parameters that define the functional and should not be modified.  Examples include
the ``$\omega$B97" class of functionals\cite{ChaHea08b,wB97X-2,wB97X-V,wB97M-V,wB97M2}
and range-separated versions of the ``Minnesota'' functionals.\cite{M11,revM11,wM05-D,wM06-D3}   
These functionals do not always afford correct the asymptotic behavior of the XC potential, however.  For the \textit{ansatz}
in Eq.~\eqref{eq:r12-partition}, the proper behavior requires $\alpha + \beta = 1$ but that constraint is sometimes violated 
(in particular, by the popular CAM-B3LYP functional\cite{CAM-B3LYP}) in the 
interest of obtaining more accurate excitation energies for localized transitions.

In contrast to this empirical approach to range separation, \textit{long-range corrected} (LRC) functionals represent a subset
of RSH functionals that are constrained to include 100\% HFX in the limit  
$r^{}_{12}\rightarrow\infty$.\cite{IikTsuYan01,RohHer08,LanRohHer08,RohMarHer09}
For a given GGA ($\cHFX=0$) or hybrid GGA ($\cHFX > 0$) functional, the corresponding LRC functional is
\begin{equation}\label{eq:LRC}
	E^\text{LRC}_\text{xc} = 
	\cHFX E_\text{x,HF}^\text{SR}
	+ E_\text{x,HF}^\text{LR}
	+ (1-\cHFX)E_\text{x,GGA}^\text{SR}
	+ E^\text{GGA}_\text{c} \; .
\end{equation}
The parameter $\wRSH$ in Eq.~\eqref{eq:r12-partition} still controls the length scale on which LR-HFX is activated, but $\alpha+\beta=1$
by construction and therefore $\vxcpot{\sigma}(r) \sim -r^{-1}$ for any $\wRSH > 0$.   The LRC strategy is thus to graft 
correct asymptotic behavior onto an existing semilocal XC functional, while doing
the least possible damage to that functional at short range.    Non-empirical adjustment (or ``tuning'')
of the parameter $\wRSH$ is often employed in this context, especially where CT states are involved, and this is discussed in 
Section~\ref{sec:TDDFT:Problems:CT}.

LRC functionals require modification of the semilocal GGA exchange functional in order to use an attenuated Coulomb potential.
(HFX integrals can be modified once and for all to separate them into LR and SR contributions.\cite{GilAda96})   There are several routes
to modify $E_\text{x,GGA}$.   The first of these, originally  
introduced by Hirao and co-workers,\cite{IikTsuYan01,TawTsuYan04,SonHirTsu07} modifies the  
exchange inhomogeneity factor that multiplies the electron-gas exchange energy density.  
This author has suggested that these functionals should be denoted LRC-$\mu$GGA,\cite{RicHer11,AlaMorHer20}
where ``GGA'' indicates the semilocal parent functional, \eg, GGA = BLYP or PBE.   Note
that ``LC'' is another common abbreviation for long-range correction so that functionals such as LC-BLYP\cite{IikTsuYan01}
might more descriptively be called LRC-$\mu$BLYP, in order to emphasize which SR-GGA exchange function ($\mu$BLYP) is being used.

For semilocal exchange functional such as
PBE that are based on a model for the exchange hole,\cite{ErnPer98,HenJanScu08a} an alternative strategy is to combine that 
model with an attenuated Coulomb potential in order to obtain $E_\text{x,GGA}^\text{SR}$.\cite{HenJanScu08a,RohMarHer09}
To distinguish this from the LRC-$\mu$PBE functional constructed using Hirao's approach, the author has suggested the 
nomenclature LRC-$\omega$PBE for the model based on the PBE exchange hole,\cite{RicHer11,AlaMorHer20}
which comports with the notation for the range-separation parameter ($\omega$) that was introduced in Ref.~\citenum{HenJanScu08a}.
The term LC-$\omega$PBE is synonymous with LRC-$\omega$PBE, and LR-$\omega$PBEh is sometimes used to indicate a short-range
hybrid ($\cHFX > 0$).   In contrast, $\omega$PBE refers only to the 
modified exchange functional, $E_\text{x,PBE}^\text{SR}$, and should not be used to mean the LRC functional because $E_\text{x,PBE}^\text{SR}$
is used in other capacities.  For example 
the HSE functional\cite{HeyScuErn03} (sometimes called HSE06\cite{KruVydIzm06}) 
uses $\omega$PBE in conjunction with SR-HFX to construct a hybrid functional that is efficient for periodic calculations.

\subsubsection{Accuracy for Vertical Excitation Energies}
\label{sec:TDDFT:Application:Benchmarks}
There have been numerous systematic surveys of the accuracy of various XC functionals for use in 
LR-TDDFT,\cite{ShaMeiSun20,KimHonHwa17,SilSchSau10,JacWatPer09,JacPerCio10,CarTruFri10,Fab10,LeaZahGor12,WanJinYu17,LooJac19b}
enough to have spawned a meta-review of the benchmark studies themselves.\cite{LauJac13}   Two of these studies are highlighted here, 
to provide some sense for how various categories of functionals can be expected to perform.  As usual
in DFT (and even more so in LR-TDDFT), for any given molecule it is likely that one could find \textit{some} XC functional 
that outperforms the statistically-best approach.  As such, it is only by understanding trends amongst functionals (and likely
trying the same calculation with more than one functional) that results can be taken seriously.   Both of the studies highlighted here
compare vertical excitation energies to experimental data, and while that has the advantage of being
a direct comparison against numbers that one might hope to simulate, it has the disadvantage that vertical excitation energies are not strictly
measurable quantities and various effects including solvatochromatic shifts and vibrational averaging are folded into the comparison.
Other studies have compared LR-TDDFT excitation energies against
correlated wave function benchmarks,\cite{SilSchSau10,KimHonHwa17,ShaMeiSun20} which makes for a much more straightforward
test of the theory although unfortunately such comparisons often tend 
to get little traction outside of quantum chemistry circles, where comparison of theory against theory is often viewed with derision.  
Fortunately, the trends that are highlighted here are reasonably similar to those obtained by comparing against \textit{ab initio} benchmarks.
In assessing the performance of various functionals, we will use the taxonomy of Jacob's ladder 
(Section~\ref{sec:TDDFT:Application:XC}) as an organizing principle.

\begin{figure}[!!ht]
\centering
\fullwidthfig{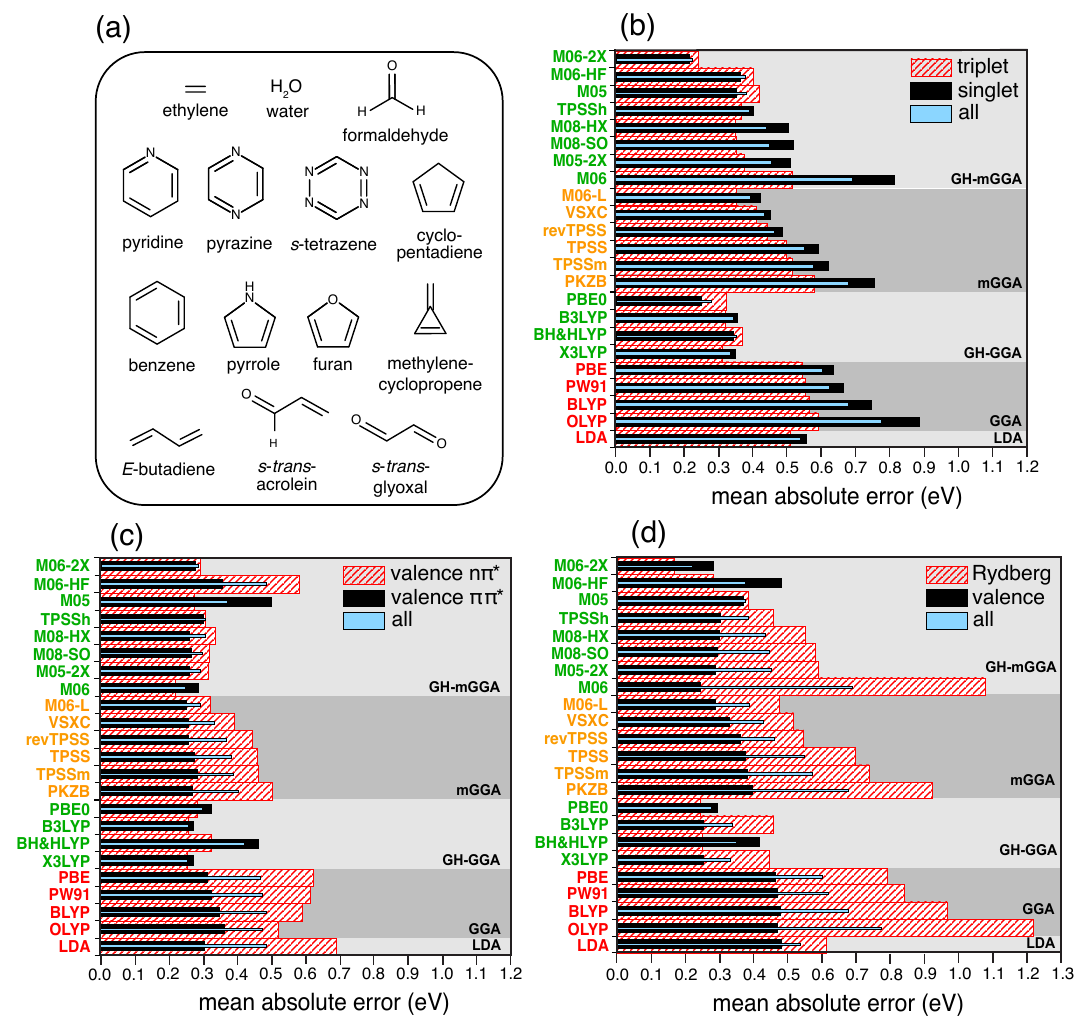}
\caption{
	Errors in TDDFT\slash 6-311++G(3df,3pd) vertical excitation energies, versus experiment.
	(a) Molecular data set, including 
	63 singlets (15 $^1\pi\pi^\ast$, 14 $^1n\pi^\ast$, 3 $^1n\sigma^\ast$, 
	1 $^1\sigma\pi^\ast$, and 30 Rydberg excitations) and 38 triplets (15 $^3\pi\pi^\ast$, 12 $^3n\pi^\ast$,
	and 11 Rydberg excitations).   Error statistics are then plotted for 
	(b) singlet versus triplet excitation energies, 
	(c) $n\pi^\ast$ versus $\pi\pi^\ast$ excitation energies, and 
	(d) Rydberg versus valence excitation energies.
	Functional names are color-coded according to the taxonomy of Jacob's ladder:
	green for global hybrids, orange for meta-GGAs, and red for GGAs and LDA.
	The global hybrids are further categorized according to whether they are based on GGAs (GH-GGAs) or meta-GGAs 
	(GH-mGGAs).   Within a given category, the functionals are ordered according to the overall MAEs for the entire data set.
	For ease of comparison, the horizontal scale is the same in each panel.
	Adapted from Ref.~\protect\citenum{LeaZahGor12}; copyright 2012 American Institute of Physics.
}\label{fig:Gordon-bench}
\end{figure}

A first set of benchmarks is depicted in Fig.~\ref{fig:Gordon-bench},
for a set of 101 transitions in 14 gas-phase molecules.\cite{LeaZahGor12}    
Error statistics are grouped and color-coded by category, including 
(GH\nbd-)GGAs and (GH\nbd-)mGGAs but not RSH functionals.   Errors are further separated into 
singlet excitations, triplet excitations, $n\pi^\ast$, $\pi\pi^\ast$, and Rydberg excitations.    Examining these data, it quickly becomes apparent
that the GH functionals significantly outperform the semilocal ones, across all types of data, although it is less clear
whether GH-mGGA functionals are categorically superior to GH-GGAs.   Perhaps surprisingly, the PBE0 and B3LYP functionals outperform
most other functionals, including much newer mGGAs and some GH-mGGAs of the Minnesota type,\cite{ZhaTru08} 
although M06-2X does exhibit slightly smaller errors.   The B3LYP and PBE0 functionals, which for many years 
have served as the closest there is to a ``default'' setting
in molecular DFT, continue to outshine many other functionals for vertical excitation energies.

The best-performing functionals (PBE0, B3LYP, and M06-2X) 
exhibit mean absolute errors  (MAEs) of $\sim0.3$~eV for the entire data set.  Unlike other functionals examined in Fig.~\ref{fig:Gordon-bench}, 
they do not seem to be systematically
worse for $n\pi^\ast$ states as compared to $\pi\pi^\ast$ states.     In contrast, none of the GGA functionals has a MAE below 0.5~eV and and
the semilocal mGGAs also have MAEs $\gtrsim 0.4$~eV, with M06-L as the best-performer amongst the latter.  All of the semilocal functionals perform
significantly worse for $n\pi^\ast$ excitations than they do for $\pi\pi^\ast$ excitations.

The comparison between Rydberg and valence excitations in Fig.~\ref{fig:Gordon-bench}(d) warrants special attention.   
With few exceptions, errors are significantly larger for the Rydberg excitations.
Significant errors in Rydberg excitation energies were noted in early molecular applications of 
LR-TDDFT,\cite{TozAmoHan99} leading to the understanding that these excitation energies are
quite sensitive to the long-range behavior of the XC potential.   That behavior is incorrect for almost all of the functionals evaluated in 
Fig.~\ref{fig:Gordon-bench}.   Later this analysis was extended to CT excitation energies in general,\cite{DreWeiHea03}
of which Rydberg excitations can be considered a special case insofar as these states involve excitation into a diffuse orbital, relatively far
from the molecular core.  As will be discussed in Section~\ref{sec:TDDFT:Problems:CT}, this observation led to the understanding that
HFX is the only component of modern functional construction that exhibits the proper asymptotic behavior for a charge-separated state, whereas 
semilocal XC potentials fall off much too rapidly with distance and thus significantly underestimate Rydberg and CT excitation energies.
It is therefore no accident that the only functionals in Fig.~\ref{fig:Gordon-bench}(d) for which the valence excitation error is larger than the Rydberg
excitation error are precisely the ones with the largest fractions of HFX:  M06-2X ($\cHFX = 0.54$),\cite{M06} M06-HF ($\cHFX = 1.0$),\cite{M06}
PBE0 ($\cHFX=0.25$),\cite{PBE0} and \bhhlyp\ ($\cHFX=0.5$).\cite{BHandH}

\begin{figure}[t]
\centering
\fullwidthfig{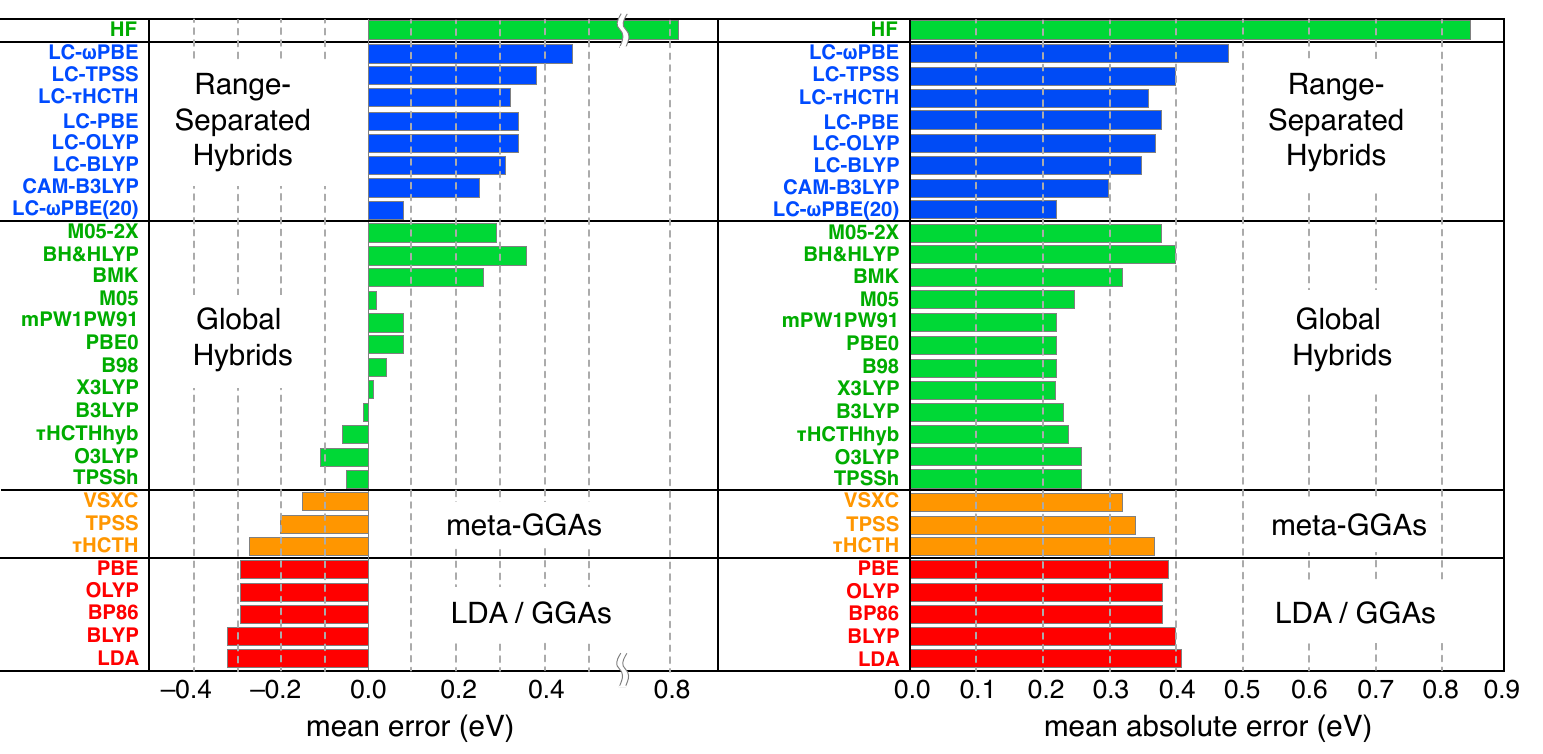}
\caption{
	(a) Mean errors and (b) mean absolute errors for 614 singlet excitation energies of 483 molecules,
	comparing LR-TDDFT\slash 6-311+G(2d,p)
	vertical excitation energies (with solvent corrections) to experimental absorption maxima.
	Original data are from Ref.~\protect\citenum{JacWatPer09} and this figure is adapted from Ref.~\protect\citenum{LauJac13};
	copyright 2013 John Wiley and Sons.
}\label{fig:LauJac13}
\end{figure}

A second statistical survey is presented in Fig.~\ref{fig:LauJac13}, taken from one of the largest statistical assessments of LR-TDDFT
to date:\cite{JacWatPer09}   614 singlet excitation energies in 483 solution-phase organic molecules.   Vertical excitation energies
have been corrected for solvent effects and compared to experimental band maxima.    (For a discussion of dielectric
continuum solvation models and their application to LR-TDDFT, see Ref.~\citenum{Her21a}.)   Functionals are once again grouped by category and
this larger data set makes it clear that the GH functionals generally outperform the semilocal mGGA functionals, which themselves outperform
the semilocal GGAs.  For most of the GH functionals, MAEs are 0.2--0.3~eV as compared to 0.4--0.5~eV for the semilocal functionals, but
the mean \textit{signed} errors [Fig.~\ref{fig:LauJac13}(a)] are much smaller for the GH functionals.
Signed errors are nearly zero for PBE0 and B3LYP, indicating no systematic error in these cases.  In contrast, errors are much larger for GH
functionals with a large fraction of HFX, including 
BMK ($\cHFX=0.42$),\cite{BMK} M05-2X ($\cHFX=0.56$),\cite{M05-2X} and \bhhlyp\ ($\cHFX=0.5$).\cite{BHandH}    
These large-$\cHFX$ functionals exhibit bias towards
overestimation of excitation energies, whereas semilocal functionals consistently underestimate them.

Also included in Fig.~\ref{fig:LauJac13} are error statistics for a set of RSH functionals.   MAEs for these functionals span a wide range from
0.2--0.5~eV and in that sense are not better than the GH functionals.   Furthermore, whereas semilocal functionals systematically underestimate
excitation energies, GH functionals are largely free of that bias except when $\cHFX > 0.4$.  Finally, RSH functionals systematically overestimate
excitation energies, which has also been observed in more recent benchmarks for biochromophores.\cite{ShaMeiSun20}
Putting these observations together, it seems that some HFX is optimal, perhaps $\cHFX \approx 0.20$--0.25, with excitation energies that are
too low for smaller values and too high for larger ones.  Included the the latter category are 
many LRC functionals that use $\cHFX=1$ in the asymptotic limit.  With that in mind, it is interesting to compare 
error statistics for LC-$\omega$PBE and LC-$\omega$PBE(20) in Fig.~\ref{fig:LauJac13}.  
The former uses a range-separation parameter $\omega = 0.4$~bohr$^{-1}$ that was optimized for ground-state properties,\cite{VydScu06}
whereas in LC-$\omega$PBE(20) this parameter is set to $\omega = 0.2$~bohr$^{-1}$, leading to significant reduction in the errors.
Attempts to fit $\omega$ using both ground-state properties as well as excitation energies typically lead to values in the range 
$\omega = 0.2$--0.3~bohr$^{-1}$, depending on whether short-range HFX is present or not.\cite{RohHer08,RohMarHer09,LanHer09}
This is consistent with the revised choice in LC-$\omega$PBE(20).

\subsubsection{Visualization}
\label{sec:TDDFT:Application:Visualization}
Having computed an excitation energy, there are a variety of tools 
available to visualize the excited state in question.   One could simply examine each pair of occupied and virtual MOs for which the coefficient
$x_{ia\sigma}$ is large, but this is often tedious due to significant configuration mixing, especially in the virtual space.  At the CIS level, it is easy
to understand why the canonical MOs are not a good basis for visualization purposes, since 
Koopmans' theorem implies that the virtual MOs are reasonable orbitals for electron attachment, not excitation.\cite{SzaOst82}  
The Hartree-Fock virtual MOs feel the full repulsive potential of the $N$-electron charge
density, whereas the occupied MOs feel only $N-1$ electrons, which has the effect of rendering the virtual MOs significantly more
diffuse than the occupied MOs.   Often the Hartree-Fock virtual MOs are simply unbound and therefore 
represent discretized continuum states,\cite{Her15} whose shapes are sensitive to small changes in basis set.\cite{BaeGrivan13}
Significant configuration mixing is therefore necessary to obtain a localized valence excitation.

In principle, exact Kohn-Sham MOs are a much better basis for excitations,\cite{RefShaGov12,BaeGrivan13,vanGriBae14} 
since both occupied and virtual MOs are subject to the same $N$-electron potential,
and in practice it is often the case that the first few Kohn-Sham virtual orbitals are bound ($\eval{a\sigma}<0$).    
Hybrid functionals, however, push the virtual orbitals and
their eigenvalues back towards the Hartree-Fock picture and even 20--25\% HFX can be enough to engender significant configuration mixing
due simply to the diffuseness of the virtual MOs.

This type of configuration mixing is artificial, in the sense that it can be removed via orbital rotation and therefore does not represent true
multiconfigurational character in the excited state.   The relevant transformations of the canonical occupied MOs is a unitary matrix
$\mathbf{U}_\text{o}$ that diagonalizes $\Delta\mathbf{P}^\text{elec}$ in Eq.~\eqref{eq:dP-elec}:
\begin{equation}\label{eq:Uo}
	\mathbf{U}_\text{o}(\Delta\mathbf{P}^\text{elec})\mathbf{U}^\dagger_\text{o} = \bm{\Lambda}^2 =
	\left(
	\begin{array}{cccc}
		\lambda^{2}_1 	& 0 				& 0 		& \cdots \\
		0 				& \lambda^{2}_2 	& 0 		& \cdots \\
						& 				& \ddots	& 0 \\
		0 				& \cdots 			& 0 		& \lambda^{2}_{\nocc}\\
	\end{array}
	\right) \; .
\end{equation}
The $\nocc\times\nocc$ diagonal matrix $\bm{\Lambda}^2$ contains the eigenvalues, which are strictly non-negative and are 
normalized such that $\sum_{i}\lambda_{i}^2=1$.   The corresponding transformation of the canonical virtual MOs is
\begin{equation}\label{eq:Uv}
	\mathbf{U}_\text{v}(\Delta\mathbf{P}^\text{hole})\mathbf{U}^\dagger_\text{v} = 
	\left(\!\!
	\begin{array}{cc}
		-\bm{\Lambda}^2 	& \!\!\bm{0} \\
		\bm{0} 			& \!\!\bm{0} \\
	\end{array}
	\!\!\right) \; .
\end{equation}
These two transformations define the \textit{natural transition orbitals} (NTOs),\cite{LuzSukUma74,Mar03,May07a,Sur07}
which are the natural orbitals (eigenfunctions) of the excited-state density matrix.\cite{Sur07}   
They can equivalently be defined based on a singular value decomposition of the $\nocc\times\nvir$ matrix of amplitudes, 
$\mathbf{x}+\mathbf{y}$:\cite{Mar03,May07a}
\begin{equation}\label{eq:SVD}
	\mathbf{U}_\text{o}(\mathbf{x}+\mathbf{y})\mathbf{U}^\dagger_\text{v}
	=\left(
	\begin{array}{cc}
		\bm{\Lambda} 	& \bm{0} \\
		\bm{0} 			& \bm{0} \\
	\end{array}
	\right) \; .
\end{equation}
This form demonstrates that no more than $\nocc$ of the singular values $\{\lambda_{i}\}$
are non-zero.    These eigenvalues appear in pairs ($\pm\lambda_{i}^2$) when 
$\Delta\mathbf{P}^\text{elec}$ and $\Delta\mathbf{P}^\text{hole}$ are diagonalized, 
because the natural occupation numbers of the excited-state density matrix are\cite{Sur07}
\begin{equation}
	n_r = \begin{cases}
		1 - \lambda_r^2, & 1 \le r \le \nocc \\ 
		\lambda_r^2, & \nocc < r \le 2 \nocc \\ 
		0, & r > 2 \nocc \\ 
	\end{cases} \; .
\end{equation}
The matrices $\mathbf{U}_\text{o}$ and $\mathbf{U}_\text{v}$ transform the canonical
occupied and virtual MOs, respectively, into a set of ``hole'' orbitals $\{\psi_{i}^\text{hole}\}$ along with corresponding ``particle'' 
(or ``electron'') orbitals, $\{\psi_{i}^\text{elec}\}$.     These are the NTOs, and their diminishing importance for describing the excitation in question 
is quantified by the values $\lambda_1^2 > \lambda_2^2 > \lambda_3^2 > \cdots$.

\begin{figure}[t]
\centering
\fig{1.0}{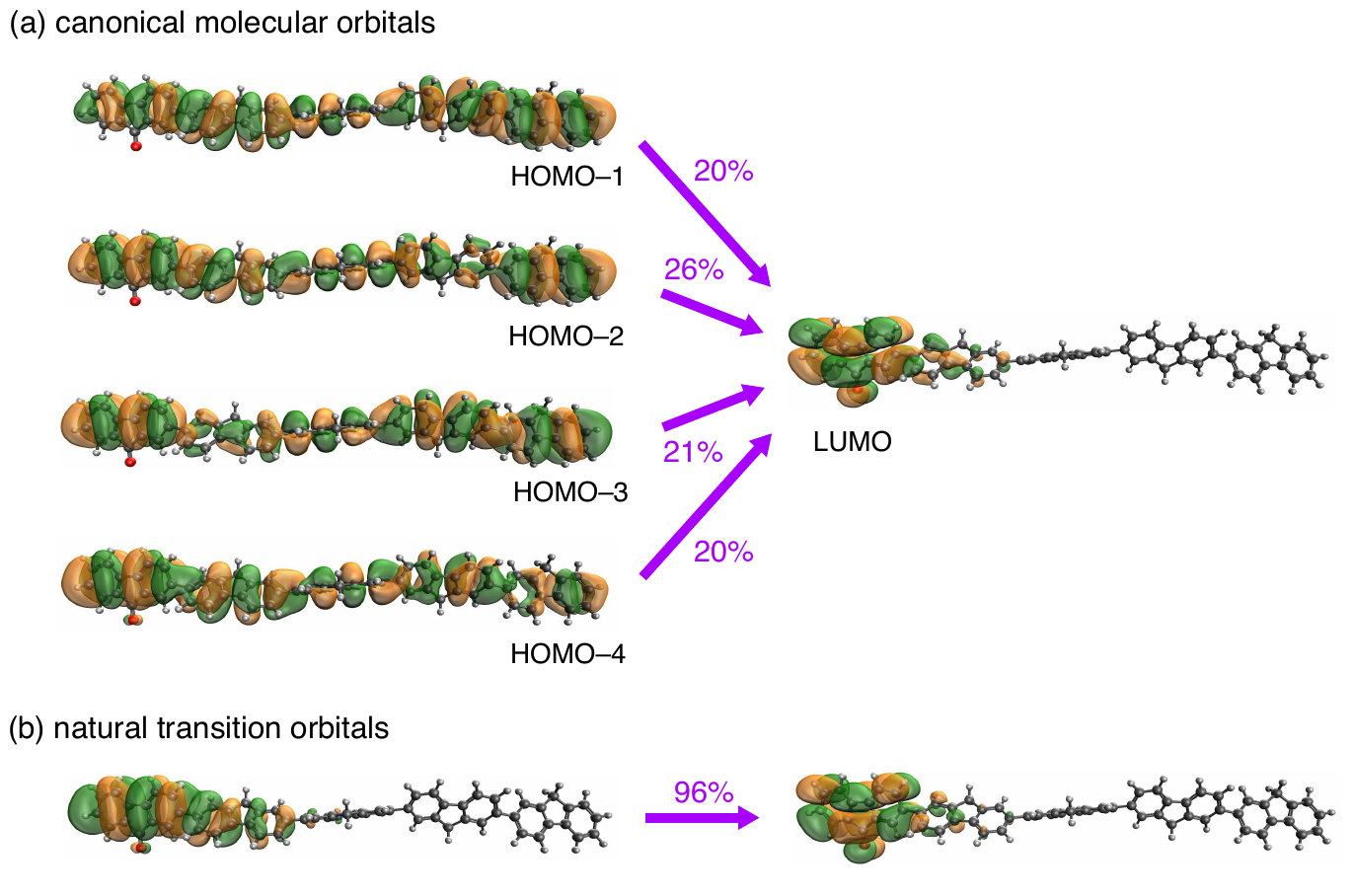}
\caption{
	(a) Canonical MO representation (with weights $x_{ia}^2$ expressed as percentages) and 
	(b) principle NTO pair (with weight $\lambda_{i}^2$) for $\text{S}_0\ra\text{S}_2$ excitation of a five-unit, fluorenone-terminated polyfluorene
	molecule in which the leftmost monomer contains a carbonyl defect that localizes the excitation.  
	LR-TDDFT calculations were performed at the CAM-B3LYP\slash 3-21G* level within the TDA and the unrelaxed density is analyzed.
}\label{fig:NTOs}
\end{figure}

NTOs often provide a much more compact description of 
the wave function as compared to the canonical MOs, yet one that preserves the phase and nodal structure that can be helpful in qualitatively 
characterizing the nature of the excitation.  This is illustrated in Fig.~\ref{fig:NTOs} for the $\text{S}_0 \ra \text{S}_2$ excitation of a 
five-unit polyfluorene molecule in which a carbonyl defect in one of the terminal fluorene monomers serves to localize the excitation.
That localization, however, is not obvious from the canonical MOs, which are delocalized across the length of the molecule, but arises from a coherent
superposition of four different occupied MOs.  Upon transformation to the NTO basis
there is only one significant singular value, with $\lambda_1^2 = 0.96$.   
The principle NTO pair, $\psi_1^\text{hole} \rightarrow \psi_1^\text{elec}$ in 
Fig.~\ref{fig:NTOs}(b), thus paints a picture that is 96\% complete.

There is an unfortunate tendency in the literature to refer to $\psi_1^\text{hole}$ as the 
``highest-occupied NTO'' (HONTO), with $\psi_1^\text{elec}$ then deemed the ``lowest-unoccupied NTO'' (LUNTO).
This terminology is incorrect insofar as ``highest'' and ``lowest'' (as in HOMO and LUMO) refer to orbital energies, which 
are not well-defined for the NTOs because the Fock matrix is not diagonal in that representation. 
The HONTO\slash LUNTO terminology should therefore be avoided so as not to conflate visual depictions of NTOs with qualitative arguments that 
might be based on one-electron energy levels, which are only well-defined in the canonical MO basis.   
The term {\em principle transition orbitals\/} (or perhaps {\em principle NTOs}) is suggested instead, 
to describe the pair of orbitals corresponding to the largest $\lambda_i$.   One might therefore describe these in sequence of principle NTOs (pNTOs)  
as $\rm pNTO, pNTO-1, pNTO-2, \ldots$ for $\lambda_1^2 > \lambda_2^2 > \lambda_3^2 > \cdots$.

Another common tool to visualize an excitation is the 
density difference as compared to the ground state.   The unrelaxed density difference
\begin{equation}
	\Delta\rho(\br) = \Delta \rho^{}_\text{elec}(\mathbf{r}) + \Delta \rho^{}_\text{hole}(\mathbf{r})
\end{equation} 
has particle and hole components that are the real-space analogues of the density matrices 
$\Delta\mathbf{P}^\text{elec}$ and $\Delta\mathbf{P}^\text{hole}$ in Eq.~\eqref{eq:dP_v1}.    Using the NTOs, the particle and hole 
densities may be expressed as 
\begin{subequations}\label{eq:rho-elec,hole}
\begin{align}\label{eq:dRho-elec}
	\Delta \rho^{}_\text{elec}(\mathbf{r}) &= \sum_{i=1}^{\nocc} \lambda_i^2 \, \bigl|\psi_i^{\rm elec}(\mathbf{r})\bigr|^2 
\\ \label{eq:dRho-hole}
	\Delta \rho^{}_\text{hole}(\mathbf{r}) &= -\sum_{i=1}^{\nocc} \lambda_i^2 \, \bigl|\psi_i^{\rm hole}(\mathbf{r})\bigr|^2 \; .
\end{align}
\end{subequations}
Note that $\Delta \rho^{}_\text{elec}(\mathbf{r})$ is positive definite and $\Delta \rho^{}_\text{hole}(\mathbf{r})$ is negative definite, consistent with 
Eq.~\eqref{eq:dP_v1}.   Because the NTOs are defined by a singular value decomposition, which distills the $\nocc\times\nvir$ 
matrix $\mathbf{x}+\mathbf{y}$ into the fewest number of non-zero parameters, 
the densities in Eq.~\eqref{eq:rho-elec,hole} are often dominated by the principle NTO pair.  
Although it is not widely recognized, the quantities $\Delta \rho^{}_\text{elec}(\mathbf{r})$ and $\Delta \rho^{}_\text{hole}(\mathbf{r})$ 
are precisely the \textit{attachment density} and the \textit{detachment density}, respectively, 
which have long been used to visualize excited states.\cite{HeaGraMau95,EtiAssMon14a,Eti18}
(These were originally introduced in a different way,\cite{HeaGraMau95} 
based on eigenvectors of $\Delta\mathbf{P}$ that afford positive or negative eigenvalues, respectively.)    
In the author's view, NTOs are still the preferable description since phase information is lost upon squaring the orbitals in
Eq.~\eqref{eq:rho-elec,hole}.

\begin{figure}
\centering
\fig{1.0}{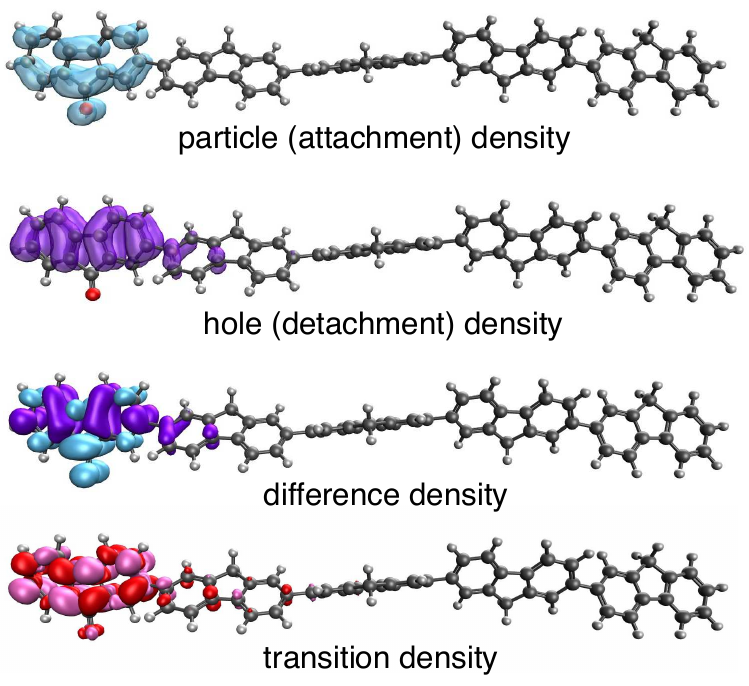}
\caption{
	Visualization of the $\text{S}_0\ra\text{S}_2$ excitation of the fluorenone-terminated polyfluorene whose orbital depiction is given in 
	Fig.~\protect\ref{fig:NTOs}, represented here in terms of difference densities.  These include the particle 
	density (or attachment density) $\Delta\rho_\text{elec}(\br)$, the hole density (or detachment density) $\Delta\rho_\text{hole}(\br)$, 
	the unrelaxed difference density $\Delta\rho(\br) = \Delta\rho_\text{elec}(\br) + \Delta\rho_\text{hole}(\br)$, and the transition density 
	$T(\br)$.  Each isosurface encompasses 95--97\% of the density in question.
}\label{fig:densities}
\end{figure}

Figure~\ref{fig:densities} illustrates these densities for the same 
$\text{S}_0\ra\text{S}_2$ excitation of polyfluorene that was examined in Fig.~\ref{fig:NTOs}.   
Because $\lambda_1^2 \approx 1$, the particle and hole (or attachment and detachment) densities have the same information
content as the principle NTO pair in Fig.~\ref{fig:NTOs}(b).
Also shown in Fig.~\ref{fig:densities} is the transition density 
$T(\br)\equiv T(\br,\br)$ where $T(\br,\br')$ is defined in Eq.~\eqref{eq:TDM}.   For an excitation $|\Psi_0\rangle \ra |\Psi\rangle$, 
the general definition of this quantity is\cite{PlaWorDre14}
 \begin{equation}
 	T(\br,\br') = N \int \Psi_0^\ast(\br',\br^{}_2,\ldots,\br^{}_N) \; \Psi(\br,\br^{}_2,\ldots,\br^{}_N) \; 
	d\br^{}_2\cdots d\br^{}_N \; ,
 \end{equation}
and for LR-TDDFT in the NTO representation it is 
 \begin{equation}\label{eq:TDM-NTO}
	T(\mathbf{r},\mathbf{r}') = \sum_{i} \lambda^{}_{i} \,
	\psi^{\rm elec}_{i}(\mathbf{r}) \, \bigl[\psi^{\rm hole}_{i}(\mathbf{r}')\bigr]^\ast \; .
\end{equation}
Thus the NTOs distill the content of the transition density into the smallest possible number of 
particle--hole pairs.  In that well-defined sense, the NTOs are the best orbitals for visualization purposes, and detection of more than one 
significant singular value $\lambda_{i}$ indicates unresolvable multideterminant character in the excited state.
For the excitation depicted in Fig.~\ref{fig:densities} there is little of that character, and 
$T(\br) \approx \psi^{\rm elec}_{1}(\mathbf{r}) \; \psi^{\rm hole}_{1}(\mathbf{r})$ is well described by the principle NTO pair.  
The nature of this product accounts for the somewhat more complicated
nodal structure as compared to $\Delta\rho_\text{elec}(\br) \approx |\psi^{\rm elec}_{1}(\mathbf{r})|^2$ or 
$\Delta\rho_\text{hole}(\br) \approx |\psi^{\rm hole}_{1}(\mathbf{r})|^2$.

\subsection{Systemic Problems}
\label{sec:TDDFT:Problems}
The utility of LR-TDDFT lies in its combination of low cost, which facilitates calculations on systems such as $\rm C_{119}H_{154}ClN_{21}O_{40}$
[Fig~\ref{fig:TDDFT-large-Besley}(b)] or conjugated polymers (Figs.~\ref{fig:NTOs} and \ref{fig:densities}), along with an accuracy 
of $\sim 0.3$~eV for localized valence excitations.   That level of accuracy requires a treatment of dynamical correlation effects, as
seen from the CIS errors in Fig.~\ref{fig:LauJac13} that exceed 0.8~eV, comparable to the $\sim1$~eV of correlation energy for a pair of electrons.
That is the good news.  In this section we discuss some of the bad news, namely, systematic errors that make certain types of problems 
extremely challenging for LR-TDDFT.    Of these, the most widely discussed is severe underestimation of excitation energies for states with
substantial CT character, ultimately manifesting in spurious CT states in a sufficiently large system
(Section~\ref{sec:TDDFT:Problems:CT}).   The second is a problem with the topology of conical intersections that involve the ground state,
which presents a problem for direct \textit{ab initio} simulation of internal conversion following photoexcitation
(Section~\ref{sec:TDDFT:Problems:CX}).

\subsubsection{Description of Charge Transfer}
\label{sec:TDDFT:Problems:CT}
Problems with the description of long-range CT excitations manifests in small, gas-phase molecules as Rydberg excitation energies that are 
systematically too low,\cite{TozAmoHan99} 
even when reasonable accuracy is obtained for valence excitations.   This was noticed in early studies of
LR-TDDFT and was quickly diagnosed as a symptom of incorrect asymptotic 
decay of the XC potential in GGA functionals that existed up to that point.\cite{LB94,TozHan98b,SchGrivan00,GruGrivan01}
The same problem was quickly recognized to affect CT excitation energies.\cite{CasGutGua00,Cas02}
Both CT and Rydberg excitations are sensitive to the long-range behavior of the potential, which 
ought to be $\vxcpot{\sigma}(r) \sim -r^{-1}$ for a charge-neutral molecule.\cite{LevPerSah84,AlmPed84,Almvon85a,CheLemRaz99,PerKur03}  
This asymptotic behavior ought to be borne by the exchange potential because correlation dies off more quickly,\cite{AlmPed84,Almvon85b}
but in practice so does semilocal exchange.

\begin{figure}
\centering
\fig{1.0}{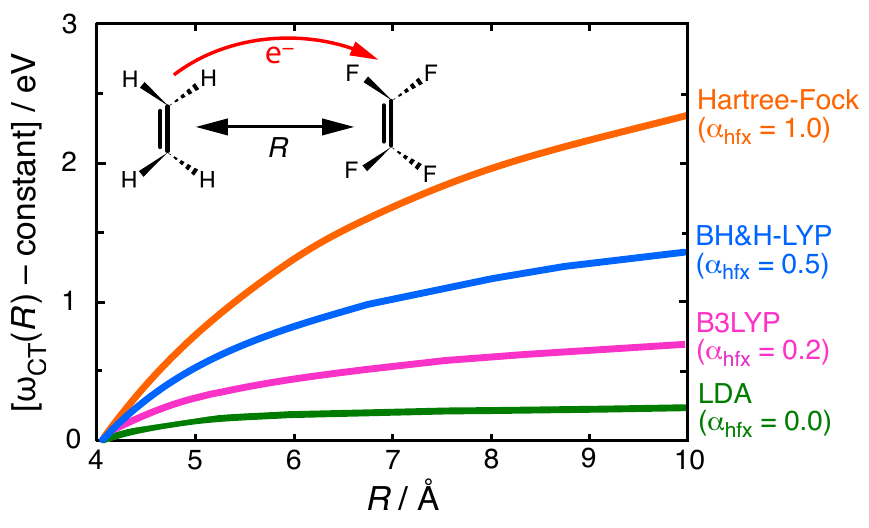}
\caption{
	Distance dependence for the lowest intermolecular CT excitation
	in $\rm (C_2H_4)\cdots (C_2F_4)$ computed using
	functionals with various fractions of HFX, as indicated.
	The curves are shifted to a common origin at $R=4$~\AA\ in order to emphasize the distance dependence of 
	$\omega^{}_{\rm CT}(R)$, which varies asymptotical as $-\cHFX/R$.
	Adapted from Ref.~\protect\citenum{DreWeiHea03}; copyright 2003 American Institute of Physics.
}\label{fig:CT_vs_R_Dreuw}
\end{figure}

Consider the form of the LR-TDDFT pseudo-eigenvalue problem for an excitation between MOs $\psi_{i\sigma}$ and $\psi_{a\sigma}$ that
are well-separated in space, such that $\psi_{i\sigma}(\br) \, \psi_{a\sigma}(\br) \approx 0$ everywhere.    A semilocal expression for 
$\vxcpot{\sigma}(\br)$ affords a semilocal XC kernel, such that the matrix elements $(ia|\kxcker{\sigma\tau}|jb)$ in $\mathbf{A}$ and $\mathbf{B}$ 
vanish in such a situation, for all $j$ and $b$. Ignoring spin by setting $\sigma=\tau$ in Eq.~\eqref{eq:A-B-matrix}, this leaves 
\begin{equation}\label{eq:A-asymp}
	A_{ia,jb} \approx (\eval{a} - \eval{i}) \delta_{ij}\delta_{ab} - \cHFX (ij|ab) \; .
\end{equation}
Only the integral $(ij|ab)$, which comes from the HFX term, survives to provide distance dependence for the $i \ra a$ excitation.
A pictorial illustration is provided in Fig.~\ref{fig:CT_vs_R_Dreuw},
which plots the distance dependence of the lowest CT excitation energy ($\omega_\text{CT}$) in the $\rm (C_2H_4)\cdots (C_2F_4)$ dimer as a function of 
intermolecular separation.\cite{DreWeiHea03}    Only Hartree-Fock theory affords the correct
distance dependence for $\omega_\text{CT}(R)$, which varies according to the Mulliken formula,\cite{Mul52,GriBae04b,DreHea05}
\begin{equation}
	\omega^{}_\text{CT}(R) = \text{IE}_\text{donor} + \text{EA}_\text{acceptor} - \frac{1}{R}
\end{equation}
in atomic units.    For hybrid functionals the last term becomes $-\cHFX/R$ rather than $-1/R$, leading to a too-small penalty for long-range CT.
For semilocal functionals where $\cHFX=0$, the CT excitation energy has no 
distance dependence whatsoever, once the donor and acceptor moieties are sufficiently far apart such that their orbitals do not overlap.  
This is reflected in the flat $\omega_\text{CT}(R)$ profile for the LDA functional in Fig.~\ref{fig:CT_vs_R_Dreuw}.  
As a result, long-range CT excitation energies are almost invariably too small in LR-TDDFT unless the functional contains 100\% HFX, 
which it usually does not because fully nonlocal exchange is somewhat unbalanced given the local nature of existing correlation functionals.
The M06-HF functional is an example that does use 100\% HFX, leading to reasonable performance for Rydberg states but larger errors 
for valence excitations [Fig.~\ref{fig:Gordon-bench}(d)].

Where small-molecule benchmarks are available, errors in CT excitation energies can exceed 3~eV,\cite{RohMarHer09} but the problem
gets worse in larger molecules so that error is likely limited only by the size of the benchmark systems for which reliable \textit{ab initio} results
are available.  A consequence of this severe underestimation of CT excitation energies 
is the appearance of completely spurious CT excited states in large systems, especially solvated 
chromophores\cite{BerSprHut03,BerSprHut04,Bes04,NeuLouBae05,BerSpr05,NeuGriBae06,LanHer07,IsbMarCur13}
but also large molecules.\cite{DreHea04,MagTre07,MasCamFri18}    When the system size is sufficiently large, there are inevitably 
well-separated occupied and virtual MOs such that the orbital energy gap $\eval{a}-\eval{i}$ is small.  For $\cHFX \approx 0$, the
electron--hole interaction vanishes and the 
diagonally-dominant $\mathbf{A}$ matrix consists of weakly-coupled blocks corresponding to these spurious CT transitions.  The kernel 
$\fxcker{\sigma\sigma}(\br,\br')$ lacks the long-range exchange (or a derivative discontinuity,\cite{GriBae04b,CasHui12,Ull12} or frequency
dependence\cite{Mai17}) that is needed to provide an energetic penalty for CT and an upshift to $\omega^{}_\text{CT} \approx \eval{a}-\eval{i}$
in Eq.~\eqref{eq:A-asymp}.

\begin{figure}
\centering
\fig{1.0}{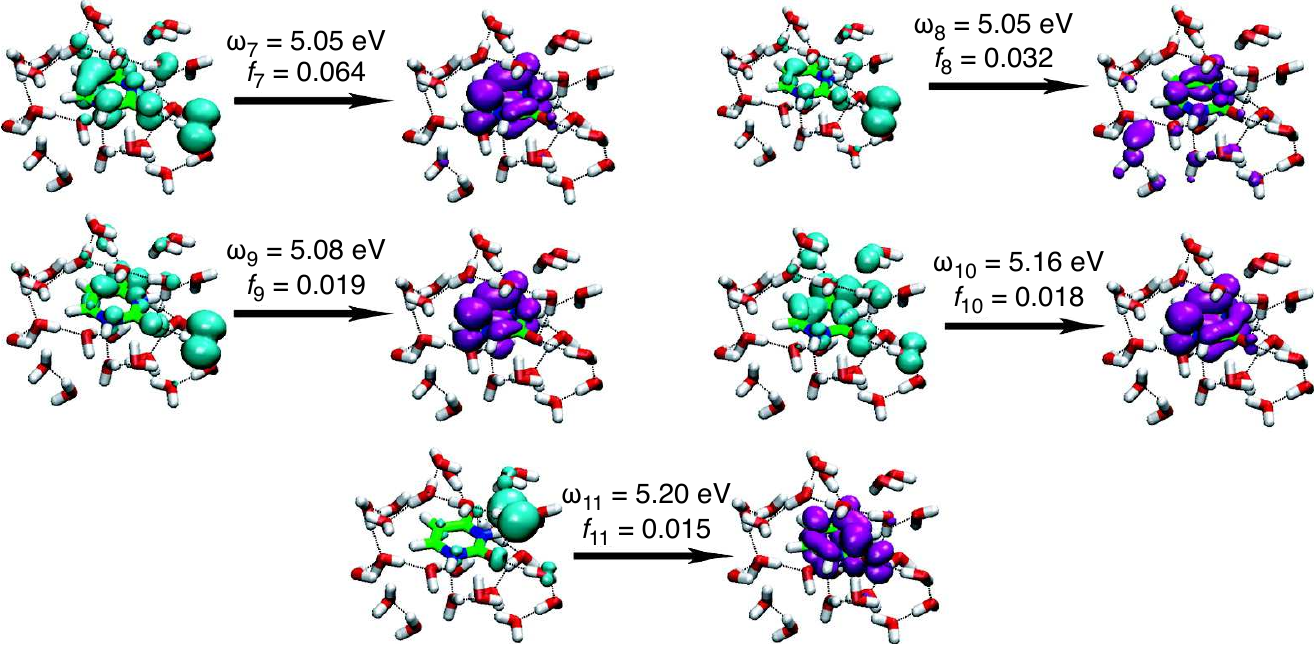}
\caption{
	Selected detachment (hole) and attachment (particle) densities, 
	for excited states of (uracil)(H$_2$O)$_{25}$ computed using LR-TDDFT at the PBE0\slash 6-31+G* level.
	These states exhibit spurious solvent-to-chromophore CT in the spectral vicinity of the $^1\pi\pi^\ast$ state
	at $\omega \approx 5.1$~eV.
	Excitation energies $\omega_n$ and oscillator strengths $f_{0n}$ are shown, illustrating intensity borrowing
	by the spurious CT states.  Reprinted from Ref.~\protect\citenum{LanHer07}; copyright 2007 American Chemical Society.  
}\label{fig:LanHer07-uracilCT}
\end{figure}

A physical example is shown in Fig.~\ref{fig:LanHer07-uracilCT} for a model of aqueously-solvated uracil.\cite{LanHer07}
Whereas this system ought to have only a $^1n\pi^\ast$ and a $^1\pi\pi^\ast$ state below 6~eV,\cite{LanRohHer08} a hybrid LR-TDDFT
calculation using the PBE0 functional results in numerous low-energy solvent-to-chromophore CT states, \eg, 27 states below 6~eV for the
(uracil)(H$_2$O)$_{25}$ cluster that is shown in Fig.~\ref{fig:LanHer07-uracilCT}
and additional states as the size of the water cluster grows.\cite{LanHer07,LanRohHer08}
Many of these states are accidentally near-degenerate with the optically-bright $^1\pi\pi^\ast$ state and as a result these nominally dark
CT states can acquire intensity from the bright state, which diminishes the intensity of the latter because total oscillator strength
is conserved by the Thomas-Reiche-Kuhn sum rule, Eq.~\eqref{eq:TRK}.
The state $\omega_9$ in Fig.~\ref{fig:LanHer07-uracilCT} exhibits the largest coefficient with $\pi\pi^\ast$ character,\cite{LanHer07}
yet due to spurious intensity borrowing it does not exhibit the largest oscillator strength and itself contains some contribution
from solvent-to-chromophore CT.    Fortunately, the same sum rule can be used to argue that the overall spectral envelope may still be valid
upon ensemble averaging and broadening, even if some fraction of the oscillator strength has been ported onto spurious CT excitations.


In large chromophores, such as conjugated polymers, spurious low-energy CT excitations can manifest as artificial delocalization of the
excitation across the length of the chromophore,\cite{IguTreChe07,MewPlaDre17,MasCamFri18} whereas the CIS method predicts that
exciton size eventually saturates even as conjugation length increases.\cite{MewPlaDre17}   As such, there is a need to develop a metric for whether a 
particular excited state has too much CT character for its excitation energy to be trusted.   
The first such CT metric to see widespread use was the quantity $\Lambda$ defined by\cite{PeaBenHel08}  
\begin{equation}\label{eq:Lambda}
	\Lambda = \frac{
		\sum_{ia\sigma} (x_{ia\sigma}+y_{ia\sigma})^2 O_{ia\sigma}
	}{
		\sum_{jb\tau} (x_{jb\tau} + y_{jb\tau})^2
	}
\end{equation}
where 
\begin{equation}
	O_{ia\sigma} = \int \big|\psi_{i\sigma}(\br)\big| \cdot \big|\psi_{a\sigma}(\br)\big| \; d\mathbf{r}
\end{equation}
measures the overlap of $|\psi_{i\sigma}(\br)|$ and $|\psi_{a\sigma}(\br)|$.   
(Absolute values are required since the occupied and virtual MOs are orthogonal.)
This overlap is then weighted by the LR-TDDFT amplitudes and normalized such that $0 \leq \Lambda \leq 1$.
For calculations that do not invoke the TDA, however, the denominator in Eq.~\eqref{eq:Lambda} is an odd choice, given the 
normalization condition in Eq.~\eqref{eq:normalization}, and this inconsistency has propagated into other CT metrics used in 
LR-TDDFT.\cite{GuiCorMen13,GuiCorAda14}   
Regarding the metric in Eq.~\eqref{eq:Lambda}, an 
early benchmark study concluded that $0.45 \leq \Lambda \leq 0.89$ for localized valence excitations, making values in this range ``safe'' for 
LR-TDDFT, whereas $0.08 \leq \Lambda \leq 0.27$ for Rydberg excitations, which are unsafe.\cite{PeaBenHel08}   It was suggested that excitation
energies for which $\Lambda \lesssim 0.3$--0.4 (depending on the functional) should not be trusted.    
Various LR-TDDFT errors have been rationalized by appeal to $\Lambda$ or similar metrics.\cite{PloTozDre10,PeaToz12,LeaZahGor12,MooSunGov15}

The point at which CT character becomes a problem is dependent on the manner in which it is quantified,\cite{MarAshCur21}
and several alternative CT metrics have been 
suggested.\cite{PeaBenHel08,LeBAdaCio11,JacLeBAda12,EhaFukAda13,GuiCorMen13,GuiCorAda14,SavGuiBre17,CamMasFri17,GarMasCam18,CamPerCio19}   
Ciofini and co-workers introduced a widely-used ``$D_\text{CT}$ metric'',\cite{LeBAdaCio11} 
originally defined in a rather complicated way but which ultimately measures 
the distance between the centroids of $\Delta\rho_\text{elec}(\br)$ and $\Delta\rho_\text{hole}(\br)$.
The centroid of $\Delta\rho_\text{elec}(\br)$ is 
\begin{equation}
	\langle\br_\text{elec}\rangle = \int \br \; \Delta\rho_\text{elec}(\br) \; d\br
\end{equation}
with an analogous definition for $\Delta\rho_\text{hole}(\br)$.   If one defines  
\begin{equation}
	d^\pm_\text{elec/hole} = \big\| 
		\langle\br_\text{elec}\rangle \pm \langle\br_\text{hole}\rangle
	\big\| \; ,
\end{equation}
then the distance between centroids of the electron and the hole is $d^-_\text{elec/hole}$, whereas $d^+_\text{elec/hole}$ is the average position 
of the center of mass of the exciton.  The quantity $d^-_\text{elec/hole}$ is equivalent to the $D_\text{CT}$ metric but is 
more directly connected to the physics of the excitation.  Other similar descriptors can be envisaged.\cite{PlaThoBap15,MewDre19} 
For example, by defining the root-mean-square size of the electron and the hole,
\begin{subequations}
\begin{align}
	\sigma_\text{elec} &= \big(
		\langle\br_\text{elec}\bm{\cdot}\br_\text{elec}\rangle - \langle\br_\text{elec}\rangle \bm{\cdot} \langle\br_\text{elec}\rangle
	\big)^{1/2}
\\
	\sigma_\text{hole} &= \big(
		\langle\br_\text{hole}\bm{\cdot}\br_\text{hole}\rangle - \langle\br_\text{hole}\rangle \bm{\cdot} \langle\br_\text{hole}\rangle
	\big)^{1/2}  \; ,
\end{align}
\end{subequations}
one may define a \textit{charge-displacement distance},
\begin{equation}
	d_\text{CD} = d^-_\text{elec/hole} - \tfrac{1}{2}(\sigma_\text{elec} + \sigma_\text{hole}) \; .
\end{equation}
The quantity $d_\text{CD}$ connects directly to the properties of the exciton and is a more physically-motivated version of the 
``electron displacement'' metric introduced by Adamo and co-workers,\cite{GuiCorAda14} and one that avoids the incorrect normalization 
in Eq.~\eqref{eq:Lambda} and is thus rigorously invariant to orbital rotations even when the TDA is not invoked.   To the best of our 
knowledge, $d_\text{CD}$ is introduced here for the first time but we suggest that $d^-_\text{elec/hole}$ and $d_\text{CD}$
should replace alternative CT metrics that serve essentially the same purpose.


\begin{figure}[t]
\centering
\fig{1.0}{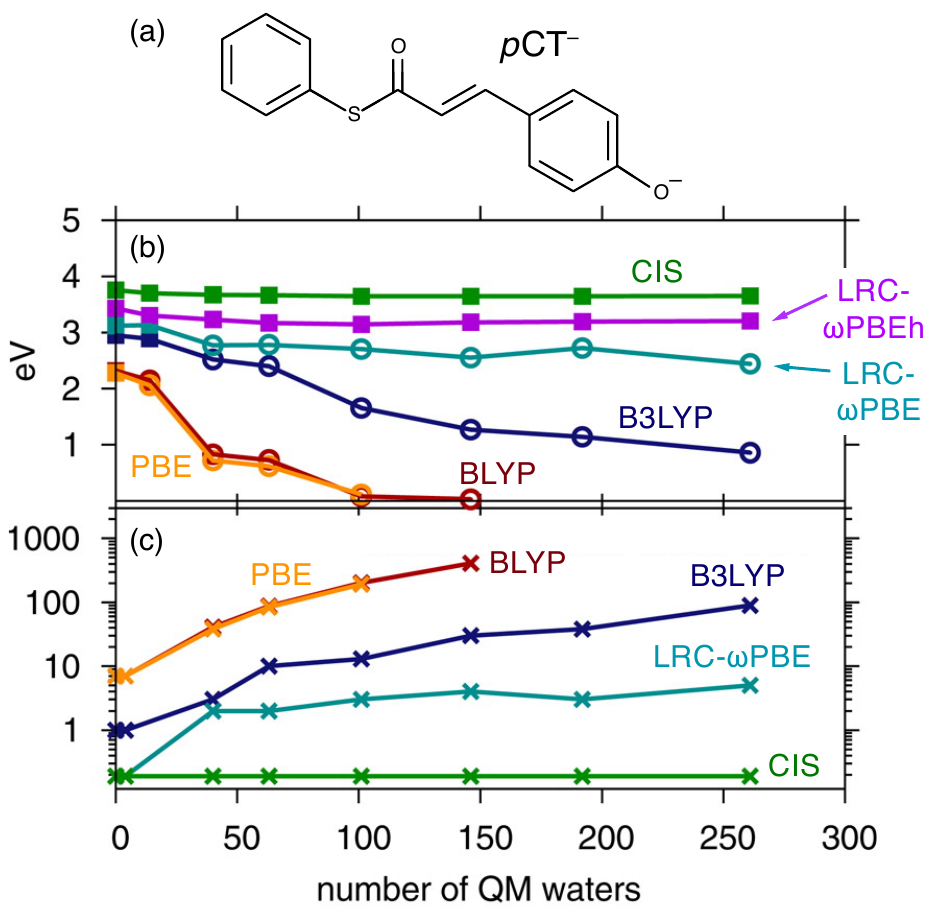}
\caption{
	(a) The chromophore \textit{trans}-thiophenyl-$p$-coumarate 
	($p$CT$^-$), along with (b) a plot of the lowest TDDFT\slash 6\nbd-31G excitation energy
	for $p$CT$^-$(aq) as a function of the number of water molecules included in the calculation, and 
	(c) the number of TDDFT states below 3~eV in this calculation.
	Adapted from Ref.~\protect\citenum{IsbMarCur13}; copyright 2013 American Chemical Society.
}\label{fig:Isborn-pCT}
\end{figure}

To combat the long-range CT problem without going beyond the adiabatic approximation, LRC functionals are used to provide an XC potential
with correct asymptotic behavior for an electron--hole pair. The LRC modification 
was introduced in Eq.~\eqref{eq:LRC} and contains an additional parameter 
that controls the separation between semilocal GGA exchange at short range and nonlocal HFX at long range.
Figure~\ref{fig:Isborn-pCT} shows an example of how these functionals can be used to mitigate the growth in spurious CT states
around a chromophore in aqueous solution.\cite{IsbMarCur13}    Whereas the number of CT states increases extremely rapidly
as water molecules are added around the system, and hybrid functionals such as B3LYP only partially mitigate this increase, the functionals
LRC-$\omega$PBE\cite{RohMarHer09} and 
LRC-$\omega$PBEh\cite{LanHer09} control this growth completely.   The LRC-$\omega$PBEh functional is a short-range 
hybrid with $\cHFX=0.2$ whereas LRC-$\omega$PBE is semilocal at short range ($\cHFX=0$), but both functionals employ 
100\% HFX in the long-range limit.
This should be contrasted with functionals such as CAM-B3LYP,\cite{CAM-B3LYP} which use range separation but sacrifice proper 
asymptotic behavior in an effort to obtain more accurate excitation energies for localized valence transitions.   Although RSH 
functionals such as CAM-B3LYP and $\omega$B97X-D are good choices
in many respects for valence excitations, neither improves the accuracy of LR-TDDFT for CT excitations.\cite{MesKal22}
Standard double-hybrid functionals contain only a fraction of HFX and thus do not improve the situation for CT states,\cite{CasGoe21a} 
unless the LRC scheme employed.\cite{CasDarGoe19} 

In the early development of LRC functionals, the range-separation parameter was often fit to minimize error in some 
benchmark thermochemical or excitation energy data.\cite{IikTsuYan01,VydScu06,SonHirTsu07,RohHer08,RohMarHer09} 
However, excitation energies were found to be quite sensitive to this parameter,\cite{RohHer08,LanRohHer08,LanHer09} 
especially for states where CT character is involved.\cite{LanRohHer08,LanHer09}
More recently, the community has increasingly turned to a more theoretically well-grounded ``optimal tuning'' 
strategy,\cite{SteKroBae09a,SteKroBae09b,BaeLivSal10,Kum17,AlaMorHer20}  which is grounded in the 
ionization energy (IE) theorem of exact DFT.\cite{PerParLev82,ZhaYan00}   That theorem simply states that 
$\text{IE} = -\eHOMO$ for the exact Kohn-Sham functional, consistent with the fact 
that the IE is set by the asymptotic decay of the wave function.\cite{Her15}   This condition is violated badly by common GGA and even hybrid 
functionals, often by several electron volts.\cite{RefBaeKro11,KroSteRef12}  
The optimal tuning (or ``IE tuning'') procedure consists in enforcing this condition for an approximate XC functional, by adjusting the range-separation
parameter $\wRSH$ such that 
\begin{equation}\label{eq:tuning}
	-\eHOMO(N,\wRSH)  = 
	\underbrace{
		E(N-1,\wRSH) - E(N,\wRSH) 
	}_{
		\text{IE}(N,\wRSH)
	} \; .
\end{equation}
Here, IE$(N,\wRSH)$ represents the $\Delta$SCF value of the IE for the $N$-electron molecule, computed using a LRC functional
with range-separation parameter $\mu$.  Alternatively, one might try to find the value of $\mu$ that comes closest to satisfying 
Eq.~\eqref{eq:tuning} for 
both the $N$-electron molecule and its $(N+1)$-electron anion, representing donor and acceptor for electron transfer.   
That procedure has been shown to reproduce not only CT excitation energies but also to afford Kohn-Sham gaps 
($\eLUMO - \eHOMO$) in good agreement with fundamental gaps ($\text{IE} - \text{EA}$).\cite{KroKum18}
The optimally-tuned value of $\wRSH$ does exhibit a strong dependence on system size,
however.\cite{KorSeaSut11,KorBre14,UhlHerCoo14,GraHer21}  
Strategies to mitigate this dependence have been suggested.\cite{ModRajCha13,LaoHer18a,GraHer21}

\subsubsection{Conical Intersections}
\label{sec:TDDFT:Problems:CX}
A different systemic problem with LR-TDDFT, which is relevant in the context of computational photochemistry, is that 
it predicts the wrong topology around any conical intersection that involves the ground state.\cite{LevKoQue06,HerZhaMor16}
The TDHF method suffers from the same deficiency, which is not a DFT artifact
\textit{per se} but rather a LR artifact, arising from an unbalanced description of the ground (reference) state and the excited (response)
states.\cite{HerMan22}  The result is that the branching space around a conical seam that involves the two lowest electronic states is necessarily 
one-dimensional rather than two-dimensional.  (For examples, see Refs.~\citenum{HerZhaMor16} or \citenum{LevKoQue06}.)  
Even the CIS method can exhibit erratic behavior when the ground state becomes quasi-degenerate with the first 
excited state,\cite{ZhaHer14b,ZhaHer21} because in the absence of double excitations the ground- and excited-state eigenvalue problems
are decoupled (according to Brillouin's theorem),\cite{SzaOst82} leading to an unbalanced description.\cite{LevKoQue06,HerMan22}
This is not a problem 
for conical intersections between two excited states because those states are coupled by the matrix $\mathbf{A}$, in both CIS
and LR-TDDFT.

\begin{figure}
\centering
\fig{1.0}{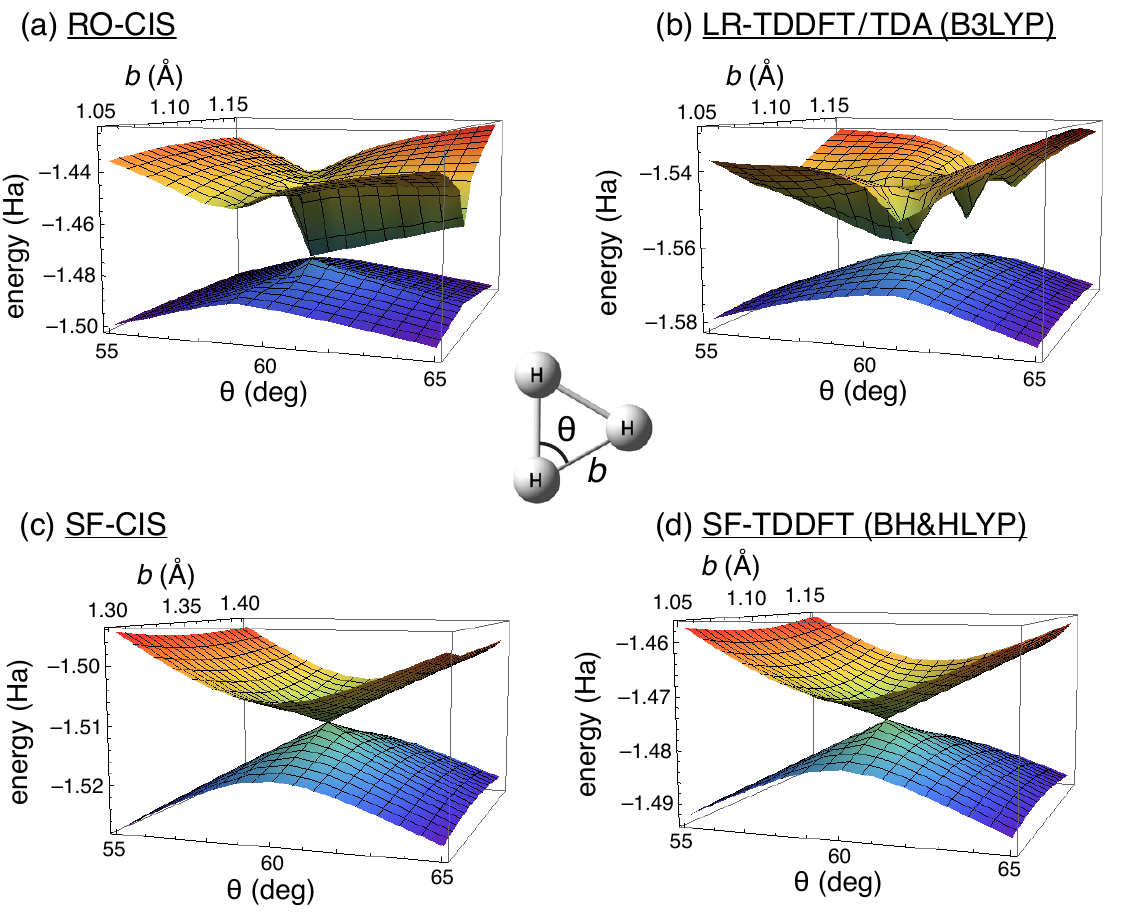}
\caption{
	Potential energy surfaces for the lowest two doublet states of H$_3$ radical
	along a bond-length coordinate $b$ and a bond-angle coordinate
	$\theta$, illustrating Jahn-Teller symmetry lowering $D_{3h}\ra C_{2v}$.   The methods are
	(a) CIS based on a restricted open-shell (RO) reference state,
	(b) LR-TDDFT\slash TDA using unrestricted B3LYP, (c) SF-CIS, and (d) SF-TDDFT using \bhhlyp.
	Reprinted from Ref.~\protect\citenum{ZhaHer14b}; copyright 2014 American Institute of Physics.
}\label{fig:H3-CX}
\end{figure}

An example of a conical intersection involving the ground state is Jahn-Teller symmetry lowering from $D_{3h}$ to $C_{2v}$,
which is illustrated for the H$_3$ radical in Fig.~\ref{fig:H3-CX}.\cite{ZhaHer14b}  In the vicinity of the $D_{3h}$ conical intersection, 
the upper-state potential surface exhibits erratic behavior at both CIS and LR-TDDFT levels of theory.   
This warping of the potential surface around a conical intersection has 
consequences in nonadiabatic molecular dynamics simulations, including SCF convergence 
difficulties\cite{YueLiuZhu18} and incorrect internal conversion timescales.\cite{ZhaHer21}   
As a result, nonadiabatic trajectory surface-hopping calculations based on LR-TDDFT 
should probably be terminated prior to internal conversion to the ground state.\cite{ZhaHer21}

The ``spin-flip'' (SF) variant of LR-TDDFT\cite{ShaHeaKry03} has been suggested as a way to overcome this problem, 
as discussed in detail in Ref.~\citenum{HerMan22}.  Briefly, SF-TDDFT uses a sacrificial reference state that is not the ground state
of interest, but rather a state with higher spin multiplicity $S+1$, for target states with total spin $S$.  By combining
single excitations with a single $\alpha\ra\beta$ spin flip, SF-TDDFT generates both ground and excited states of multiplicity $2S+1$ 
as spin-flipping excitations, meaning that both are obtained as solutions to a common eigenvalue
problem.    This eliminates the imbalance and restores correct topology to conical intersections involving the ground state,
as seen for H$_3$ in Figs.~\ref{fig:H3-CX}(c) and \ref{fig:H3-CX}(d).
Functionals with $\approx 50$\% HFX perform well in the context of SF-TDDFT,\cite{HuiNatIpa10,HerMan22} 
and the Becke ``half-and-half'' functional \bhhlyp\ (with $\cHFX=0.5$) has become the \textit{de facto} standard for SF-TDDFT.\cite{HerMan22}

An unfortunate side effect of SF-TDDFT is that it tends to exacerbate spin contamination,\cite{ZhaHer15b,HerMan22}
especially as one moves away from the Franck-Condon point on the potential surface and starts to enter regions of photochemical
interest.  This necessitates the use of state-tracking algorithms to maintain a consistent spin multiplicity.\cite{HarKeiZah14,ZhaHer15b}
There have been various attempts to find a more theoretically appealing solution to this conundrum by adding additional determinants
to the excitation space in order to restore $\hat{S}^2$ symmetry.\cite{HerMan22}   Methods developed along these lines 
include a fully spin-complete version of 
SF-TDDFT,\cite{ZhaHer15b} which adds the minimal number of additional determinants needed to obtain $\hat{S}^2$ eigenstates
(based on the equation-of-motion formalism),\cite{Row68}
and also a ``mixed-reference'' spin-flip (MRSF) approach, which uses a combination of high-spin and low-spin $S+1$ reference states to generate
target states with spin $S$.\cite{LeeFilLee18,LeeShoFil19,HorLeeFil19,LeeKimNak19,LeeHorFil21,HorSadLee21}   
Although the MRSF-TDDFT excitation manifold is not formally spin-complete, in practice the spin contamination 
is very small.\cite{LeeFilLee18}
The analytic gradient\cite{LeeKimNak19} and nonadiabatic derivative couplings\cite{LeeHorFil21} for MRSF-TDDFT 
have recently been formulated, facilitating nonadiabatic molecular dynamics simulations.

\section{Excited-State Kohn-Sham Theory:  The \texorpdfstring{$\bm\Delta$}{$\Delta$}SCF Approach}
\label{sec:DeltaSCF}
For periodic DFT calculations, LR-TDDFT is theoretically ill-posed if semilocal functionals are used within the 
adiabatic approximation.\cite{HirHeaBar99,Cas09,UllYan16}   
Specifically, the too-rapid asymptotic decay of $\vxcpot{\sigma}(r)$ causes the lowest LR-TDDFT 
excitation energy to collapse to the Kohn-Sham gap, $\hbar\omega = \eLUMO - \eHOMO$.\cite{HirHeaBar99,Cas09}
Semilocal LR-TDDFT also does not produce bound excitons in periodic systems,\cite{UllYan16}
and in large (but finite) conjugated polymers, the exciton delocalization length typically extends to the length of the entire 
molecule.\cite{TreIguChe05,MewPlaDre17}  This observation can be conceptualized as incomplete cancellation of self-interaction
that grows worse with system size, and infinitely worse under periodic boundary conditions.\cite{HirHeaBar99}  Equivalently, it is 
a manifestation of the systematic underestimation of CT energies that was discussed in Section~\ref{sec:TDDFT:Problems:CT}.

In recognition of these and other systemic problems exhibited by LR-TDDFT, 
there has been growing interest in ``$\Delta$SCF'' approaches that attempt to determine excited-state solutions
to the Kohn-Sham SCF equation.\cite{HaiHea21,VanMalLub22}    Having found such a solution, the excitation 
energy is computed simply as the difference relative to the ground-state energy, hence ``$\Delta$SCF".
In contrast to the well-automated machinery of LR-TDDFT, these methods are less ``black-box'', involving more effort and finesse
on the part of the user, because each excited state requires a separate calculation.  On the other hand, 
the $\Delta$SCF approach can exploit ground-state gradient technology for geometry optimizations and vibrational
frequency calculations.\cite{HanGeoBes13} 
For this reason, the $\Delta$SCF procedure has sometimes been called excited-state Kohn-Sham theory.\cite{HanGeoBes13,HaiHea21} 

In cases where LR-TDDFT exhibits known deficiencies, the $\Delta$SCF approach may be more accurate and more reliable even if the formal
justification (based on the Hohenberg-Kohn theorems\cite{KocHol01,Cap06}) is absent because the system is not in its ground state.  
The method therefore rests upon the assumption that the description of short-range dynamical
correlation depends upon the local environment of an electron and can be ported to a ``non-\textit{aufbau}''
solution of the SCF equations, in which an electron has been promoted into a virtual MO.   Such a state does not formally
satisfy the non-interacting $v$-representability requirement of ground-state DFT.\cite{Mor02,Cap06,CasHui12,Mai16}

Excited-state SCF solutions do contain full orbital relaxation, yet these solutions are
inherently unstable because they are saddle points 
rather than local minima in the space of orbital rotations.   Attempts to locate these non-\aufbau\ solutions, each characterized  
by a virtual (empty) level that is lower in energy than the HOMO level, may suffer ``variational collapse" to the
ground state or to a lower-lying SCF solution.  It is up to the user to determine that the SCF solution 
corresponds to the state of interest; if not, then the search must be begun anew, using a different SCF
convergence algorithm or a different initial guess.  Several modified SCF algorithms have been
developed to try to locate non-\textit{aufbau} solutions, based on overlaps with a set of user-specified 
MOs\cite{GilBesGil08,BesGilGil09,BarGilGil18b,CorTakPri22} or else based on direct search.\cite{YeWelRic17,HaiHea20a,CarHer20b}
These algorithms are described in Section~\ref{sec:DeltaSCF:Theory}.
Examples of the $\Delta$SCF methodology in action are presented in Section~\ref{sec:DeltaSCF:Examples}.


\subsection{Theory}
\label{sec:DeltaSCF:Theory}
\subsubsection{General Considerations}
\begin{figure}
\centering
\fig{1.0}{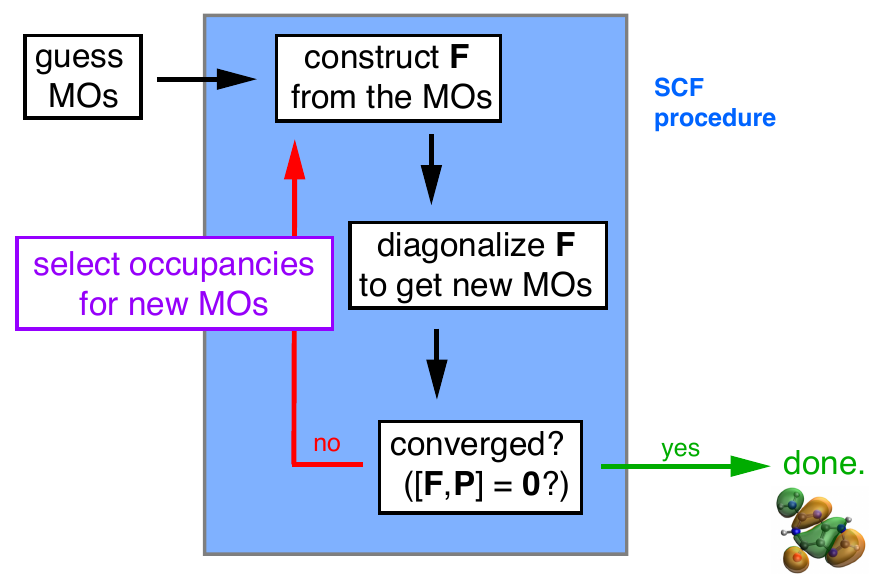}
\caption{
	Flowchart illustration of the SCF algorithm.   In the usual approach, occupancy selection is done according to the \aufbau\ 
	criterion, with the lowest-energy MOs chosen as the occupied set.  For $\Delta$SCF calculations a different choice is required.
}\label{fig:SCF}
\end{figure}

A flowchart of the SCF procedure is illustrated in Fig.~\ref{fig:SCF}.  At each iteration, the occupied MOs $\{\psi_{i\sigma}\}$ 
are used to construct the Fock matrix $\mathbf{F}_\sigma$, which is then diagonalized to obtain new MOs.  Notably, diagonalization results
in $\nbasis = \nocc + \nvir$ MOs and one must decide how to choose the occupied set.  Ordinarily, the lowest eigenvalues $\eval{r\sigma}$ are 
selected (\textit{aufbau} principle), resulting in the ground-state determinant upon SCF convergence.  
To locate an excited-state SCF solution instead, one seeds the procedure with initial-guess MOs from a
ground-state calculation but with non-\aufbau\ occupancies, promoting an electron from HOMO to LUMO, for example. 
This makes the LUMO into an occupied level and the HOMO into a virtual level, resulting in a ``hole'', \ie, a virtual level whose energy lies below
that of the highest occupied level.    When the Fock matrix is constructed from this new set of occupied MOs and then diagonalized, the
question becomes which of the new MOs should be the occupied ones, since energy levels may have shifted.
The SCF procedure therefore deviates from the usual one only when it comes to selecting the occupied subset from among the 
$\nbasis$ MOs.
Several different options have been explored, as discussed in the next section.

Before reviewing algorithms for locating non-\aufbau\ SCF solutions, however, it is important to note some properties of those solutions
that are different from ground-state properties.  First, because the effective Hamiltonian $\hat{F}_\sigma[\{\psi_{i\sigma}\}]$ depends on the MOs themselves,
the ground- and excited-state Slater determinants are eigenfunctions of different Hamiltonians and are therefore not orthogonal.  One 
consequence is that the formula for oscillator strengths in terms of transition dipole matrix elements [Eq.~\eqref{eq:f0n}] is not strictly
valid,\cite{JacHer11c} as that formula is derived using the assumption that the eigenfunctions
of the Hamiltonian form a complete orthonormal set.\cite{McH99}  
In small-molecule tests, however, overlap integrals between ground- and excited-state determinants are found to be $\lesssim 0.1$.\cite{GilBesGil08}

Another general concern is that excited states are always open-shell species, even if the ground state is closed-shell, so any 
single-determinant approximation is certain to be spin-contaminated, perhaps badly so.  Indeed, single-determinant approximations 
for open-shell singlet states are often characterized by $\langle\hat{S}^2\rangle \approx 1$ (in atomic units of $\hbar^2$), which is equal to the average
of pure-state singlet and triplet values.   
 A similar phenomenon occurs, for similar reasons, in the case of the spin-unrestricted Hartree-Fock 
wave function in the separated-atom limit,\cite{SzaOst82} because 
a spin-pure state with two half-filled orbitals can be described using
a minimum of two Slater determinants.   The same two determinants (with different relative signs) are needed to describe both singlet spin-coupling 
(total $S=0$) as well as the $M_S=0$ component of triplet spin-coupling ($S=1$).

In practice, $\Delta$SCF excitation energies for open-shell singlet excited states are often surprisingly accurate despite significant 
spin contamination,\cite{GilBesGil08,BarGilGil18b,BarGilGil18a} 
although there are exceptions.  One such exception is the $^1$B$_{1u}$ state of ethylene, whose underestimation by almost 2~eV is 
attributed to severe spin contamination.\cite{GilBesGil08}    Yamaguchi and co-workers have developed spin-projection 
techniques that can be used to recover spin-pure states in such cases,\cite{KitSaiIto07b,ThoHra19,KitSaiYam18,NaiSchPol08} 
and approximate spin purification is often used as a practical workaround in ``broken-symmetry''
DFT calculations of transition metal complexes.\cite{Dau94,NaiSchPol08,KitSaiYam18}
For an open-shell singlet, the most common approach is to approximate the singlet energy as 
\begin{equation}\label{eq:ASP}
	E_\text{singlet} \approx 2 E_\text{mix} - E_\text{triplet}  \; .
\end{equation}
Here, $E_\text{mix}$ is the energy of the contaminated (mixed-spin or broken-symmetry) state that is obtained in searching for a singlet solution, whereas
$E_\text{triplet}$ is the triplet energy for the same system, for which spin contamination is typically less severe.  This procedure has a 
long history,\cite{ZieRauBae77,Dau94} and Eq.~\eqref{eq:ASP} can be viewed as an approximate form of spin projection, generalizable 
to cases where the target state has spin $S > 0$.\cite{ThoHra19}  The formula in Eq.~\eqref{eq:ASP} is sometimes 
implemented in a self-consistent way, \ie, using Eq.~\eqref{eq:ASP} as the \textit{ansatz} and minimizing with respect to orbital rotations.
That method is known as \textit{restricted open-shell Kohn-Sham} (ROKS) theory,\cite{FraHutMar98,FilSha99a,KowTsuChe13}
and it affords a common set of orbitals for both multiplicites.  More often, however, Eq.~\eqref{eq:ASP} is used as 
an \textit{a posteriori} correction scheme.\cite{KowYosVan11,HanGeoBes13,Bes14,BriBesRob13,BriBes15,HanWriUro15,JiaYuWan15,KowLeIrl16}
Even then, Eq.~\eqref{eq:ASP} can easily be used in geometry optimizations (at the cost of two energy and gradient 
evaluations per step) and in vibrational frequency calculations.\cite{HanGeoBes13}
For the aforementioned $^1$B$_{1u}$ state of C$_2$H$_4$, application of Eq.~\eqref{eq:ASP} reduces the $\Delta$SCF error (as compared to
experiment) from 1.8~eV to 0.3~eV.\cite{GilBesGil08}

\subsubsection{Orbital-Optimized Non-{\em Aufbau} SCF Solutions}
\label{sec:DeltaSCF:Theory:Algorithms}
The simplest means to construct a non-\aufbau\ occupied set is known as the \textit{maximum overlap method} 
(MOM).\cite{GilBesGil08,BesGilGil09,BarGilGil18b,CorTakPri22}   Starting from an initial guess corresponding to non-\aufbau\ occupation of
the ground-state MOs, this approach uses an overlap criterion to identify the new MOs at each subsequent SCF iteration.
To do this, one must compute the projections $p_{r\sigma}$ of the MOs $\psi_{r\sigma}^{(n)}$ at the $n$th iteration onto a reference
set of MOs.   The reference set might be the MOs at the previous iteration, in which case 
\begin{equation}\label{eq:MOM}
	p_{r\sigma} = \bigg(
		\sum_i^\text{occ} \big\langle\psi_{i\sigma}^{(n-1)} \big|\psi_{r\sigma}^{(n)}\big\rangle^2 
	\bigg)^{\!1/2} \; ,
\end{equation}
or else it could be the initial set of ground-state MOs, $\{ \psi_{i\sigma}^{(0)} \}$:
\begin{equation}\label{eq:IMOM}	
	p_{r\sigma} = \bigg(
		\sum_i^\text{occ} \big\langle\psi_{i\sigma}^{(0)} \big|\psi_{r\sigma}^{(n)}\big\rangle^2	
	\bigg)^{\!1/2} \; .
\end{equation}
The first choice [Eq.~\eqref{eq:MOM}] represents the original version of the algorithm,\cite{GilBesGil08} whereas Eq.~\eqref{eq:IMOM} has been
called the ``initial MOM'' (IMOM) algorithm and tends to have better success at converging orbital-relaxed non-\aufbau\ 
states.\cite{BarGilGil18b}    The signature of success is a ``hole below the Fermi level'', \ie, a virtual MO whose energy is lower than
the HOMO energy.

The MOM and IMOM algorithms consist simply in replacing the \aufbau\ selection of occupied MOs with a selection based on
the $\nocc$ largest values of the overlaps $p_{r\sigma}$.   All other aspects of the SCF algorithm remain the same.  
This approach exhibits the same cost per SCF iteration as the ground-state algorithm and when it succeeds, the rate of convergence
(measured in number of SCF cycles) is typically on par with a conventional ground-state calculation.  
There are certainly cases where MOM and IMOM fail,\cite{HaiHea20a,CarHer20b} however, 
typically resulting in variational collapse to the ground-state SCF solution.  In such cases, more robust
SCF convergence algorithms are required.

One such approach is the ``$\sigma$-SCF'' method,\cite{YeWelRic17} which is based on minimizing the functional
\begin{equation}
	\sigma^2_\omega[\Psi] = \langle\Psi|(\omega - \hat{F})^2|\Psi\rangle 
\end{equation}
for a specified energy $\omega$.  This idea stems 
from recognizing that eigenstates $\Hat{F}|\Psi\rangle = \omega|\Psi\rangle$ satisfy the zero-variance condition
$\langle\Hat{F}^2\rangle = \langle\Hat{F}\rangle^2$.
The $\sigma$-SCF approach avoids variational collapse by solving a proper minimization problem, but the appearance
of $\hat{F}^2$ means that four-particle operators are required and the requisite transformations endow this method with  
$\mathcal{O}(\nbasis^5)$ scaling.\cite{YeWelRic17}   This makes the $\sigma$-SCF approach 
much more expensive than conventional SCF theory.

An alternative approach with the same formal scaling as the ground-state SCF problem is 
\textit{squared-gradient minimization} (SGM).\cite{HaiHea20a} Here, the idea is to convert an
inherently unstable saddle-point optimization into a search for a local minimum by optimizing an objective function equal to the 
squared gradient of the energy with respect to orbital rotations.  A local minimum can always be converged (if slowly), 
whereas a saddle point can be missed, and this makes SGM more robust as compared to MOM or IMOM.
While the cost remains $\mathcal{O}(\nbasis^3)$, it is 
2--3 times more expensive per SCF iteration as compared to a conventional SCF calculation, 
due to the cost of constructing the objective function.\cite{HaiHea20a}   It is also known that the squared gradient
$\|\hat{\bm{\nabla}}V(\mathbf{x})\|^2$ of a function $V(\mathbf{x})$ may contain minima that do not correspond to stationary points
of the original function.\cite{AngDiLuo00,AngDiLRuo02,DoyWal02,DoyWal03}
From the standpoint of trying to locate an orbital-relaxed excited-state Slater determinant, these are spurious solutions.

A middle way between MOM and SGM is \textit{state-targeted energy projection} (STEP),\cite{CarHer20b}
which does not increase the cost per SCF iteration yet shows much more robust convergence behavior as compared to MOM or IMOM.
The STEP approach constructs a projection operator onto the virtual space,
\begin{equation}\label{eq:Q}
	\hat{Q}_\sigma = \sum_{a}^\text{vir} |\psi_{a\sigma}\rangle\langle\psi_{a\sigma}| \; ,
\end{equation}
where the summation runs over some or all of the virtual MOs.     The matrix representation of $\hat{Q}_\sigma$ is 
$\mathbf{Q}_\sigma = \mathbf{C}_\sigma\mathbf{C}_\sigma^\dagger$, 
where $\mathbf{C}_\sigma$ consists of column vectors corresponding to whichever
MOs are included in Eq.~\eqref{eq:Q}.  The Fock matrix is then modified according to 
\begin{equation}
	\mathbf{F}'_\sigma = \mathbf{F}_\sigma + \eta\mathbf{SQ}_\sigma\mathbf{S}
\end{equation}
where $\mathbf{S}$ is the atomic orbital (AO) overlap matrix.   The effect of the additional term is to shift all of the orbitals that are included
in Eq.~\eqref{eq:Q} by an energy $\eta$.   By pre-selecting a virtual MO from the ground-state calculation that will be occupied in 
the first iteration of STEP, one can modify the Fock matrix to shift other virtual orbitals (including a lower-energy one that was
occupied in the ground state but whose electron was promoted) to energies above the non-\aufbau\ orbital.   
For example, upon $\text{HOMO}\rightarrow\text{LUMO}$ promotion, the original HOMO is unoccupied and should be included in Eq.~\eqref{eq:Q},
whereas the LUMO becomes occupied and should be excluded from $\hat{Q}_\sigma$.   The STEP algorithm is a form of level-shifting
that tends to ensure that the SCF algorithm converges to the ``closest'' stationary point in the space of MO coefficients, which therefore
resembles the initial guess.\cite{CarHer20b}
Like MOM and IMOM, STEP can be used in conjunction with ground-state gradient technology to perform geometry optimizations
and vibrational frequency calculations.

\subsubsection{Transition Potential Methods}
\label{sec:DeltaSCF:Theory:TP}
The methods described in Section~\ref{sec:DeltaSCF:Theory:Algorithms} involve state-specific orbital optimization, meaning that the 
SCF procedure must be iterated to convergence separately for each excited state of interest.  
This has the advantage of including full orbital relaxation effects (beyond LR theory), but the disadvantage that there
is no guarantee that an excited state resembling the one of interest can actually be found.   A simpler (if cruder) approach was devised
long ago by Slater,\cite{SlaWoo71,Sla72} and forms the basis of several popular techniques for estimating x-ray excitation energies from
Kohn-Sham eigenvalues.\cite{CavOdeNor05,LeeLjuLyu10,FraZhoCor16,ZhaHuaBen16,MicReu19}   

To understand Slater's method, imagine that $E(\{n_i\})$ is the energy of a single-determinant wave function with orbital occupation numbers
$\{n_i\}$, some of which might be fractional.   Expanding the energy as a Taylor series around a reference energy $E_0 = E(\{n_i^0\})$, 
keeping the orbitals fixed, one obtains 
\begin{equation}\label{eq:Slater1}
	E = E_0 + \sum_i (n_i - n_i^0) \frac{\partial E}{\partial n_i}
	+ \frac{1}{2}\sum_{i,j} (n_i - n_i^0)(n_j - n_j^0) \frac{\partial^2 E}{\partial n_i \partial n_j} + \cdots \; .
\end{equation}
According to the Slater-Janak theorem,\cite{Jan78} the first derivative is an orbital eigenvalue: $\eval{i} = \partial E/\partial n_i$.
Now consider promotion of one electron from an occupied MO to a virtual MO.  It suffices to deal with just a pair of occupancies
$(n_i,n_a)$, in terms of which the transition in question can be abbreviated as $(1,0)\ra (0,1)$.    If a fractional-occupancy state with 
$n_i = 1/2 = n_a$ is used for the reference $\{n_i^0\}$, then using Eq.~\eqref{eq:Slater1} to compute the excitation energy 
$\Delta E = E(0,1) - E(1,0)$ leads to an estimate 
\begin{equation}\label{eq:Slater2}
	\Delta E_\text{STM} \approx \eval{a}(\tfrac{1}{2},\tfrac{1}{2}) - \eval{i}(\tfrac{1}{2},\tfrac{1}{2}) \; ,
\end{equation}
where the approximation neglects terms of order $(n_j-n_j^0)^3$.\cite{LeeLjuLyu10}   
This forms the basis of the \textit{Slater transition method} (STM), 
wherein an SCF calculation is carried out for the fractional-occupancy state $(n_i=1/2,n_a=1/2)$ and then 
the energy difference $\Delta E=\eval{a}-\eval{i}$ affords an estimate of the excitation energy.

Variants of STM have historically been popular for x-ray 
spectroscopy,\cite{CavOdeNor05,LeeLjuLyu10,FraZhoCor16,ZhaHuaBen16,MicReu19} 
particularly in the context of periodic DFT calculations where LR-TDDFT with semilocal functionals is problematic.\cite{Cas09}
In principle this method requires a separate SCF calculation for each excitation of interest, and while it is generally 
easy to converge the x-ray ``edge'' in this way (\ie, a core $\ra$ LUMO transition), 
higher-lying states will require a convergence algorithm that can avoid variational collapse.
Moreover, this state-by-state approach leads to nonorthogonal MOs 
and therefore exhibits the same ambiguities regarding oscillator strengths as the 
$\Delta$SCF method.\cite{JacHer11c}  For these reasons, it is common to omit the $1/2$ electron in the virtual space
(with only pragmatic justification), leaving $n_i=1/2$ in the core-excited MO.  This variant of the procedure has
been called the \textit{transition potential} (TP) method.\cite{SteLisDec95,HuCho96,TriPetAgr98a,TriPetAgr98b,MicReu19}
By neglecting to occupy any core-excited virtual states at all, this approach sidesteps the issue of nonorthogonality, at least for a given choice of $n_i$.
Oscillator strengths can be computed in a straightforward way from 
matrix elements $|\langle\Psi_0|\hat{\bm\mu}|\Psi_i^a\rangle|^2$ constructed from the orbitals obtained from the fractional-occupancy
SCF calculation.

Modifications to the formula in Eq.~\eqref{eq:Slater2} 
have also been proposed,\cite{WildeGSom75,Cho95a,MicReu19,HirNakCha21} 
sometimes involving more than one SCF calculation with differing fractional occupancies,
or by combining eigenvalues of both the neutral molecule and its cation or anion.\cite{HirChaSon20,ChaDawNak22a}
These modifications represent attempts to eliminate higher-order errors in Eq.~\eqref{eq:Slater1}.
An example is the ``generalized STM'' (gSTM) method,\cite{WildeGSom75,MicReu19} which replaces Eq.~\eqref{eq:Slater2} with 
\begin{equation}\label{eq:gSTM}
	\Delta E_\text{gSTM} = 
	\frac{1}{4}\big[\eval{a}(1,0) - \eval{i}(1,0)\big]
	+ \frac{3}{4}\big[\eval{a}(\tfrac{1}{3},\tfrac{2}{3}) - \eval{i}(\tfrac{1}{3},\tfrac{2}{3})\big] \; .
\end{equation}
This is based on an alternative approximation for the integral 
\begin{align}
\begin{aligned}
	\Delta E &= \int_1^0 \frac{dE(n_i = \xi, n_a = 1-\xi)}{d\xi} \; d\xi
\\ &
	= \int_1^0 \big[
		\eval{i}(n_i=\xi,n_a=1-\xi) - \eval{a}(n_i=\xi,n_a=1-\xi)
	\big] d\xi \; .
\end{aligned}
\end{align}
The original STM in Eq.~\eqref{eq:Slater2} corresponds to a midpoint approximation for this integral.\cite{WildeGSom75,MicReu19} 
The gSTM approach requires two separate SCF calculations, one with $(n_i = 1, n_a = 0)$ and the other with $(n_i=1/3,n_a=2/3)$.

Variants that set $n_i=0$ (removing the entirety of the core electron) have also been suggested and are sometimes called
``full core hole'' (FCH) methods.\cite{CavOdeNor05,LeeLjuLyu10,FraZhoCor16,ZhaHuaBen16,MicReu19} 
The TP approach is then a ``half core hole'' (HCH) method.   Although the FCH approach deviates significantly from 
Slater's original idea, it can be conceptualized as an attempt to restore charge balance, once the 1/2 electron in the virtual space
has been abandoned for reasons of convenience.     The \textit{excited core hole} (XCH) approach\cite{PreGal06} is yet another 
variant that creates a charge-neutral state (which is important for periodic DFT calculations) by placing the excited
electron in the LUMO and using the full virtual spectrum from that calculation:\cite{PreGal06,MicReu19}
\begin{equation}\label{eq:XCH}
	\Delta E_\text{XCH} = \eval{a}(n_i=0,n^{}_\text{LUMO}=1) - \eval{i}(n_i=0,n^{}_\text{LUMO}=1) \; .
\end{equation}
Together, these STM- and TP-type procedures are known as \textit{occupancy-constrained} $\Delta$SCF methods.  In that context 
there has been some discussion of ``many-electron'' effects on oscillator strengths for 
x-ray transitions.\cite{LiaVinPem17,LiaPre19}   What ``many-electron'' means in this context is multi-determinant
character in the final state, which is of course included automatically in a LR-TDDFT calculation.

\subsection{Examples}
\label{sec:DeltaSCF:Examples}
The primary purpose of this chapter is to survey methods rather than applications but we will highlight a
few recent applications of the $\Delta$SCF approach in order demonstrate that it 
can be an elegant and low-cost alternative in cases where LR-TDDFT performs poorly, such as CT states.\cite{LiuZhaBao17,HaiHea21}  
Whereas LR-TDDFT systematically (and sometimes dramatically)
underestimates CT excitation energies, the same excitation energies are systematically overestimated by the uncorrelated CIS method.\cite{Sub11}
At the CIS level, a long-range excitation uses up the one occupied $\ra$ virtual excitation that is included in the 
\textit{ansatz} and leaves no excitations to facilitate orbital relaxation around either the electron or the hole, hence the overestimation.
LR-TDDFT and CIS may therefore bracket the correct answer for a CT state but these upper and lower bounds can be several electron volts
apart!\cite{LanHer09}  The $\Delta$SCF approach includes full orbital relaxation and is also less sensitive to the asymptotic behavior of the XC potential.
There has also been some preliminary work on the description of conical intersections and nonadiabatic dynamics using $\Delta$SCF 
methods.\cite{VanMalLub22,KumLub22}

\begin{figure}
\centering
\fig{1.0}{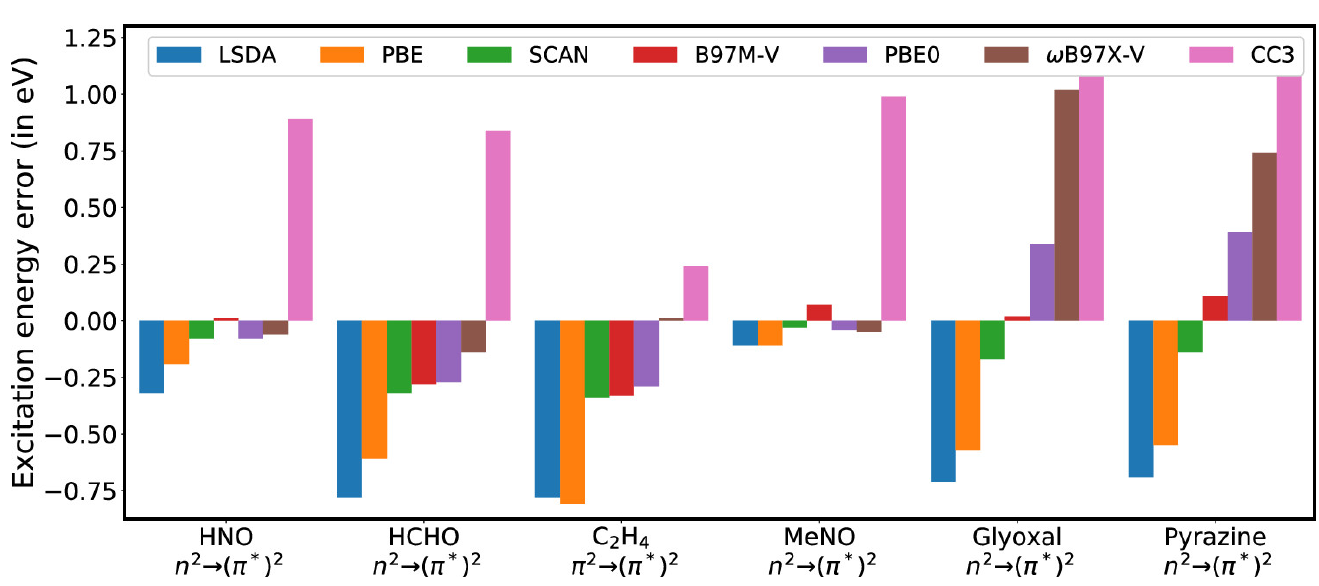}
\caption{
	Errors in doubly-excited states at the $\Delta$SCF\slash aug-cc-pVTZ level, versus benchmarks from 
	Ref.~\protect\citenum{LooBogSce19}.   CC3 values are also provided, for comparison.
	Reproduced from Ref.~\protect\citenum{HaiHea21}; copyright 2021 American Chemical Society.
}\label{fig:2x}
\end{figure}

States with double-excitation character represent another categorical failure of LR-TDDFT within the adiabatic approximation,\cite{EllGolCan11}  
with the most famous example being 
the optically-dark S$_1$(2$^1$A$_g^-$) state in carotenoids,\cite{BalAbrPol16,LlaPasRob17,HasUraYuk18} or the analogous 
2$^1$A$_g^-$ state in butadiene and other conjugated polyenes.\cite{SerMerNeb93,NakNakHir98,CavZhaMai04,StaWorSch06,WatCha12}
Doubly-excited states can be captured accurately using $\Delta$SCF methods,\cite{HaiHea20a,CarHer20b,HaiHea21}
as shown for a few examples in Fig.~\ref{fig:2x}.  For these challenging cases, taken from a benchmark data set of double excitations,\cite{LooBogSce19} 
several mGGA and hybrid functionals prove to be significantly more accurate than the CC3 method, which includes 
triple excitations and is generally close to CCSD(T) in quality,\cite{CC3} with similar scaling.\cite{PauMyhKoc21}
For the full data set from Ref.~\citenum{LooBogSce19}, the hybrid GGA functional $\omega$B97X-V achieves a mean absolute
error (MAE) of 0.6~eV and a maximum error of 1.1~eV, whereas for CC3 the MAE is 1.0~eV and the maximum error is 1.8~eV.\cite{CarHer20b}
The mGGA functional B97M-V does even better, with a MAE of 0.15~eV and a maximum error of 0.46~eV.\cite{CarHer20b}


\begin{figure}[ht]
\centering
\fig{1.0}{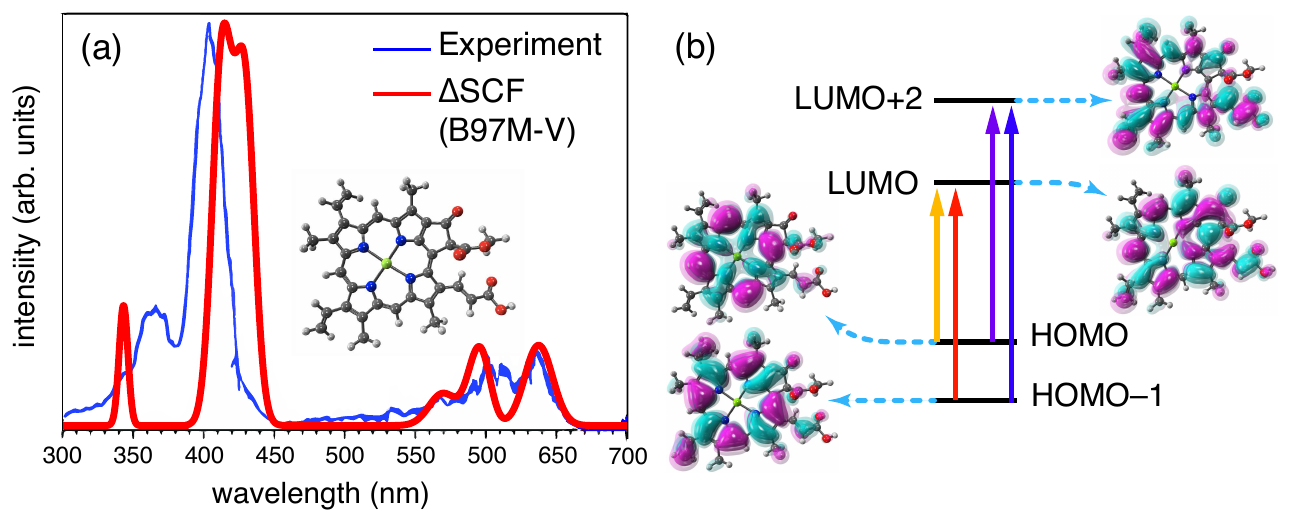}
\caption{
	(a) Absorption spectrum of chlorophyll~$a$ computed via STEP-based $\Delta$SCF calculations at the B97M-V\slash def2-TZVP
	level, spin-purified according to Eq.~\protect\eqref{eq:ASP}  
	and superimposed on a gas-phase experimental spectrum from Ref.~\protect\citenum{StoMusKja15}.
	(b) Pictorial representation of Gouterman's four-orbital model.
	Reprinted from Ref.~\protect\citenum{CarHer20b}; copyright 2020 American Chemical Society.
}\label{fig:Chl}
\end{figure}

The $\Delta$SCF methodology can also be used to compute an electronic absorption spectrum, although this must be done one state at a time
by converging a sequence of non-\aufbau\ determinants representing each excited state, and there is 
no guarantee that some states are not accidentally omitted.  A successful example is shown in Fig.~\ref{fig:Chl},
reproducing the absorption spectrum of the chlorophyll~$a$ molecule that was only recently
measured in the gas phase.\cite{MilTokRub15,StoMusKja15}    Using a STEP-based $\Delta$SCF procedure, the major peaks in that spectrum can be identified 
with transitions amongst the frontier MOs,\cite{CarHer20b} confirming the basic picture of Gouterman's four-orbital model.\cite{GouWagSny63}   
LR-TDDFT calculations of the same molecule require twice as many states in order to 
resolve the spectrum up to 300~nm.  Many of these states have near-zero oscillator strengths,\cite{StoMusKja15} 
suggesting possible contamination by spurious CT states.


\begin{table}[t]
\centering
\caption{
	Error statistics (in eV) for ROKS calculations of core-level excitation energies, including relativistic corrections.$^a$
}\label{table:xray1}
\begin{threeparttable}
\begin{tabular}{l .c c .c }
\hline\hline
	\tworow{Functional~~~~} & \mc{2}{c}{K edge$^b$} && \mc{2}{c}{L$_{2,3}$ edges$^c$}  
\\ \cline{2-3}\cline{5-6}
	& \mc{1}{c}{mean} & RMSE && \mc{1}{c}{mean} & RMSE  
\\ \hline
	LDA				& -4.3	& 4.4 \\
	PBE				& -0.9	& 0.9 \\
	B97M-V			&   1.8	& 1.8 \\
	SCAN			&    0.1	& 0.2 	&&  0.1	& 0.2 \\
	PBE0			& -0.6	& 0.6 \\
	$\omega$B97X-V 	&  0.3	& 0.4 	&& -0.2	& 0.4 \\
\hline\hline	
\end{tabular}
\begin{tablenotes}[flushleft]\fns \item
	$^a$Data from Ref.~\protect\citenum{HaiHea20b}, using the SGM algorithm and aug-cc-pCVTZ basis set.
	$^b$Data set includes 40 transitions for C, N, O, and F atoms.  Error is defined with respect to experiment,
	with atom-specific scalar relativistic effects included in the calculation.
	$^c$Including spin-orbit effects.
\end{tablenotes}
\end{threeparttable}
\end{table}

Core--valence excitation energies are fertile ground for $\Delta$SCF techniques.  These states appear at 
photon energies $\hbar\omega > 200$~eV and therefore it is not feasible to reach them by iterative solution of an eigenvalue 
problem starting from the lowest excitation energies.   The frozen-valence approximation is one way to reach these states in LR-TDDFT,
which is very accurate for K-edge transitions\cite{HerFra20} but may be questionable for L- or M-edge excitations.   Fortunately, 
core-to-LUMO excitations are relatively easy to locate using MOM.\cite{BesGilGil09}  
Table~\ref{table:xray1} shows some error
statistics for a benchmark set of K- and L-edge transitions.\cite{HaiHea20b}  Except for the LDA functional, all of the errors are $<2$~eV and 
several functionals achieve errors $<0.5$~eV, in excitation energies that are hundreds of electron volts.   Notably, some of these same functionals
also afford accurate L-edge transition energies, if spin-orbit interactions are included in order to describe the splitting of
the 2p subshell into 2p$_{1/2}$ and 2p$_{3/2}$ states.\cite{KasSteJen20} 
This splitting can be quite large, \eg, $\approx 13$~eV for Fe(II).\cite{GupSen75,YamHay08} 
Errors of $<0.5$~eV are also possible for heavier elements using ROKS with relativistic corrections.\cite{CunHaiKan22}

This excellent performance is perhaps somewhat surprising due to the substantially different
self-interaction errors in core versus valence orbitals.\cite{ImaNak07,ImaNak12}   The ``optimal tuning'' of LRC functionals that was 
described in Section~\ref{sec:TDDFT:Problems:CT}, in which the range-separation parameter is adjusted to set $\eHOMO =-\text{IE}$, 
can be understood as an attempt to cancel the self-interaction error associated with the HOMO, but that is likely to leave residual self-interaction 
in the much more compact core orbitals.  These errors are exposed in $\Delta$SCF calculations of core-level electron binding energies
(for x-ray photoelectron spectroscopy), where many functionals afford errors $\gtrsim10$~eV for transition metals.\cite{JorXieMor22} 
Even the SCAN functional, which performs well for core-excited states of second-row atoms (Table~\ref{table:xray1}), affords errors of
$\sim1$~eV for core-level binding energies.\cite{KahLis22}    In several cases, Hartree-Fock theory proves to be more accurate than standard functionals that include
correlation, even upon accounting for relativistic corrections.\cite{BelBagIll15,JorXieMor22}    
This is consistent with other results indicating that the restricted open-shell (RO\nbd-)CIS method
is a reasonable level of theory for M- and L-edge spectra of solid-state transition metal oxides, despite its lack of correlation effects,
provided that spin-orbit corrections are included.\cite{KubVerVur18}   
Resolution of this apparent paradox remains an open question.

\begin{table}[t]
\centering
\caption{
	K-edge excitation energies (in eV) computed in various ways.$^a$
}\label{table:xray2}
\begin{threeparttable}
\begin{tabular}{ll cc c ..}
\hline\hline
	\tworow{Method} & \tworow{Functional} & \mc{2}{c}{Molecule} && \mc{2}{c}{Error$^c$} \\ \cline{3-4}\cline{6-7}
	&& HF & CH$_4$ && \mc{1}{c}{HF} & \mc{1}{c}{CH$_4$} 
\\ \hline
	LR-TDDFT			& SCAN				& 666.1	& 273.8 	&& -21.3 	& -14.2 	\\
	$\Delta$SCF +ASP$^b$	& SCAN				& 687.1	& 287.9 	&&  -0.3	&  -0.1	\\
	ROKS				& SCAN				& 687.0	& 288.0 	&& -0.4	&   0.0	\\
	LR-TDDFT			& $\omega$B97X-V 	& 668.7	& 276.5 	&& -18.7	&  -11.5	\\
	$\Delta$SCF +ASP$^b$	& $\omega$B97X-V		& 687.2	& 288.5 	&& -0.2 	& 0.5	\\
	ROKS				& $\omega$B97X-V		& 687.1	& 288.5 	&& -0.3	& 0.5	\\
	Experiment 			&					& 687.4	& 288.0 	&& \mbox{---}	& \mbox{---}		\\
\hline\hline	
\end{tabular}
\begin{tablenotes}[flushleft]\fns \item 
	$^a$Data are from Ref.~\protect\citenum{HaiHea20b}, aug-cc-pCVTZ basis set.
	$^b$Using approximate spin projection (ASP), Eq.~\protect\eqref{eq:ASP}.
	$^c$With respect to experiment.
\end{tablenotes}
\end{threeparttable}
\end{table}

For second-row atoms, errors in both core-level IEs\cite{BesGilGil09,BelBagIll15,BelSaiVin16} 
and also core-level excitation energies\cite{BesGilGil09,HaiHea20b,CarHer20b} 
are comparatively small when using the $\Delta$SCF approach, although even for these elements Hartree-Fock theory is competitive with 
DFT,\cite{BelSaiVin16} suggesting that orbital relaxation is much more important than correlation.
These rather small errors should be contrasted with much larger ones encountered when LR-TDDFT is applied to the same
states using the frozen-valence approximation.   Table~\ref{table:xray2} shows results for two different molecules (HF and CH$_4$) 
using two different functionals (SCAN and $\omega$B97X-V) that both perform well in $\Delta$SCF benchmarks.   In contrast to the sub-eV
errors obtained using the $\Delta$SCF approach, 
LR-TDDFT calculations exhibit errors in excess of 10~eV for the carbon K-edge transition and $\sim20$~eV for the fluorine K-edge
transition.  For K-edge transitions of Mn(II) at $\hbar\omega \sim 6540$~eV, LR-TDDFT errors of 
$\approx 32$~eV are obtained using B3LYP, and errors using the GGA functional BP86 are 
$\approx 62$~eV.\cite{RoeBecDub12}  Errors are even larger for heavier elements.\cite{ChaKowMag18}
Notably, the sign of the errors in Table~\ref{table:xray2}
points to underestimation of the excitation energy, consistent with too-soft asymptotic
decay of the potential for a transition with CT character from a very compact 1s orbital to a radially-diffuse LUMO.  
LRC functionals perform much better in this capacity.\cite{doCHolSla15,FraBruVid21}   

That said, the \textit{precision} of core excitation energies computed using LR-TDDFT 
is rather good even if the accuracy is not,\cite{FraBruVid21} 
meaning that chemical shifts can be obtained even if absolute excitation energies must be shifted to match experiment.  It is the magnitude
of the required shifts that is somewhat unnerving.  This inspired work on RSH functionals that partition $r_{12}^{-1}$ into short\nbd-, 
middle\nbd-, and long-range components,\cite{SonWatNak08,SonWatHir09,BesPeaToz09,BesAsm10} with the intention to use a
larger fraction of exact exchange ($\cHFX\approx 0.87$) at a length scale $\wRSH^{-1} \approx 0.24$~\AA.\cite{CapPenBes13}
These ``short-range corrected'' (SRC) functionals
work rather well for LR-TDDFT x-ray calculations,\cite{BesPeaToz09} although they are  
empirically parameterized specifically for that purpose and may not be good functionals for other applications such as 
ground-state thermochemistry or valence excitation energies. 
Furthermore, the SRC functionals are \textit{not} LRC functionals in the sense of Eq.~\eqref{eq:LRC}, because 
they do not go to a limit of 100\% HFX as $r^{}_{12}\ra \infty$.   Proper LRC functionals have been developed to describe
core-level excitations.\cite{SonWatNak08}

\begin{figure}
\centering
\fig{1.0}{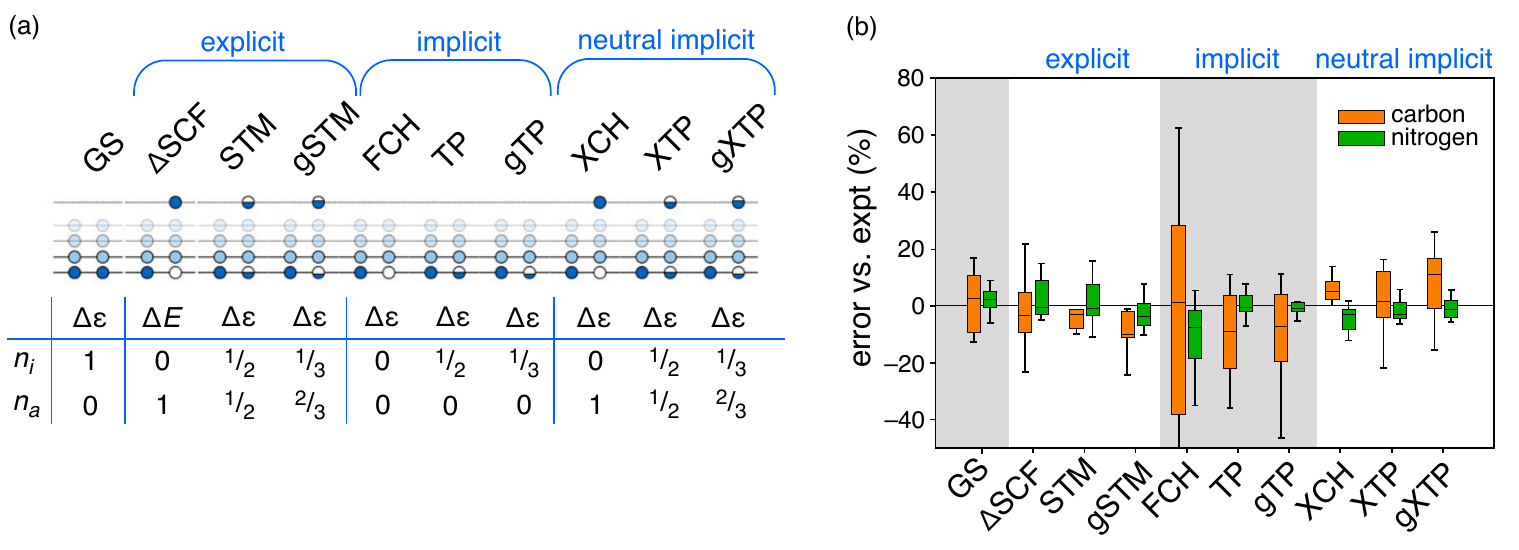}
\caption{
	(a) Pictorial view of TP-type methods based on Kohn-Sham eigenvalues, classified into ``explicit'' methods that 
	require a separate SCF calculation for each excited state, versus ``implicit'' methods that do not.
	(b) Percent error for each method (versus experiment), applying the PBE functional to a data set of K-edge excitation energies.
	Box plots extend from the 
	25th to the 75th error percentile with the median value indicated, while the whiskers show the largest outliers.
	Adapted from Ref.~\protect\citenum{MicReu19}; copyright 2019 American Institute of Physics.
}\label{fig:MicReu19}
\end{figure}

As illustrated by the porphyrin spectrum in Fig.~\ref{fig:Chl} and other examples,\cite{HaiHea21} accurate electronic absorption spectra
can be computed via $\Delta$SCF calculations that are carried out in a state-by-state manner.  However, this does present something 
of a nuisance as compared to automatic generation of numerous excitation energies in a LR-TDDFT calculation.
Some of the TP or occupancy-constrained $\Delta$SCF approaches that were described in Section~\ref{sec:DeltaSCF:Theory:TP}
bypass this annoyance by using orbital eigenvalues from one or two fractional-occupancy SCF calculations to obtain the entire spectrum
of excitation energies.
These methods are assessed (versus experiment) in Fig.~\ref{fig:MicReu19}, for a data set of K-edge transitions.\cite{MicReu19}
Following the notation of Ref.~\citenum{MicReu19}, these methods are characterized as either ``explicit'' or ``implicit'', with explicit methods
involving occupancy constraints that are applied state-by-state, as suggested by the original Slater method, 
whereas the implicit methods place no electron at all into the virtual space, which makes for a very simple computational scheme.
The ``neutral implicit'' methods follow the paradigm of the XCH approach,\cite{PreGal06} placing an electron (or a fraction of an
electron) into the LUMO only; it is often possible to optimize a core-to-LUMO SCF solution without specialized
algorithms.    Finally, the ``ground state'' results in Fig.~\ref{fig:MicReu19} represent a control experiment in which the ground-state eigenvalue
difference $\eLUMO-\eval{\text{1s}}$ is used to approximate the K-edge transition energy.   That method fares 
surprisingly well, or perhaps the other approaches should be said to fare surprisingly poorly.   If the user desires to avoid state-by-state
optimization of an occupancy-constrained determinant, then the XCH approach would seem to be the best option.

\section{Time-Dependent Kohn-Sham Theory:  ``Real-Time" TDDFT}
\label{sec:TDKS}
The cost of the $\Delta$SCF approaches described in Section~\ref{sec:DeltaSCF} is typically no more than 
a few times the cost of ground-state DFT, depending 
somewhat on the algorithm that is used to converge the non-\aufbau\ SCF solution.   With the exception of the $\sigma$-SCF method,
the algorithms described in that section each possess 
the same $\mathcal{O}(\nbasis^3)$ scaling as the ground-state calculation.  However, these methods must be applied 
in a state-by-state manner, constructing a different initial guess for each state of interest, which limits
their applicability to problems where only a small number of states is desired or required.  In contrast, using LR-TDDFT it is possible
to obtain a large number of states in an automated fashion, at least for medium-sized molecules.  For large molecules, the cost 
of computing a very large number of excited states can become prohibitively expensive, especially in systems where the 
density of states is large. The appearance of spurious CT states in the spectrum exacerbates the cost, even if their
affect on the overall spectral envelope is nil.\cite{LanHer07,CarMunHer21}    In situations such as these, where proliferation of states
(whether real or spurious) makes the iterative diagonalization cost-prohibitive, the TDKS approach may be advantageous.   
Using this method, a broadband spectrum can be computed from the oscillating dipole moment function obtained from time-dependent
electron dynamics.

\subsection{Theory}
\label{sec:TDKS:Theory}

The TDKS approach is also known as ``real-time'' (RT\nbd-)TDDFT,\cite{ProIsb16,LiGovIsb20} to distinguish it from LR-TDDFT.  
Starting from the ground-state Kohn-Sham determinant, an external electric field $\bm{\mathcal{E}}(\br,t)$ is turned on at $t=0$, either as an impulse
or as a continuous wave, and the resulting perturbation creates a time-evolving superposition state whose Fourier components encode
the excitation energies.   This is analogous to propagation of a non-stationary wave packet according to the time-dependent Schr\"odinger equation.
Similar to that situation, the time-dependent MOs $\psi_{k\sigma}(\br,t)$ are complex-valued for $t>0$.  
Unlike the many-electron time-dependent Schr\"odinger equation, the effective Hamiltonian $\hat{F}$ in Kohn-Sham theory depends 
on its own time-evolving eigenfunctions, and $\hat{F}(t)$
does not commute with $\hat{F}(t')$ for $t \neq t'$.   Therefore the time evolution operator
\begin{equation}
	\hat{U}(t_2,t_1) = \exp\left[
		\frac{-\cmplxi(t_2-t_1)}{\hbar} \hat{H}
	\right]
\end{equation}
for time-independent Hamiltonian $\hat{H}$ must be generalized to
\begin{equation}
	\hat{U}(t_2,t_1) = \hat{\cal T}\exp\left[
		-\cmplxi\int_{t_1}^{t_2} \hat{F}(t) \; dt
	\right]
\end{equation}
for TDKS calculations, 
where $\hat{\cal T}$ is a time-ordering operator.  This leads to so-called Magnus expansion that generalizes the time-independent
Baker-Campbell-Hausdorff expansion.\cite{CasIse06,BlaCasOte09,ZhuHer18}   

A variety of time-propagation algorithms have been developed for TDKS 
simulations,\cite{CasMarRub04,LiSmiMar05,WilGoiLi16,PueMarRub18,ZhuHer18,LiGovIsb20} the simplest of which is a 
``modified-midpoint'' algorithm,\cite{LiSmiMar05}  a type of explicit Euler integration scheme 
that requires only one Fock matrix construction per time step.   Denoting the Fock and density matrices in the orthonormal
MO basis as $\mathbf{F}$ and $\mathbf{P}$, respectively, the Fock matrix $\mathbf{F}$ is first propagated forward in time by a half step,
$(\Delta t)/2$.  At each subsequent instant in time, $t_n = n\Delta t$, the matrix representation of the propagator is constructed according to
\begin{equation}\label{eq:U-MMUT}
	\mathbf{U}_n = \exp[-\cmplxi (\Delta t) \mathbf{F}_{n+1/2}] \; ,
\end{equation}
again in the MO basis.\cite{ZhuHer18}   The MO density matrix is then propagated from $t_{n-1/2}$ to $t_{n+1/2}$ according to
\begin{equation}\label{eq:P-MMUT}
	\mathbf{P}_{n+1/2} = \mathbf{U}_n \; \mathbf{P}_{n-1/2}\; \mathbf{U}_n^\dagger \; .
\end{equation}
(Spin indices are omitted as these equations are valid for either spin.)
Construction of $\mathbf{U}_n$ requires diagonalization of the Fock matrix, which is not a problem in Gaussian basis sets but is not feasible
in plane-wave basis sets or on a grid.   See Ref.~\citenum{CasMarRub04} for a discussion of alternatives
when $\mathbf{F}$ is too large to diagonalize.   For Gaussian basis calculations, 
self-consistent propagators based on predictor--corrector algorithms have also been developed.\cite{ZhuHer18,YeWanZha22} 
These may require more than one Fock build per time step but allow for the use
of somewhat larger time steps as well as for automatic detection of time steps that are too long, which is not always obvious from the 
usual criterion of checking to make sure that fluctuations in the total energy are bounded.\cite{ZhuHer18} 

For the modified-midpoint algorithm that is encapsulated by Eqs.~\eqref{eq:U-MMUT} and \eqref{eq:P-MMUT}, the cost 
of a single time step is comparable to the cost of a single SCF cycle of the ground-state calculation.   The storage requirement is also modest, 
amounting to a few complex-valued matrices of dimension $\nbasis\times\nbasis$.   This should be contrasted with the storage requirement 
for iterative solution of the LR-TDDFT pseudo-eigenvalue problem, which is $\mathcal{O}(\nroots\nocc\nvir)$ with a prefactor that reflects the
number of iterations and therefore the size of the iterative subspace.  The modified-midpoint
approach works well provided that the time step $\Delta t$ is sufficiently small; values ranging from 0.01--0.50~a.u.\ are typical,
where $\text{1~a.u.} \approx 2.42 \times 10^{-17}$~s = 24.2~attoseconds.   (The time for one orbit in the Bohr model of the hydrogen atom
is $2\pi$ times this value, which establishes a timescale for electronic motion.)   The maximum acceptable value of $\Delta t$ is limited not only by stability of
the time integration but also by the excitation energies that one desires to access, as discussed below.

Unlike LR-TDDFT, which operates by definition in the limit of a vanishingly weak external field, the TDKS approach is non-perturbative
and in principle can be used to simulate electron dynamics in strong laser fields, \eg, to simulate nonlinear optical properties of 
materials,\cite{YabSugShi12}
or to make contact with emerging attosecond spectroscopies\cite{RamLeoNeu16} 
that create electronic wave packets that are out of equilibrium with the nuclei and thus outside of the Born-Oppenheimer
approximation.\cite{Lep12,UllBan12,NisDecCal17,PalMar20}    In practice, there are various issues related to the use of the adiabatic
approximation (Section~\ref{sec:TDDFT:Formalism:AA}),\cite{FukEllRub13,LuoFukMai16,Mai17} 
meaning the use of ground-state functionals with no memory, such that the time dependence
is carried solely by the time-evolving density, $E_\xc[\rho(\br,t)]$.  On the other hand, within the adiabatic approximation the
initial-state dependence vanishes since the XC kernel is fully specified in terms of the instantaneous time-evolving density.\cite{FisCraGov15}
The topic of strong-field electron dynamics and how it can be described using TDKS calculations is not considered 
here, except to note that there have been successful TDKS simulations of strong-field 
photoionization,\cite{KliSaaKla09,KarPabChe14,KraSch14,KraSonSch14,KraSch15a,KraSch15b,
HoeSch17a,HoeSch17b,HoeSch18,HoeLiSch20,LeeHoeLi20,HoeLiSch21}
and also of high harmonic 
generation,\cite{LupHea13,WhiHeiSaa16,CocMusLab16,LabZapCoc18,CocLup19,PauCocLup21,WitParBou21,CocLup22,ZhuHer22}
both in Gaussian-orbital representations of the density.
Unlike the grid-based treatments that are common in atomic physics, Gaussian-based methods are scalable to molecules. 
However, self-interaction error is known to significantly suppress strong-field ionization rates,\cite{KawNakYab09} 
therefore much of the aforementioned work has been performed at the TD-CIS level, 
where self-interaction is not a concern and the exchange potential has the correct asymptotic form.

This section focuses on the use of TDKS simulations to obtain broadband spectra.   Within the electric dipole approximation
(which is also invoked in LR-TDDFT insofar as oscillator strengths are proportional to transition dipole matrix elements), the absorption
spectrum corresponds to the dipole strength function\cite{TusGovLop15,ZhuAlaHer21}
\begin{equation}\label{eq:S(w)}
	S(\omega) = \left(\frac{4\pi\omega}{3c}\right)\mathfrak{Im}\big[
		\alpha_{xx}(\omega) + \alpha_{yy}(\omega) + \alpha_{zz}(\omega)
	\big]
\end{equation}
where for example 
\begin{equation}\label{eq:alpha}
	\alpha_{xy}(\omega) = 
	\frac{\partial \mu_x(\omega)}{\partial \mathcal{E}_y(\omega)}
\end{equation}
is an element of the dynamic polarizability tensor, $\bm{\alpha}(\omega)$.
The quantities $\mu_x(\omega)$ and $\mathcal{E}_y(\omega)$ are the Fourier transforms of the time-dependent
dipole moment and the external electric field, respectively, although for an impulsive $\delta$-function pulse the denominator in 
Eq.~\eqref{eq:alpha} can be replaced by the field amplitude while the numerator is replaced by 
\begin{equation}\label{eq:mu(omega)}
	\mu(\omega) = \int_0^\infty w(t) \; \mu(t) \; e^{-\cmplxi \omega t}\; dt \; .
\end{equation}
Here, $w(t)$ is a windowing or padding function, as in standard signal processing.\cite{LiGovIsb20}
To obtain a linear absorption spectrum the external field $\bm{\mathcal{E}}(\br,t)$ should be weak, impulsive, and off-resonant, 
and should contain components in all three
Cartesian directions in order to excite states of all symmetries, effectively averaging over molecular orientations.
With an appropriately chosen integration time step, 
this procedure reproduces the same spectrum that is obtained using LR-TDDFT,
if \textit{all} of the LR-TDDFT excitations within the energy window of interest are included.\cite{TusGovLop15,ZhuHer18}
The TDKS approach affords the entire broadband spectrum from a single Fourier transform [Eq.~\eqref{eq:mu(omega)}] following 
sufficient time propagation, but it is not straightforward to assign the features in the TDKS spectrum 
to transitions between specific MOs.   Techniques to do so have been developed, based on identifying individual
Fourier modes in the dipole moment matrix (expressed in the MO basis)
at a specific transition frequency.\cite{RepKonKad15,KadKonGao15,BruLaMLop16,RosKuiPus17,SinGarLoz18} 
This does require some post-processing and some insight regarding the important MOs. 
Time-dependent generalization of 
the NTO basis (Section~\ref{sec:TDDFT:Application:Visualization}) have also been proposed.\cite{ZhoKan21}

The cost of the time steps needed to propagate the Kohn-Sham MOs in time, and thus to obtain the time-evolving density
and dipole moment function, is comparable to the cost of a single ground-state SCF cycle but many time steps are required.  
A typical simulation time might be 30~fs to obtain a fully-converged spectrum,\cite{ZhuHer18}
but with $\Delta t = 0.1~\text{a.u.} = 2.4$~as (1~as = $10^{-18}$~s),
this represents $> 10^5$ time steps.    Recently, Pad\'e approximant techniques have been introduced in order 
to obtain $\mu(\omega)$ based on a short time series of input data $\mu(t)$.\cite{BruLaMLop16,LiGovIsb20}
Using this approach, spectra that are well-converged (with respect to LR-TDDFT results) can be obtained in $<10$~fs of time propagation and
rough spectra can be obtained with as little as 3--5~fs.\cite{ZhuAlaHer21}   The time step $\Delta t$
dictates the spectral window that can be accessed, via the usual time-energy uncertainty relationship $(\Delta E)(\Delta t) \gtrsim \hbar$
that comes from the Fourier transform.  In practice, the spectrum is only reliably converged up the Nyquist frequency 
$f_\text{Ny}=\pi/(\Delta t)$, and perhaps only a fraction of that value.\cite{ZhuHer18,ZhuAlaHer21}
This implies that especially small time steps are required for x-ray applications.   
A time step $\Delta t = 0.1$~a.u., for example, corresponds to a Nyquist frequency $\hbar f_\text{Ny} = 854$~eV, which is well above the K~edge 
for second-row elements (C, N, O, etc.) but not for third-row elements.  The K~edge for elements Al--Cl lies above 1,500~eV.

\subsection{Examples}
\label{sec:TDKS:Examples}

\begin{figure}[t]
\centering
\fig{1.0}{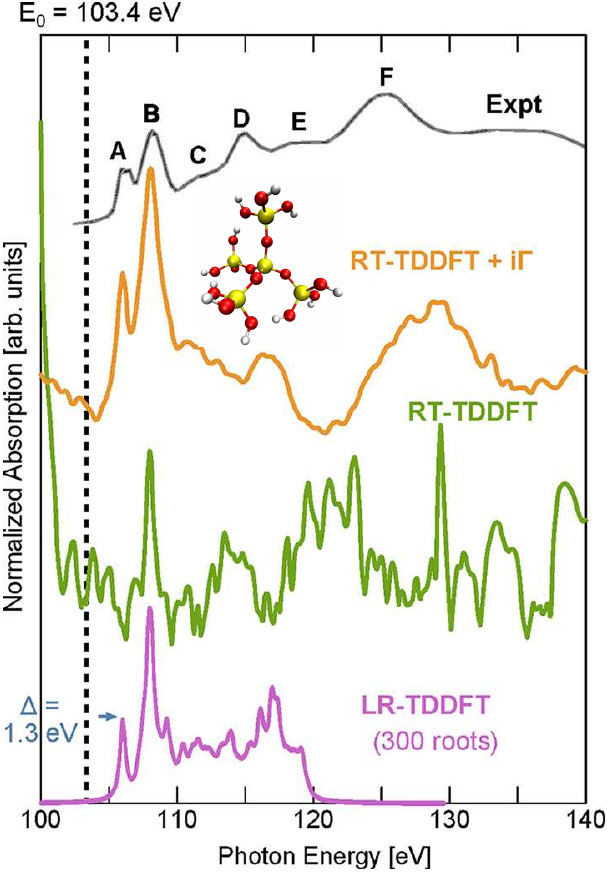}
\caption{
	Experimental x-ray absorption spectra of $\alpha$-quartz at the L$_{2,3}$ edge,\protect\cite{LiBanKas94} 
	along with LR- and RT-TDDFT calculations using a cluster model Si$_5$O$_{16}$H$_{12}$ of the bulk material (as shown), with 
	modified hydrogen charges to enforce charge neutrality.\protect\cite{FerBalLop15}
	The spectrum labeled ``RT-TDDFT + $\cmplxi\Gamma$'' includes
	phenomenological lifetime parameters for the virtual orbitals.
	All calculations were performed with an optimally-tuned version of LRC-$\omega$PBEh.  
	The dashed line labeled $E_0$ is the experimental ionization energy of Si(2p).
	Reproduced from Ref.~\protect\citenum{FerBalLop15}; copyright 2015 American Chemical Society.
}\label{fig:SiO2}
\end{figure}

As an example of a broadband spectrum of interest in materials science, Fig.~\ref{fig:SiO2} presents x-ray spectra
at the L$_{2,3}$ edge of $\alpha$-quartz, computed using several different TDDFT methods.\cite{FerBalLop15}
Starting from the lowest valence excitations, a LR-TDDFT calculation with $\nroots = 300$ reproduces the first two features in the
experimental spectrum (labeled ``A'' and ``B'') but is unable to resolve the higher-energy features.    (The calculations do not include
spin-orbit coupling and thus do not reproduce the doublet for peak~A, which arises due to the 0.6~eV splitting of the 2p$_{3/2}$ and
2p$_{1/2}$ levels.\cite{LiBanKas94})    Although the most intense feature (peak~B) is evident in the RT-TDDFT spectrum, at a peak position 
that precisely matches the corresponding LR-TDDFT spectrum, the RT-TDDFT spectrum is quite noisy and other features in that spectrum are obscured
by this noise.   

That noise is actually a basis-set artifact arising from the absence of proper continuum states (or the inability to describe ionization within a
finite-basis approximation), which has the effect of artificially trapping 
metastable excitations that lie above the ionization threshold.\cite{LopGov13,FerBalLop15}   A solution to this problem
is to incorporate phenomenological lifetimes for the unbound MOs, meaning those with $\eval{p\sigma} > 0$.   This procedure is described in
Ref.~\citenum{LopGov13} and corresponds to a modification $\eval{p\sigma}\rightarrow \eval{p\sigma} + \cmplxi\Gamma$ in the MO basis, 
where $\Gamma^{-1}$ is a phenomenological lifetime that is modeled as a function of energy, decreasing exponentially above the vacuum level.  As seen
in Fig.~\ref{fig:SiO2}, this ``$\cmplxi\Gamma$'' modification removes the noise from the RT-TDDFT spectrum, such that most of the experimental
features become evident even at energies far above what can feasibly be reached with LR-TDDFT.   This does require some phenomenological
modeling, however. 

\begin{figure}
\centering
\fig{1.0}{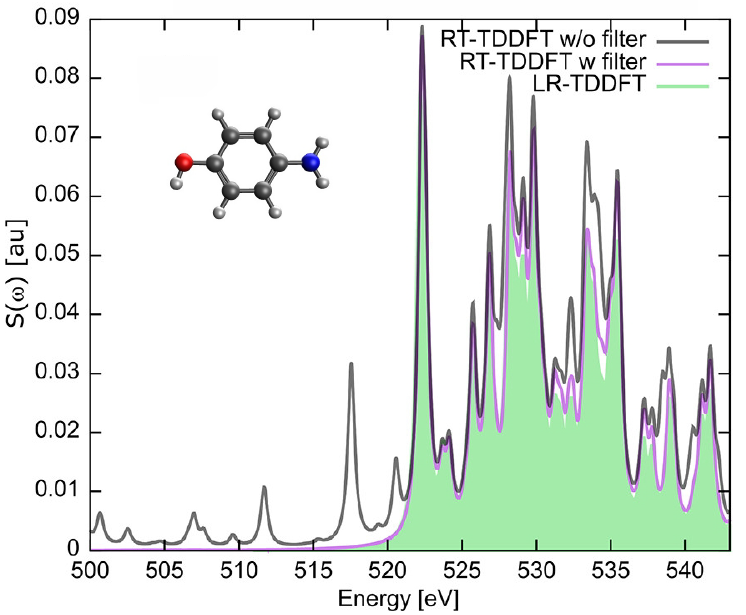}
\caption{
	Absorption spectrum of 4-aminophenol at the oxygen K edge (PBE0\slash def2-TZVP level), illustrating the appearance of pre-edge
	intruder peaks in the RT-TDDFT spectrum that are not present in the LR-TDDFT spectrum.  These can be suppressed by filtering
	the dipole moment function.
	Reproduced from Ref.~\protect\citenum{YanSisChe22}; copyright 2022 American Chemical Society.
}\label{fig:intruders}
\end{figure}

When computing K-edge x-ray spectra using high-quality basis sets, similar artifacts can manifest as spurious pre-edge features
that are not seen in compact basis sets where the density of levels $\eval{p\sigma}$ is more sparse.  
In a molecule that contains both nitrogen and oxygen, for example, 
excitations from N(1s) core orbitals to the highest-energy virtual MOs can manifest as spurious pre-edge features at the 
\textit{oxygen} K-edge,\cite{ZhuAlaHer21,YanSisChe22} as shown in Fig.~\ref{fig:intruders}.
These artifacts appear despite the fact that the nitrogen K-edge lies $>100$~eV below the oxygen K-edge!
These intruder peaks could potentially be mitigated via heuristic lifetime models for the unbound states, as described above, 
although a simpler fix is to modify the time-dependent dipole moment matrix in the MO basis.  That matrix is
\begin{equation}
	D_{jk\sigma,x}(t) = -e \langle\psi_{j\sigma}(t)|  \hat{x} |\psi_{k\sigma}(t)\rangle 
\end{equation}
for the $x$ component.  The time-dependent dipole moment function that is 
needed in Eq.~\eqref{eq:mu(omega)} is then 
\begin{equation}
	\mu_x(t) = \mbox{tr}(\mathbf{P}_{\!\alpha} \mathbf{D}_{\alpha,x} + \mathbf{P}_{\!\beta} \mathbf{D}_{\beta,x} ) \; .
\end{equation}
By eliminating the rows and columns of $\mathbf{D}_{\sigma,x}$ 
that correspond to occupied MOs other than the ones of interest, prior to computing the Fourier transform in Eq.~\eqref{eq:mu(omega)},
the undesired resonances can be removed from the spectrum.\cite{YanSisChe22}    For the oxygen K~edge, this means retaining only those
rows and columns where $j$ and $k$ refer to the O(1s) orbitals.  This is precisely analogous to the frozen-valence 
truncation of the LR-TDDFT excitation manifold that is used to obtain core-level spectra (Section~\ref{sec:TDDFT:Application:Trunc}), 
and may have similar limitations for L- and M-edge spectra.  For the oxygen K~edge, 
Fig.~\ref{fig:intruders} shows that this procedure affords a spectrum 
in good agreement with LR-TDDFT, free of contamination by N(1s) excitations.    This filtering procedure does require 
the user to decide in advance which edges are of interest, so that multiple edges can no longer be computed in a single calculation
unless the entire trajectory of dipole moment matrices is stored.

The TDKS approach has been extended to compute excited-state absorption spectra,\cite{FisCraGov15,FisCraGov16} 
using an excited-state density prepared via LR-TDDFT as the initial density at $t=0$.  Because the initial
state is non-stationary, this requires that the ``field on'' simulation be referenced to a time-evolving ``field-off'' simulation.\cite{FisCraGov15}
This approach has recently been applied to simulate emerging 
transient x-ray experiments,\cite{CavNasZha21,LieHoWea21,LoeLieGov21,LieFoxAnd22} carried out at free-electron laser facilities using
x-ray pulses with femtosecond time resolution.    As an example, it is possible to follow metal-to-metal CT dynamics in the mixed-valence 
$\rm [(CN)_5Fe^{II}CNRu^{III}(NH_3)_5]^-$ compound, which occur on a $\sim60$~fs timescale following excitation at 800~nm, using
time-resolved x-ray emission spectroscopy at the iron K~edge ($\rm 2p_{3/2} \ra 1s$ transition at 7,114~eV).\cite{LieFoxAnd22}

\section*{Acknowledgements}
The author's work on excited-state DFT has been supported for 
many years by the U.S. National Science Foundation, at present under grant no.\ CHE-1955282.

\addcontentsline{toc}{section}{References}

\end{document}